\documentclass[twocolumn,aps,pra,superscriptaddress]{revtex4-1}

\usepackage{lineno}
\usepackage{amsmath,amssymb}
\usepackage{enumitem}
\newcommand\litem[1]{\item{\bfseries #1}}

\usepackage{bm}
\usepackage{bbm}
\usepackage{color}
\usepackage{overpic}
\usepackage{latexsym}
\usepackage{color}
\usepackage[english]{babel}
\usepackage{times}
\usepackage{latexsym}
\usepackage{stmaryrd}
\usepackage{psfrag,graphicx}
\usepackage{epsf}
\usepackage{amsmath}
\usepackage{amssymb}
\usepackage{amsfonts}
\usepackage{natbib}
\usepackage{epstopdf}
\usepackage{appendix}
\usepackage{enumitem}
\usepackage{braket} 				
\usepackage{amsthm}					
\usepackage{soul}
\usepackage[caption=false]{subfig}

\definecolor{mygrey}{gray}{0.35}
\definecolor{myblue}{rgb}{0.2,0.2,0.8}
\definecolor{myzard}{cmyk}{0,0,0.05,0}
\definecolor{mywhite}{rgb}{1,1,1}
\definecolor{myred}{rgb}{0.9,0.1,0.}
\usepackage[colorlinks=true,citecolor=myblue,linkcolor=myblue,urlcolor=myblue]{hyperref}

\newtheoremstyle{customStyle1}  
{0pt}       
{0pt}       
{\normalfont}   
{\parindent}        
{\em}  
{. --}   	 
{.5em}       
{\thmname{#1}\thmnumber{ #2}\thmnote{ (#3)}}  

\newtheorem{theorem}{Theorem}
\newtheorem{lemma}[theorem]{Lemma}
\newtheorem{definition}[theorem]{Definition}

\newtheorem{proposition}[theorem]{Proposition}

\makeatletter
\newcommand\flausr{\@fleqntrue}
\makeatother

\newcommand{\tr}{\operatorname{\bf{tr}}} 

\renewcommand{\vec}[1]{\text{\boldmath$#1$}}

\begin{document}

\title{A resource theory of superposition}

\author{T. Theurer}
\affiliation{Institut f\"{u}r Theoretische Physik, Albert-Einstein-Allee 11,
	Universit\"{a}t Ulm, 89069 Ulm, Germany}
\author{N. Killoran}
\affiliation{Institut f\"{u}r Theoretische Physik, Albert-Einstein-Allee 11,
	Universit\"{a}t Ulm, 89069 Ulm, Germany}
\affiliation{Department of Electrical and Computer Engineering, University of Toronto, Toronto, 
Canada}
\author{D. Egloff}
\affiliation{Institut f\"{u}r Theoretische Physik, Albert-Einstein-Allee 11,
	Universit\"{a}t Ulm, 89069 Ulm, Germany}
\author{M.B. Plenio}
\affiliation{Institut f\"{u}r Theoretische Physik, Albert-Einstein-Allee 11,
	Universit\"{a}t Ulm, 89069 Ulm, Germany}

\date{\today}

\begin{abstract}
	
The superposition principle lies at the heart of many non-classical properties of quantum mechanics. 
Motivated by this, we introduce a rigorous resource theory framework for the quantification of superposition 
of a finite number of linear independent states. This theory is a generalization of resource theories 
of coherence. We determine the general structure of operations which do not create superposition, find 
a fundamental connection to unambiguous state discrimination, and propose several quantitative
superposition measures. Using this theory, we show that trace decreasing operations can be completed for free which, when 
specialised to the theory of coherence, resolves an outstanding open question and is used to address the free probabilistic transformation between pure states. Finally, we prove 
that linearly independent superposition is a necessary and sufficient condition for the faithful creation 
of entanglement in discrete settings, establishing a strong structural connection between our theory 
of superposition and entanglement theory.
\end{abstract}
\maketitle

{{\em Introduction.} -- \phantomsection{}\addcontentsline{toc}{section}{Introduction}}
During the last decades, there has been an increasing interest in quantum technologies.
The main reason for this is the operational advantages of protocols or devices 
working in the quantum regime over those relying on classical physics. Early examples 
include entanglement-based quantum cryptography \cite{ekert1991quantum}, quantum 
dense coding \cite{bennett1992communication} and quantum teleportation \cite{bennett1993teleporting}, 
where entanglement is a resource which is consumed and manipulated. Therefore the 
detection, manipulation and quantification of entanglement was investigated, leading to 
the resource theory of entanglement \cite{plenio2007introduction}. Typical quantum resource 
theories (QRTs) are built by imposing an additional restriction to the laws of quantum 
mechanics \cite{gour2008resource,brandao2008entanglement,horodecki2013quantumness}. In 
the case of entanglement theory, this is the restriction to local operations and classical 
communication (LOCC). From  such a restriction, the two main ingredients of QRTs emerge: 
The free operations and the free states (which are LOCC and separable states in the case 
of entanglement theory). All states which are not free contain the resource under investigation 
and are considered costly. Therefore free operations must transform free states to free 
states, allowing for the resource to be manipulated but not freely created. Once these 
main ingredients are defined, a resource theory investigates the manipulation, detection, 
quantification and usage of the resource.

In principle, not only entanglement but every property of quantum mechanics not present 
in classical physics could lead to an operational advantage \cite{MEKP.16,hillery2016coherence}.
This motivates the considerable interest in the rigorous quantification of non-classicality 
\cite{adesso2016discord,streltsov2016quantum,killoran2016converting,regula2017converting,vogel2014unified,sperling2015convex}.
The superposition principle underlies many non-classical properties of quantum mechanics 
including entanglement or coherence. Recently resource theories of coherence \cite{baumgratz2014quantifying,aberg2006quantifying,streltsov2016quantum} and their 
role in fields as diverse as quantum computation \cite{PhysRevA.93.012111,PhysRevLett.116.160407,MEKP.16}, 
quantum phase discrimination \cite{napoli2016robustness} and quantum thermodynamics 
\cite{PhysRevA.93.052335} attracted considerable attention. In these settings, 
the free states form a finite orthonormal basis of the system under consideration and the 
resource is the superposition of these, called coherence. Here we present a generalization 
of coherence theories and relax the requirement of orthogonality of the free states to linear 
independence. To be precise, we construct a resource theory in which the pure free states are 
a finite linearly independent set and their non-trivial superpositions are resource states. 
Mixed states are free if and only if they can be represented as statistical mixtures of free 
pure states. Thus our framework contains coherence theory as a special case. For obvious 
reasons, we call the free states superposition-free and the resource states superposition 
states.

Such a generalization of coherence theory is interesting for several reasons. Linear independence 
relaxes the convenient but restrictive requirement of orthogonality, yet still provides a fundamental 
framework in which the notion of superposition is unambiguous and self-consistent. From a conceptual 
point of view, our theory helps to clarify the role of orthogonality versus linear independence.
We show that many of the results of coherence theory are just special cases of their counterparts 
in our non-orthogonal setting. This indicates that linearly independent superposition, rather
than the stronger requirement of orthogonality, is a major underlying factor in such quantum 
resource theories. In addition, superposition states can be faithfully converted into 
entanglement, which implies a fundamental connection between entanglement and single-system 
non-classicality \cite{killoran2016converting, regula2017converting}. Thus our resource theory 
can give new insights into the resource theory of entanglement and, vice versa, the faithful 
mapping between these theories allows for an investigation of the controversial notion of 
non-classicality based on the well-founded principles of entanglement \cite{PhysRevLett.94.173602}. As an application, the 
theory presented here can quantify the non-classicality in the superposition of a finite number 
of optical coherent states. This is not possible using the framework of coherence theory, since 
the optical coherent states are not orthogonal. Our theory can thus be seen as a starting point 
for more general resource theories with less restrictive, yet still physically meaningful constraints 
on the free states. Mastering these further generalizations will allow to quantify optical and 
other forms of non-classicality rigorously and to unify their description with entanglement theory 
(see also \cite{vogel2014unified,brandao2008entanglement,tan2017quantifying}).

This Letter is structured as follows. In the next section, we define our free states and operations 
formally. To validate the choice of linear independent free states, we prove that linear 
independence is a necessary and sufficient ingredient for the faithful creation of entanglement, 
completing earlier results from \cite{killoran2016converting}. Then we characterize the free operations 
using the concept of reciprocal states  known from unambiguous state discrimination \cite{chefles1998unambiguous,chefles2000quantum}. This leads to a proof that any
trace-decreasing operation can be completed for free to a trace-preserving operation 
in the theory of superposition and hence in the special case of coherence theory. We proceed to
address the quantification of superposition and propose several measures. For free transformations 
between pure states we show that generically the maximal probability of
success is the solution of a semidefinite program. Finally, we investigate states with
maximal superposition and the operational advantages they allow for, before concluding with a discussion 
on future research directions. Proofs and some additional results are given in the Appendices, including a game in which access to superposition turns certain loss into certain win.


{{\em Basic framework.} -- \phantomsection{}\addcontentsline{toc}{section}{Basic framework}}
In this section, we  give the formal definition of the free states and operations that we consider.
\begin{definition} \label{def:freeStates}
	Let $\{\ket{c_i}\}_{i=1}^{d}$ be a normalized, linear independent and not necessarily orthogonal basis of the Hilbert space represented by $\mathbb{C}^d, d \in \mathbb{N}$. Those basis states are called pure superposition-free states. All density operators $\rho$ of the form
	\begin{align}
	\rho=\sum_{i=1}^{d} \rho_i \ket{c_i} \bra{c_i},
	\end{align}
	where the $\rho_i$ form a probability distribution,
	are called superposition-free. The set of superposition-free density operators is denoted by $\mathcal{F}$ and forms the set of free states.
	All density operators which are not superposition-free are called superposition states and form the set of resource states.
\end{definition}
For $d=1$, the concept of superposition is empty, thus all following results assume $d\ge 2$. In \cite{killoran2016converting,sperling2015convex}, the \emph{classical rank} of a state has been introduced as the minimum number of free states we need to superpose in order to represent the state. We will say that an isometry $\Lambda$ is a faithful conversion operation (to and from entanglement) when the Schmidt rank of $\Lambda\ket{\psi}$ is equal to the classical rank of $\ket{\psi}$.
The relevance of linear independence for our resource theory is based on the following theorem.
\begin{theorem}
	If the free states in a finite dimensional Hilbert space form a countable set, then linear independence of the free states is a necessary and sufficient condition for the existence of a faithful conversion operation.
	In case the free states are a finite set of optical coherent states, the faithful conversion can be implemented by a beam splitter \cite{vogel2014unified}.
\end{theorem}
Sufficiency is proved in an earlier theorem from \cite{killoran2016converting} (and extended in \cite{regula2017converting}). In the Appendices, we prove the converse result, thus completing the original theorem.
\begin{definition}
	\label{def:freeOperations}
	A Kraus operator $K_n$ is called su\-per\-position-free if $K_n\rho K_n^\dagger\in\mathcal{F}$ for all $\rho \in \mathcal{F}$. Quantum operations $\Phi(\rho)$ are called superposition-free if they are trace preserving and can be written as
	\begin{equation}
		\Phi(\rho)=\sum_n K_n \rho K_n^\dagger,
	\end{equation}
	where all $K_n$ are free.
	The set of superposition-free operations forms the free operations and is denoted by $\mathcal{FO}$.
\end{definition}
At this point, let us highlight that the definition of the free
operations is not unique. This is a common trait of QRTs. The
biggest possible class of free operations for our choice of the
free states is given by those quantum operations that map the 
free states onto themselves which are denoted by $\mathcal{MFO}$ 
(maximally superposition-free operations). However, in general, 
these operations do not possess a representation in terms of 
superposition-free Kraus operators.
\begin{proposition} \label{Proposition:SetOfClassicalMaps}
	$\mathcal{MFO}$ is strictly larger than $\mathcal{FO}$. This is also valid in the special case of coherence theory.
\end{proposition}
Hence someone who has access to measurement outcomes of an element of $\mathcal{MFO}$ and can thus do post-selection could conclude that a superposition-free operation generated superposition from a superposition-free state.
Our definition of the free operations guarantees that one cannot create resources for free by obtaining measurement results. On the other hand, it is not as restricted as other definitions demanding for example a free dilation \cite{chitambar2016critical,marvian2016quantify}.
For a discussion of alternative choices, see the Appendices.

{{\em Free operations.} -- \phantomsection{}\addcontentsline{toc}{section}{Free operations}}
In order to describe the general structure of $\mathcal{FO}$, we need to introduce some notation.
Since the pure superposition-free states form a basis of $\mathbb{C}^d, d \in \mathbb{N}$, there exist vectors $\ket{c_i^\perp}, i=1,...,d$ such that
\begin{equation} \label{def d}
	\braket{c_i^\perp|c_j}=\delta_{i,j}, 
\end{equation}
which are not normalized but form a basis as well.
In the context of unambiguous state discrimination, the states one gets by normalizing  $\ket{c_i^\perp}$ are called reciprocal states \cite{chefles1998unambiguous,chefles2000quantum}.
For explicit calculations, it is convenient to express both $\{\ket{c_i}\}_{i=1}^{d}$ and $\{\ket{c_i^\perp}\}_{i=1}^{d}$ with respect to an orthonormal basis $\{\ket{i}\}_{i=1}^{d}$ which will be called computational.
Now we can introduce two linear operators $V$ and $W$ such that $V\ket{i}=\ket{c_i}$ and $W\ket{i}=\ket{c_i^\perp}$.
Notice that both $V$ and $W$ are full rank since they correspond to basis transformations.
From (\ref{def d}), it follows that $\delta_{i,j}=\braket{c_i^\perp |c_j}=\bra{i}W^\dagger V\ket{j}$
and thus $W=(V^\dagger)^{-1}$.
With this notation at hand, the explicit form of a superposition-free Kraus operator can be given, which is done in the following theorem.
\begin{theorem}\label{theoremKrausOperators}
	A Kraus operator $K_n$ is superposition-free if and only if it is of the form
	\begin{align} \label{classicalKrausOperator}
		K_n&=\sum_{k} c_{k,n}\ket{c_{f_n(k)}}\bra{c_k^\perp},
	\end{align}
	where the $c_{k,n}\in\mathbb{C}$ and the $f_n(k)$ are index functions.
\end{theorem}
Incoherent Kraus operators $\tilde{K}_n$ as defined in the limit of coherence theory \cite{baumgratz2014quantifying} are thus given by $\tilde{K}_n=\sum_{k} c_{k,n}\ket{f_n(k)}\bra{k}$ \cite{winter2016operational,yao2015quantum}.
If we choose the incoherent states $\{\ket{k}\}$ as the computational basis, the operator $K_n=V \tilde{K}_n V^{-1}$ has the form of a superposition-free Kraus operator.
In order to have a valid, trace non-increasing quantum operation, we need
\begin{equation}
	\mathbbm{1} \ge \sum_{n} K_n^\dagger K_n =\sum_{n} (V^\dagger)^{-1} \tilde{K}_n^\dagger V^\dagger V \tilde{K}_n V^{-1}.
\end{equation}
If the pure superposition-free states are not orthogonal, $V^\dagger\ne V^{-1}$ and in general it is therefore not possible to transform a trace non-increasing set of incoherent Kraus operators by a basis transformation $V$ into a superposition-free one.

Intuitively, the introduction of additional systems in free states is for free. With the above theorem at hand, we can show that this is indeed the case.
\begin{proposition}\label{prop:easyDecomp}
	If both $\sigma_B$ and all $K_n$ are free,
	the quantum operation $\Phi(\rho_A)=\tr_B\sum_n K_n \rho_A \otimes \sigma_B K_n^\dagger$ is free.
\end{proposition} 
When dealing with trace decreasing operations that can be decomposed into superposition-free 
Kraus operators, the question arises whether they are part of a (trace preserving) 
superposition-free operation. If this was not the case, it would imply that one cannot call the trace decreasing operation free because one disregards a part that can only be done 
in a non-free way \footnote{This can happen in the case of entanglement theory. Let  $\{\ket{\psi_i}=\ket{\phi_i}\otimes\ket{\xi_i}\}_i$ be an unextendable separable product basis \cite{bennett1999unextendible} of a bipartite system and define a trace-decreasing separable operation $\Lambda[\rho]=\sum_i \ket{0}\bra{\phi_i}\otimes\ket{0}\bra{\xi_i} \rho \ket{\phi_i}\bra{0}\otimes\ket{\xi_i}\bra{0}$ where $\ket{0}$ is an arbitrary reference state. This operation cannot be completed by separable Kraus operators by construction.}. This leads us to our first main result.
\begin{theorem} \label{Theorem:CompletionOfMaps}
	Assume we have an (incomplete) set of Kraus operators $\{K_m\}$ such that $\sum_m K_m^\dagger K_m \le \mathbbm{1}$. Then there always exist superposition-free Kraus operators $\{F_n\}$ with $\sum_m K_m^\dagger K_m +\sum_n F_n^\dagger F_n=\mathbbm{1}$.
\end{theorem}
From here on we will call trace-decreasing operations with a decomposition into superposition-free Kraus operators superposition-free as well, since we can always complete them for free. Note that this is also valid in the special case of coherence theory.

{{\em Superposition measures.} -- \phantomsection{}\addcontentsline{toc}{section}{Superposition measures}}
In this section, we address the quantification of superposition, extending the method used in \cite{baumgratz2014quantifying} to quantify coherence.
\begin{definition} \label{C}\allowdisplaybreaks
	A function $M$ mapping all quantum states to the non-negative real numbers is called a superposition measure if it is 
\begin{description}[style=multiline,leftmargin=1.1cm,font=\normalfont,noitemsep,topsep=0pt]
	\item[$(S1)$]  Faithful: $M(\rho)=0 \text{ if and only if } \rho \in \mathcal{F}.$
	\item[$(S2a)$] Monotonic under $\mathcal{FO}$: $M(\rho) \ge M(\Phi(\rho))\ \text{for all } \Phi \in \mathcal{FO}.$
	\item[$(S2b)$]  Monotonic under superposition-free selective measurements on average: $M(\rho) \ge \sum_n p_n M(\rho_n):\ p_n=\mathrm{tr}(K_n \rho K_n^\dagger),\ \rho_n= \left(K_n\rho K_n^\dagger\right)/p_n$ for all $\left\{K_n\right\}:$ $\sum_n K_n^\dagger K_n =\mathbbm{1},\ K_n\mathcal{F}K_n^\dagger \subset \mathcal{F}.$
	\item[$(S3)$] Convex: $\sum_n p_n M(\sigma_n)\ge M\left(\sum_n p_n \sigma_n\right)$ for all $\left\{\sigma_n\right\}, \ p_n\ge 0,\ \sum_n p_n =1.$
\end{description}\par
\noindent If only condition $(S1)$ and  $(S2a)$ or  $(S2b)$ are satisfied, we call $M$ a superposition monotone.
\end{definition}
Property $(S1)$ demands that a state has zero superposition if and only if the state is superposition-free.
As stated in $(S2a)$, the application of a superposition-free operation to a state should not increase its superposition. If one does superposition-free selective measurements, one does not expect the superposition to increase on average which is exactly the point of $(S2b)$. The convexity condition $(S3)$ enforces that mixing states cannot increase the average superposition.
It can be shown easily that $(S2a)$ follows from $(S2b)$ and $(S3)$.
As in coherence theory \cite{baumgratz2014quantifying}, some distance measures $\mathcal{D}$ can be used to define superposition measures and monotones. We define a candidate $M_\mathcal{D}$ by
\begin{equation}
M_\mathcal{D}(\rho)  =\min_{\sigma \in \mathcal{F}} \mathcal{D}(\rho,\sigma).
\end{equation}
If $\mathcal{D}$ is a metric, $M_\mathcal{D}$ fulfills $(S1)$. If it is furthermore contractive under completely positive and trace preserving (CPTP) maps, it fulfills $(S2a)$ \cite{QuantifyingEntanglement,baumgratz2014quantifying} and for $\mathcal{D}$ being jointly convex \cite{wehrl1978general}, the induced $M_\mathcal{D}$ fulfills condition $(S3)$.

In accordance with \cite{sperling2015convex,killoran2016converting,winter2016operational}, we define the superposition rank $r_S(\ket{\psi})$ for a state $\ket{\psi}= \sum_{j} \psi_j \ket{c_j}$ as the number of
$\psi_i \ne 0$.
Assume a state $\ket{\varphi}=\sum_{j} \varphi_j \ket{c_j}$ can be transformed (with some probability $p>0$) to a state $\ket{\xi}=\sum_{j} \xi_j \ket{c_j}$ by $\mathcal{FO}$.
According to proposition \ref{prop:easyDecomp}, this is possible if and only if there exists a superposition-free Kraus operator $K=\sum_{i} c_i \ket{c_{f(i)}}\bra{c_i^\perp}$ with the properties
\begin{align}
\sqrt{p} \sum_{i} \xi_{i} \ket{c_{i}}=\sqrt{p}\ket{\xi}= K \ket{\varphi}= \sum_{i} \varphi_i c_i \ket{c_{f(i)}}
\end{align}
and $K^\dagger K\le \mathbbm{1}$. Hence the number of $\xi_i \ne 0$ is at most as large as the number of $\varphi_i \ne 0$. This proves that the superposition rank cannot increase under the action of a superposition-free Kraus operator. With the definition of the superposition rank at hand, we present some explicit superposition measures.
\begin{proposition}\label{prop:Measures}
	The following functions are superposition measures as defined in Definition \ref{C}.
	
		\noindent 1. The relative entropy of superposition 
		\begin{align}
		M_{\mathrm{rel.ent}}(\rho)&=\min_{\sigma \in \mathcal{F}} S(\rho||\sigma),
		\end{align}
		where $S(\rho||\sigma)=\mathrm{tr}\left[\rho \log \rho\right]-\mathrm{tr}\left[\rho \log \sigma\right]$ denotes the quantum relative entropy. See \cite{baumgratz2014quantifying} for the case of coherence theory.
		
		\noindent2. The $l_1$-measure of superposition 
		\begin{equation}
		M_{l_1} (\rho)=\sum_{i\ne j} |\rho_{ij}|
		\end{equation}
		for $\rho=\sum_{ij} \rho_{ij} \ket{c_i} \bra{c_j}$. See again \cite{baumgratz2014quantifying} for the case of coherence theory.
		
		\noindent3. The rank-measure of superposition 
		\begin{align}
		&M_\text{rank}(\ket{\psi})=\log(r_S(\ket{\psi})), \nonumber \\
		&M_\text{rank}(\rho)=\min_{\rho=\sum_i\lambda_i\ \ket{\psi_i}\bra{\psi_i}}\sum_i \lambda_i M_\text{rank}(\ket{\psi_i}).
		\end{align}
		
		\noindent4. The robustness of superposition 
		\begin{align}
		M_R(\rho)&=\min_{\tau \text{  density matrix}} \left\lbrace s\ge0 : \frac{\rho+s\tau}{1+s}\in\mathcal{F} \right\rbrace.
		\end{align}
		This quantity has an operational interpretation in the limit of coherence theory: the robustness of coherence quantifies the advantage enabled by a quantum state in a phase discrimination task \cite{napoli2016robustness}.
\end{proposition}

{{\em State transformations and resources.} -- \phantomsection{}\addcontentsline{toc}{section}{State transformations and resources}}
In resource theories, it is an important question to which other states a given state can be transformed  under the free operations because this leads to a hierarchy of ``usefulness'' in protocols. Here we consider the transformation between single copies of pure states. Let us first clarify when probabilistic conversions are possible at all.
As already mentioned, there is no possibility to increase the superposition rank
of a pure state by applying a superposition-free Kraus operator. On the other hand, if two states $\ket{\psi}= \sum_{j \in R} \psi_j \ket{c_j}$ and $\ket{\varphi}= \sum_{j\in S} \varphi_j \ket{c_j}$ have the same superposition rank $r=|S|=|R|$, then there exists a superposition-free transformation that transforms one to the other with probability larger than zero.
To see this, interpret $R$ and $S$ as (arbitrarily) ordered indexing sets. Define a function $f$ that maps the $n$-th element of $R$ to the $n$-th element of $S$ and a superposition-free Kraus operator
\begin{align}
    K=\sqrt{p} \sum_{j\in R} \frac{\varphi_{f(j)}}{\psi_j} \ket{c_{f(j)}} \bra{c_j^\perp}.
\end{align}
Hence
$K \ket{\psi}=\sqrt{p} \ket{\varphi}$
and since $\psi_j \ne 0$ for all $j \in R$ and the pure superposition-free states $\{\ket{c_j}\}$ are linear independent, $p$ can always be chosen such that $p > 0$ and $K^\dagger K \le \mathbbm{1}$.
With the help of theorem \ref{Theorem:CompletionOfMaps}, this proves that there exists a probabilistic superposition-free transformation.
Different orderings of $S$ leads to $r!$ different functions $f_n$ and thus Kraus operators $K_n$.
For convenience, we define
\begin{align}
    &F_n= \sum_{j} \frac{\varphi_{f_n(j)}}{\psi_j} \ket{c_{f_n(j)}} \bra{c_j^\perp}
\end{align}
with $F_n\ket{\psi}=\ket{\phi}$ and $K_n=\sqrt{p_n} F_n$.
This allows us to state our second main result:
The optimum free conversion probability between two pure states of the same superposition rank is the solution of the semidefinite program
\begin{alignat}{2} \label{primalProblem}
&\text{maximize} \qquad&& \sum_n p_n \nonumber \\
&\text{subject to} &&\sum_n p_n F_n^\dagger F_n \le \mathbbm{1}, \ p_n\ge0 \quad \text{ for all } n,
\end{alignat}
which can be solved efficiently using numerical algorithms \cite{boyd2004convex,jarre2013optimierung}.
Doing so, our investigations indicate that deterministic superposition-free transformations are rare in the case of non-orthogonal bases. Already for qubits, the probability for the existence of a deterministic transformation between two randomly picked states seems to be zero. For qubits, this is investigated analytically for a specific initial state in  the Appendices.
If we consider superposition-free transformations to a target state with lower superposition rank than the initial state, a probabilistic transformation is still possible by the same arguments. The optimization problem however is more troublesome since we have to include Kraus operators where different pure superposition-free states are mapped to the same superposition-free target state. Therefore the optimization problem is no longer semidefinite.

If a $d$-dimensional superposition state can be used to generate all other $d$-dimensional states deterministically by means of $\mathcal{FO}$, it can be used for all applications. These states are said to have maximal superposition.
This definition is independent of a specific superposition measure and can serve to normalize measures. Such golden units exist in coherence theory for all dimensions \cite{baumgratz2014quantifying}, but only for qubits in our case.
\begin{proposition}\label{Prop:MaximallyNonClassicalQubitState}
	For qubit systems with $\braket{c_1|c_2}\ne 0$, there exists a single state with maximal superposition.
	For higher dimensions, there exists no state with maximal superposition in general.
\end{proposition}
This is different to coherence theory where in dimension $d$, all states of the form $\ket{m_d}=1/d\sum_{n=1}^{d} \exp(i\phi_n)\ket{n}\ (\phi_n \in \mathbb{R})$ are maximally coherent \cite{baumgratz2014quantifying}. A reason for this seems to be that in our more general setting, one loses entire classes of deterministic free transformations, for example diagonal unitaries which change the phases $\phi_n$.

On the other hand, as in coherence theory \cite{baumgratz2014quantifying}, the consumption of a qubit state with maximal superposition allows to implement any unitary qubit gate by means of $\mathcal{FO}$.
\begin{theorem}\label{theorem:UnitaryReali}
	Any unitary operation $U$ on a qubit can be implemented by means of $\mathcal{FO}$ and the consumption of an additional qubit state with maximal superposition $\ket{m_2}$ provided both qubits posses the same superposition-free basis. This means that for every $U$ there exists a fixed $\Psi\in \mathcal{FO}$ independent of $\rho_s$ acting on two qubits such that
	\begin{equation}
	\Psi\left(\rho_s\otimes\ket{m_2}\bra{m_2}\right)=\left(U\rho_s U^\dagger\right)\otimes \rho_h,
	\end{equation}
	where $\rho_h$ is a superposition-free qubit state.
\end{theorem}
This means that consuming enough qubits with maximal superposition, one can perform any unitary and thus any operation \footnote{We can express the unitary operation as a matrix with respect to the orthonormal basis obtained when applying the Gram-Schmidt process on the pure superposition-free states. As shown in \cite{reck1994experimental}, we can then decompose the unitary into unitaries $U_2$ acting on two-dimensional subspaces spanned by two pure free states. With the help of a qubit state with maximal superposition (with respect to the two free states spanning the two-dimensional subspace under consideration) every $U_2$ can be implemented.}

{{\em Conclusions.} -- \phantomsection{}\addcontentsline{toc}{section}{Conclusions}}
We introduced a resource theory of superposition, which is a generalization of coherence theory \cite{baumgratz2014quantifying} and we showed that in a non-continuous setting, this is 
the only generalization that allows for a faithful conversion to entanglement. Using the tools 
of quantum resource theories, we defined superposition-free states and operations. This allowed 
us to prove that several measures are good quantifiers of superposition, in particular the relative 
entropy of superposition and the easy to compute $l_1$ -measure of superposition. We 
also uncovered an important partial order structure for pure superposition states: a state can 
be probabilistically converted to another target state via superposition-free operations only when 
the target has an equal or lower superposition rank. The maximal probability for successful transformations 
between states of the same superposition rank is the solution of a semidefinite program. Contrasting 
with coherence theory, we find that only in two dimensions is there a state with maximal superposition 
content which can be consumed to implement an arbitrary unitary using only free operations.

Our results can help to investigate phenomena such as catalytic transformations \cite{jonathan1999entanglement,du2015conditions,aaberg2014catalytic,duarte2016self,bu2016catalytic}, and act as a starting point for the investigation of mixed state transformations, transformations in the asymptotic limit \cite{winter2016operational} or approximate transformations \cite{renes2015relative}. Akin to developments in coherence theory, we can also incorporate further physical restrictions \cite{streltsov2016quantum} such as conservation of energy \cite{chiribella2015optimal}, or restrictions for distributed scenarios such as local superposition-free operations and classical communication \cite{chitambar2015relating,chitambar2016assisted,streltsov2015hierarchies,streltsov2016entanglement}. As in coherence theory \cite{winter2016operational,chitambar2016assisted}, there are also connections to entanglement theory \cite{killoran2016converting, regula2017converting} to be further understood. As potential next steps, our results could be extended to infinite dimensional states, continuous settings, or linearly dependent free states (like those found in magic state quantum computation \cite{1367-2630-16-1-013009,PhysRevLett.118.090501}).
This leads towards the ultimate goal of a fully general theory of non-classicality which puts superposition, coherence, entanglement, and quantum optical coherence on a unified standing.

We thank J. M. Matera for useful comments.
This work was supported by the ERC Synergy grant BioQ, the EU project QUCHIP and an Alexander von Humboldt Professorship.

\newpage
\begin{appendices}
	\appendixpage
	\begingroup
	\allowdisplaybreaks
	In these Appendices, we give the proofs of the results in the main text and some further results. For readability, we use the short-cuts $s_\theta:=\sin \theta$ and $c_\theta:=\cos \theta$.
	
	\section{Example for the value of superposition}
	The following game is an example that superposition is a resource in a channel discrimination task  (a branch of quantum metrology) and inspired by \cite{napoli2016robustness}. Assume we have two players, say Alice and Bob. Alice performs a selective quantum operation (which is known to Bob) with outcomes $n=0,1,...,d <\infty$ on a state she received from Bob. If the result is $n=0$, they start a new turn and Bob has to hand in a new state.  In case the result was $n\ne0$ she returns the post-measurement state to Bob, who is allowed to apply an arbitrary quantum operation on it. Then he has two choices: he either tells Alice his guess about the outcome $n$ or he asks for a new turn. He has lost immediately if he gives a wrong answer and he wins if his answer is correct. 
	
	Now, using the notation from the main text, we will construct a superposition free selective quantum operation for which access to a given superposition state turns certain failure into certain success. We identify outcomes $n=1,...,d$ with the free Kraus operators 
	\begin{align}
		K_n=\sqrt{p/d}\sum_{j=1}^d e^{\frac{2\pi i j n}{d}}\ket{c_j}\bra{c_j^\perp}
	\end{align}
	with $0<p\le1$ such that
	\begin{align}
		\sum_n K_n^\dagger K_n=p \sum_{j}\ket{c_j^\perp}\bra{c_j^\perp}\le \mathbbm{1}.
	\end{align}
	In addition, we identify $n=0$ with the free Kraus operator $K_0$ which makes the operation trace preserving.
	
	In case Bob can hand in only free states 
	\begin{align}
		\rho_{f}=\sum_j \rho_j \ket{c_j} \bra{c_j},
	\end{align}
	the states 
	\begin{align}
		\rho_n=\frac{K_n \rho_f K_n^\dagger}{\tr K_n \rho_f K_n^\dagger}=\rho_f
	\end{align}
	he retrieves carry no information about $n$. 
	In addition, the probability 
	\begin{align}
		p_n=\tr K_n \rho_f K_n^\dagger=p/d 
	\end{align}
	for outcome $n$ to occur is independent of $n$, too. Thus his best choice is to make a random guess. Since $p\le1$, Bob will lose with certainty for $d$ against infinity. 
	However, if he has access to the superposition state
	\begin{align}
		\ket{\phi}=\frac{1}{N}\sum_j\ket{c_j},
	\end{align}
	then the state he retrieves in case of outcome $n$ is given by
	\begin{align}
		\ket{\phi_n}=\frac{1}{N} \sum_l e^{\frac{2\pi i l n}{d}} \ket{c_l}.
	\end{align}
	As we will show, these states are linearly independent. Since the free states are linearly independent, 
	\begin{align}
		0=\sum_n x_n \ket{\phi_n} 
	\end{align}
	is equivalent to
	\begin{align}
		\sum_n e^{\frac{2\pi i j n}{d}} x_n = 0 \quad \forall l 
	\end{align}
	or 
	\begin{align}
		\sum_n u_n x_n=0
	\end{align}
	with
	\begin{align}
		u_n=\begin{pmatrix}
			e^{\frac{2\pi i 1 n}{d}} \\
			e^{\frac{2\pi i 2 n}{d}} \\
			e^{\frac{2\pi i 3 n}{d}} \\
			...\\
			e^{\frac{2\pi i d n}{d}} \\
		\end{pmatrix}.
	\end{align}
	Since
	\begin{align}
		u_n^\dagger u_m = \delta_{mn} d,
	\end{align}
	the only solution is $x_n=0$ $\forall n$ which finishes the proof. Thus Bob can do unambiguous state discrimination \cite{chefles1998unambiguous,chefles2000quantum} on the states $\{\ket{\phi_n}\}$ and will, after enough repetitions, win with certainty.

	\section{Choice of the free operations} \label{appendix:Choice of the free operations}
	In this section, we discuss alternative choices of the free operations defined in the main text and  their relation to the free operations in coherence theory.
	
	In the case of entanglement theory, the restriction to local operations and classical communication (LOCC) is very well motivated from a practical point of view \cite{plenio2007introduction}. The distant parties are allowed to perform arbitrary local quantum operations and exchange classical information but they are not allowed to transfer any quantum systems between the labs.
	Since classical bits cannot create entanglement, entanglement remains a resource that can be manipulated but not created.
	In addition, it is much cheaper to send classical information than quantum information because it can be amplified easily. However, this choice of the free operations is not unique. Different classes of free operations have been considered such as one way (forward or backward) classical communication, two way classical communication or the class of separable operations \cite{horodecki2009quantum}. They all have their justification, either in a practical scenario or for their comparably simple mathematical structure which allows to find bound for protocols using LOCC.
	
	Thus a debate about the choice of the free operations is necessary 
	in every resource theory.
	Recently, this happened extensively in the case of coherence theory, especially since it seems difficult to justify restrictions by practical considerations (such as spacial separation in LOCC). In \cite{streltsov2016quantum}, nine different definitions of incoherent operations are collected and inclusion relations are given. The analogue of $\mathcal{FO}$ is denoted by $\mathcal{IO}$ and $\mathcal{MFO}$ is equivalent to $\mathcal{MIO}$. 
	One of the major concerns about $\mathcal{IO}$ is that these operations do not posses a \textit{free dilation} in general \cite{chitambar2016critical,marvian2016quantify}. 
	Every quantum operation on a system $A$ in a state $\rho$ can be obtained from a Stinespring dilation \cite{nielsen2010quantum}: An auxiliary system $B$ in a state $\sigma$ is introduced followed by a global unitary operation $U$ on $A$ and $B$. 
	After a projective measurement by projectors $P_m$ and a classical processing of the outcome, system $B$ is discarded. 
	According to \cite{chitambar2016critical}, an operation possesses a free dilation if it can be obtained via a Stinespring dilation where $\sigma$ and $U$ are free and the projective measurement is a complete set of projectors on the free states. 
	In coherence theory, the set of operations with free dilation is denoted by $\mathcal{PIO}$ (physically incoherent operations) and has been introduced in \cite{chitambar2016critical}. They also showed that $\mathcal{IO}$ is strictly larger than $\mathcal{PIO}$.
	
	Whilst  $\mathcal{PIO}$ has a strong physical motivation, its power is severely reduced in comparison to $\mathcal{IO}$. Even the asymptotic conversion rate of the maximally coherent qubit state to any other  coherent qubit state is strictly zero \cite{chitambar2016critical}. The generalization of $\mathcal{PIO}$ to our framework is even more restricted. If the pure free states are not orthogonal, \textit{no} complete set of projectors on the free states exists.
	In addition, the set of free unitary operations is further limited as can be seen at the example of unitary operations on qubits. Unitary operations on the Bloch sphere are represented by rotations about a given axis through the origin. If the two pure free states are orthogonal and represented by $(0,0,-1)$ and $(0,0,1)$, a free unitary can be decomposed into an arbitrary rotation around the $z$-axis and a NOT gate. If the pure free states are not orthogonal, only the equivalent to the NOT gate remains. Thus this set of free operations seems too restricted in our case to give rise to an interesting resource theory. 
	
	\section{Free operations on qubits}\label{Free operations on qubits}
	{\it  Geometrical interpretation of quantum operations on the Bloch sphere --}
	For some of the proofs of the results in the main text, we make use of the geometrical interpretation of quantum operations on the Bloch sphere presented in \cite{pasieka2009geometric}. Therefore we give a short review on this topic.
	Every qubit state $\rho$ can be expanded into the Pauli basis
	\begin{equation}
		\rho=\frac{1}{2} \begin{pmatrix}  
			1 \\
			\vec{r}
		\end{pmatrix} \vec{\sigma}
	\end{equation}
	with $\vec{\sigma}=(\mathbbm{1},\sigma_x,\sigma_y,\sigma_z)^t$. Here $\sigma_i$ denotes the Pauli matrices and $\vec{r}:|\vec{r}|\le 1$ is a 3-component real column vector. In addition, every matrix of this form describes a valid qubit state. Every quantum operation $\Psi$ (a linear, completely positive and trace preserving map) on the qubit  can be expressed as a matrix acting on the vector of expansion coefficients. This matrix representation of $\Psi$ is then necessarily of the form 
	\begin{equation}
		\Psi = \begin{pmatrix}  
			1  &\begin{matrix} 0 & 0& 0	\end{matrix} \\
			\vec{t} & T  
		\end{pmatrix},
	\end{equation}
	where $\vec{t}$ is a 3-component real column vector and $T$ is a $3\times3$ real matrix. However, not every operation of this form has to be a quantum operation. 
	
	The qubit operations can be decomposed into the following four geometric operations on the Bloch sphere:
	\begin{enumerate}
		\item Rotation $\tilde{W}^t$ 
		\item Compression along x-,y- and z-axis to an ellipsoid with possible reflection through the y-z plane $D$
		\item Rotation $W$
		\item Translation $\vec{t}$
	\end{enumerate}
	with the effect
	\begin{equation}
		\vec{r}\rightarrow T\vec{r}+\vec{t}=WD\tilde{W}^t \vec{r}+\vec{t}.
	\end{equation}
	If those operations map the Bloch sphere into itself, $\Psi$ is positive semidefinite but not necessarily completely positive.
	
	{\it  Superposition for qubits --}
	Considering qubits, one can always choose a computational basis such that the superposition-free states are given by the Bloch vectors $\vec{r}_c=(a,0,c)^t$ with $0\le a<1$ fixed and 
	\begin{equation} \label{cond c}
		|c|\le \sqrt{1-a^2}. 
	\end{equation}
	We will use this computational basis for some of the proofs in the remainder of these Appendices.
	The pure superposition-free states $\ket{c_1},\ket{c_2}$ are then given by the Bloch vectors
	\begin{equation}
		\vec{r}_{c_1}=\begin{pmatrix}  
			a \\
			0 \\
			\sqrt{1-a^2}
		\end{pmatrix} \qquad \mathrm{and} \qquad  \vec{r}_{c_2}=\begin{pmatrix}  
		a \\
		0 \\
		-\sqrt{1-a^2}
	\end{pmatrix}.
\end{equation} 
This is equivalent to 
\begin{align} \label{bloch}
	&\ket{c_1}=\frac{1}{2}\begin{pmatrix}  
		\sqrt{1+a}+\sqrt{1-a} \\
		\sqrt{1+a}-\sqrt{1-a} 
	\end{pmatrix}, \qquad \nonumber \\
	& \ket{c_2}=\frac{1}{2}\begin{pmatrix}  
		\sqrt{1+a}-\sqrt{1-a} \\
		\sqrt{1+a}+\sqrt{1-a} 
	\end{pmatrix}.
\end{align}
Since $\braket{c_1|c_2}=a$, $a$ is a measure of the overlap of the two pure superposition-free states.
To prove a difference between $\mathcal{FO}$ and $\mathcal{MFO}$, we will use a certain quantum operation $\Phi$ with a matrix representation in the geometrical picture. This matrix will be defined here and in the following lemma it will be shown that this is indeed a quantum operation. 

\begin{definition}\label{def:Phi}
	The matrix $\Phi=\Phi(a,\theta,\phi)$ is defined by
	\begin{align} \label{Phi}
		\Phi&=\begin{pmatrix}  
			1  & 0 & 0& 0	 \\
			\vec{t} & \vec{w}&\vec{0}&\vec{0}  
		\end{pmatrix}, \nonumber \\
		\vec{w}&=\frac{1}{1+a}\begin{pmatrix}  
			a-c_\phi s_\theta \\
			-s_\phi s_\theta \\
			-\frac{c_\theta}{2}(1+a) \\
		\end{pmatrix} ,\nonumber \\
		\vec{t}&=\frac{a}{1+a} \begin{pmatrix}  
			1+c_\phi s_\theta\\
			s_\phi s_\theta \\
			\frac{c_\theta}{2a}(1+a) \\
		\end{pmatrix}.
	\end{align}
\end{definition}

\begin{lemma} \label{Lemma:Phi}
	The matrix $\Phi$ 
	represents a completely positive and trace preserving map in the geometrical picture. 
	With the superposition-free states as defined above, it maps superposition-free states to superposition-free states.
\end{lemma}

\begin{proof} Since the Pauli matrices are traceless, $\Phi$ is trace preserving. To show that $\Phi$ is completely positive, we will use 
	the Choi-Jamiolkowski isomorphism \cite{choi1975completely,jiang2013channel} which states that 
	\begin{equation}
		C_\Psi:=\sum_{i,j} \ket{i}\bra{j}\otimes\Psi(\ket{i}\bra{j})\ge 0 \Leftrightarrow \Psi\  \mathrm{completely\ positive}.
	\end{equation}
	We have
	\begin{alignat}{2}
		\ket{1}\bra{1}&=\frac{1}{2} \begin{pmatrix}  
			1\\
			0 \\
			0  \\
			1\\
		\end{pmatrix}\vec{\sigma},  &\qquad 
		\ket{2}\bra{1}&=\frac{1}{2} \begin{pmatrix}  
			0\\
			1 \\
			-i  \\
			0\\
		\end{pmatrix}\vec{\sigma}, \nonumber \\
		\ket{2}\bra{2}&=\frac{1}{2} \begin{pmatrix}  
			1\\
			0 \\
			0  \\
			-1\\
		\end{pmatrix}\vec{\sigma},  &\qquad 
		\ket{1}\bra{2}&=\frac{1}{2} \begin{pmatrix}  
			0\\
			1 \\
			i  \\
			0\\
		\end{pmatrix}\vec{\sigma} 
	\end{alignat}
	and thus
	\begin{alignat}{2} \label{phiOnOrthogonalStates}
		\Phi\ket{1}\bra{1}&=\frac{1}{2}\begin{pmatrix}  
			1\\
			\vec{t}
		\end{pmatrix}\vec{\sigma} , \nonumber &\qquad 
		\Phi\ket{2}\bra{1}&= \frac{1}{2}\begin{pmatrix}  
			0\\
			\vec{w}
		\end{pmatrix}\vec{\sigma}, \nonumber \\
		\Phi\ket{2}\bra{2}&= \frac{1}{2}\begin{pmatrix}  
			1\\
			\vec{t}
		\end{pmatrix}\vec{\sigma},  &\qquad 
		\Phi\ket{1}\bra{2}&= \frac{1}{2}\begin{pmatrix}  
			0\\
			\vec{w}
		\end{pmatrix}\vec{\sigma}.
	\end{alignat}
	This allows to calculate $C_\Phi$ which is given by
	\begin{equation}
		\frac{1}{2}\begin{pmatrix}  
			\frac{1}{2}\left(2+c_\theta\right)  &&  \frac{a+ae^{-i\phi}s_\theta}{1+a}&& -\frac{c_\theta}{2} && \frac{a-e^{-i\phi}s_\theta}{1+a}\\
			\frac{a+ae^{i\phi}s_\theta}{1+a} && \frac{1}{2}\left(2-c_\theta\right) && \frac{a-e^{i\phi}s_\theta}{1+a} && \frac{c_\theta}{2} \\
			-\frac{c_\theta}{2} &&  \frac{a-e^{-i\phi}s_\theta}{1+a}   && \frac{1}{2}\left(2+c_\theta\right)      && \frac{a+ae^{-i\phi}s_\theta}{1+a} \\
			\frac{a-e^{i\phi}s_\theta}{1+a}&&   \frac{c_\theta}{2}  &&  \frac{a+ae^{i\phi}s_\theta}{1+a}  && \frac{1}{2}\left(2-c_\theta\right) \\
		\end{pmatrix}.
	\end{equation}
	The eigenvalues of $C_\Phi$ are $0,1$ and $\frac{2+2a\pm R}{4(1+a)}$ with $R$ given by
	\begin{align}
		&\sqrt{2}\sqrt{1-2a+9a^2-(-1+a)^2c_{2\theta}+8(-1+a)ac_\Phi s_\theta} \nonumber \\
		&\le \sqrt{2}\sqrt{1-2a+9a^2+(1-a)^2+8(1-a)a} \nonumber \\
		&=2(1-a).	
	\end{align}
	Using $2+2a-2(1-a)\ge 0$, all eigenvalues are larger or equal zero and thus $\Phi$ is completely positive. 
	As a last step, it is easy to check that superposition-free states are mapped to superposition-free states since
	\begin{equation}
		\Phi \begin{pmatrix}  
			1\\
			a \\
			0  \\
			c\\
		\end{pmatrix}=\begin{pmatrix}  
		1\\
		\vec{t}+a\vec{w}
	\end{pmatrix}=\begin{pmatrix}  
	1\\
	a\\
	0 \\
	\frac{c_\theta}{2}(1-a) \\
\end{pmatrix} .
\end{equation}
\end{proof}

\begin{lemma} \label{LemmaQubitMaxNonClCand}
	If the superposition-free states are chosen as above, the state $\ket{m_2}$ corresponding to the Bloch vector $\vec{r}_m=(-1,0,0)^t$ is for $a\ne 0$ the only candidate to have maximal superposition. The operation $\Phi$  defined in definition \ref{def:Phi} can be used to generate all other qubit states deterministically from  $\ket{m_2}$.  
\end{lemma}

\begin{proof} 
	First we will only consider the generation of pure states from $\ket{m_2}$. 
	The states we want to generate will be called target states. Since all pure qubit states are represented by a unit length Bloch vector, their Bloch vectors can be parametrized in polar coordinates,
	\begin{equation}
		\vec{r}_{t}=\begin{pmatrix}  
			c_\phi s_\theta\\
			s_\phi s_\theta \\
			c_\theta \\
		\end{pmatrix}.
	\end{equation} 
	In fact it is now easy to check that
	\begin{equation}
		\Phi \begin{pmatrix}  
			1\\
			-1 \\
			0  \\
			0\\
		\end{pmatrix}= \begin{pmatrix}  
		1\\
		\vec{t}-\vec{w}
	\end{pmatrix}=\begin{pmatrix}  
	1\\
	c_\phi s_\theta\\
	s_\phi s_\theta \\
	c_\theta \\
\end{pmatrix}
\end{equation}
and thus $\Phi$ transforms $\vec{r}_m=(-1,0,0)$ to the desired target state. 

The generation of mixed target states $\rho_M$ is also possible due to linearity. Since $\rho_M$ can be decomposed into pure states through $\rho_M=\sum_{i} p_i \ket{\phi_i}\bra{\phi_i}$, we can just apply the operation $\Phi_M=\sum_{i} p_i \Phi_i$ to $\ket{m_2}$ where $\Phi_i$ generates $\ket{\phi_i}$ from $\ket{m_2}$. 

Finally we need to show that $\ket{m_2}$ is the only candidate to have maximal superposition.
This can be seen using again the geometrical interpretation of quantum operations on the Bloch sphere. The euclidean distance between a quantum state and the set of superposition-free states is never smaller than the euclidean distance between their images under any quantum operation. The rotations and the translation preserve the distance, the compression can only reduce it.	

In the case of $\mathcal{MFO}$, the image of the superposition-free states has to be a subset of the superposition-free states. Thus the euclidean distance between a quantum state and the set of superposition-free states cannot increase under the action of $\mathcal{MFO}$.
Since the euclidean distance between $\ket{m_2}$ and the  superposition-free states is for $a\ne0$ larger than the euclidean distance between any other state and the superposition-free states, $\ket{m_2}$ cannot be generated with certainty from any other state by means of $\mathcal{MFO}$ and thus not by means of $\mathcal{FO}$. 
\end{proof}

{\it  Superposition-free Kraus operators for qubits --}
Here we will use the results from the main text to find sufficient and necessary conditions for deterministic superposition-free operations on qubits.
Remember that a superposition-free Kraus operator can be derived from an incoherent one via the transformation matrix $V$. Further remember that incoherent Kraus operators have at most one non-zero entry per column. 
Thus for qubits, there are four different types of incoherent Kraus operators given by
\begin{alignat}{2}\label{qubitIncoherentKrausOperators}
	&\tilde{K}_1=\begin{pmatrix}  
		\alpha & \beta \\
		0& 0 
	\end{pmatrix}, \
	&& \tilde{K}_2=\begin{pmatrix}  
		\gamma & 0 \\
		0& \delta
	\end{pmatrix}, \ \nonumber \\
	&\tilde{K}_3=\begin{pmatrix}  
		0 & 0 \\
		\mu& \nu
	\end{pmatrix}, \ 
	&&\tilde{K}_4=\begin{pmatrix}  
		0 & \xi \\
		\epsilon& 0 
	\end{pmatrix} 
\end{alignat}
with $\alpha, \beta, \gamma, \delta, \mu, \nu$ complex numbers. Since for Kraus operators an overall phase can be neglected, it is possible to choose in every Kraus operator one of the two non-zero entries to be real.
If one chooses the computational basis $\{\ket{1},\ket{2}\}$ in a way that 
\begin{align}
	&\ket{c_1}=\ket{1},\nonumber \nonumber \\
	&\ket{c_2}=s_\theta\ket{1}+c_\theta\ket{2} ,\nonumber  \\ 
	&0\le\theta < \frac{\pi}{2},
\end{align}
the transformation matrix $V$ is given by
\begin{align}
	V=\begin{pmatrix}  
		1 & s_\theta \\
		0& c_\theta
	\end{pmatrix}
\end{align}
and
\begin{align}
	\ V^{-1}= \begin{pmatrix}  
		1 & -s_\theta/c_\theta \\
		0& 1/c_\theta 
	\end{pmatrix},\ V^\dagger V=\begin{pmatrix}  
	1 & s_\theta\\
	s_\theta& 1 
\end{pmatrix}.
\end{align}
First consider the case of a deterministic superposition-free operation in which every type of superposition-free Kraus operator $K_n=V \tilde{K}_n V^{-1}$ occurs only once.
This results in the condition
\begin{widetext}
	\begin{align}
		\mathbbm{1}&\overset{!}{=}\sum_{n} K_n^\dagger K_n=\sum_{n} (V^{-1})^\dagger \tilde{K}_n^\dagger V^\dagger V \tilde{K}_n V^{-1} \nonumber \\
		&=\begin{pmatrix}  
			|\alpha|^2+|\gamma|^2+|\mu|^2+|\epsilon|^2 & \begin{matrix}  -s_\theta/c_\theta \left(|\alpha|^2+|\gamma|^2+|\mu|^2+|\epsilon|^2-\gamma^*\delta-\epsilon^*\xi\right) \\ + 1/c_\theta (\alpha^*\beta+\mu^* \nu)  \end{matrix}\\
			c.c& \begin{matrix}  s^2_\theta/c^2_\theta \left(|\alpha|^2 +|\gamma|^2+|\mu|^2+|\epsilon|^2-\gamma^*\delta-\gamma\delta^*-\epsilon^*\xi-\epsilon\xi^*\right) \\
				-s_\theta/c^2_\theta\left(\alpha^*\beta+\alpha\beta^*+\mu^* \nu+\mu \nu^*\right) 
				+1/c^2_\theta \left(|\beta|^2+|\delta|^2+|\nu|^2+|\xi|^2 \right)\end{matrix}
		\end{pmatrix}
	\end{align}
\end{widetext}
where $c.c$ means the complex conjugate of the upper right matrix entry. A straight forward simplification leads to the three equations
\begin{align} 
	1&\overset{!}{=} |\alpha|^2+|\gamma|^2+|\mu|^2+|\epsilon|^2, \nonumber \\
	1&\overset{!}{=} |\beta|^2+|\delta|^2+|\nu|^2+|\xi|^2, \nonumber \\
	0&\overset{!}{=} \alpha^*\beta+\mu^*\nu+s_\theta \left(\gamma^* \delta +\epsilon^*\xi-1\right).
\end{align}
Since these equations contain $\theta$, they seem to depend on the explicit choice of the computational basis. Now assume we had chosen another computational basis. Then it can be transformed by a unitary into the one we considered and (neglecting a physically unimportant phase) $s_\theta$ is given by $|\braket{c_2|c_1}|$. Thus in general we have
\begin{align} \label{qubitKrausConditions}
	1&\overset{!}{=} |\alpha|^2+|\gamma|^2+|\mu|^2+|\epsilon|^2, \nonumber \\
	1&\overset{!}{=} |\beta|^2+|\delta|^2+|\nu|^2+|\xi|^2, \nonumber \\
	0&\overset{!}{=} \alpha^*\beta+\mu^*\nu+|\braket{c_2|c_1}| \left(\gamma^* \delta +\epsilon^*\xi-1\right).
\end{align}
Until now we only considered operations containing one Kraus operator of each type. 
In a more general scenario we can consider multiple Kraus operators of the same type and denote them by 
\begin{alignat}{2} \label{multipleKrausOfSameType}
	&\tilde{K}_{1,i}=\begin{pmatrix}  
		\alpha_i & \beta_i \\
		0& 0 
	\end{pmatrix}, \
	&&\tilde{K}_{2,j}=\begin{pmatrix}  
		\gamma_j & 0 \\
		0& \delta_j
	\end{pmatrix}, \ \nonumber \\
	&\tilde{K}_{3,k}=\begin{pmatrix}  
		0 & 0 \\
		\mu_k& \nu_k
	\end{pmatrix}, \ 
	&&\tilde{K}_{4,l}=\begin{pmatrix}  
		0 & \xi_l \\
		\epsilon_l& 0 
	\end{pmatrix}.
\end{alignat}
Then the above equations are modified by linearity to
\begin{align} \label{ConditionsKrausOperators}
	1\overset{!}{=}& \sum_{i}|\alpha_i|^2+\sum_{j}|\gamma_j|^2+\sum_{k}|\mu_k|^2+\sum_{l}|\epsilon_l|^2, \nonumber \\
	1\overset{!}{=}& \sum_{i}|\beta_i|^2+\sum_{j}|\delta_j|^2+\sum_{k}|\nu_k|^2+\sum_{l}|\xi_l|^2, \nonumber \\
	0\overset{!}{=}& \sum_{i}\alpha_i^*\beta_i+\sum_{k}\mu_k^*\nu_k \nonumber \\
	&+|\braket{c_2|c_1}| \left(\sum_{j}\gamma_j^* \delta_j +\sum_{l}\epsilon_l^*\xi_l-1\right).
\end{align}
That it can be useful to consider more than one superposition-free Kraus operator of the same type can be seen at the example of a quantum operation
with decomposition into the two Kraus operators 
\begin{align} 
	&\tilde{K}_{1,1}=\frac{1}{\sqrt{2}}\begin{pmatrix}  
		1 & 1\\
		0& 0 
	\end{pmatrix}, \
	\tilde{K}_{1,2}=\frac{1}{\sqrt{2}}\begin{pmatrix}  
		1 & -1 \\
		0& 0
	\end{pmatrix}.
\end{align}
Notice that this operation is trace preserving, incoherent and uses two Kraus operators of the first type.

\section{Proofs}
\setcounter{theorem}{1}
Here we provide the proofs of the results in the main text which we restate for readability.
\begin{theorem}
	If the free states in a finite dimensional Hilbert space form a countable set, then linear independence of the free states is a necessary and sufficient condition for the existence of a faithful conversion operation.
	In case the free states are a finite set of optical coherent states, the faithful conversion can be implemented by a beam splitter \cite{vogel2014unified}.
\end{theorem}

\begin{proof}
	That linear independence is a sufficient condition for faithful entanglement conversion has been shown by construction in \cite{killoran2016converting}. We will prove here that linear independence is also a necessary condition if the free states are countable.
	
	Suppose the set of free states $\mathcal{C}$ is countable and linearly dependent. 
	
	We begin by a preliminary consideration showing that with these conditions, almost all states have maximal classical rank.
	Without loss of generality, we write $\mathcal{C}=\{\ket{c_i} \}_{i=1}^\infty$ where $\{\ket{c_i} \}_{i=1}^d$ forms a basis of span($\mathcal{C}$). Then the set $\mathcal{E}$ of subsets $\pi_i$ of $\mathcal{C}$ with $d$ elements is countable as well. We split this set into two sets, the set $\mathcal{B}=\{b_i \}_{i=1}^\infty$ which contains the $\mathcal{\pi}_i$ which form a basis of span($\mathcal{C}$) and the set $\mathcal{S}=\{s_i \}_{i=1}^\infty$ which contains the $\pi_i$ that only span a subspace.
	
	Since we can identify all pure states, up to a phase, with a point on the surface of the unit sphere of $\mathbb{C}^d$, we can use the uniform measure on the surface of this unit sphere to measure the amount of states in a given set. The set $S(s_i)$ of pure states $\ket{\psi}$ representable by the span of $s_i$ is a set of measure zero since it is a subspace with lower dimension. 
	With
	\begin{align}
		\tilde{r}_i(\ket{\psi})=\min \left\{r \left| \ket{\psi}=\sum_{j=1}^r \psi_j \ket{c_j} : \ket{c_j}\in b_i  \right. \right\},
	\end{align}
	the set  $B(b_i)=\left\{\ket{\psi} \left| \tilde{r}_i(\ket{\psi}) < d   \right. \right\}$ is also a set of measure zero, since it consists of a countable union of lower dimensional objects as well.
	
	Since $\mathcal{E}$ is countable, the set
	\begin{align}
		E=\bigcup\limits_{b_i\in \mathcal{B}} B(b_i)  \cup \bigcup\limits_{s_i\in \mathcal{S}} S(s_i)
	\end{align}
	is the countable union of set of measure zeros and thus a set of measure zero itself. All pure states that are not in $E$ and thus almost all states have classical rank $d$. 
	
	This allows us to prove the first part of our theorem.
	Assume we can find a faithful conversion operation $\Lambda$ 
	for $\{\ket{c_i} \}_{i=1}^d$, i.e., 
	\begin{align}
		\Lambda\ket{c_i}=\ket{e_i}\otimes \ket{f_i}
	\end{align}
	for some local states $\{\ket{e_i}\}_{i=1}^d$, $\{\ket{f_i}\}_{i=1}^d$.
	Consider next the free state 
	\begin{align}
		|c_{d+1}\rangle=\sum_i^d\alpha_i\ket{c_i}.
	\end{align}
	We will not allow trivial dependences, so we must have at least two nonzero $\alpha_i$. Without loss of generality, assume $\alpha_1, \alpha_2 \ne 0$. The action of $\Lambda$ on 
	this state is
	\begin{align}
		\Lambda\ket{c_{d+1}}=\sum_{i=1}^d \alpha_i\ket{e_i}\otimes\ket{f_i}.
	\end{align}
	In order for $\Lambda$ to be faithful, the transformed state $\Lambda\ket{c_{d+1}}$ must have a Schmidt rank of 1.
	Equivalently, the matrix 
	\begin{align}
		M_{d+1}:=\sum_{i=1}^d\alpha_i \ket{e_i}\bra{f_i}
	\end{align}
	must have a rank of 1.
	We now consider two exhaustive cases.
	\begin{itemize}
		\item dim(span$\{\ket{f_i}\}_{i=1}^d) = 1$: In this case, $\ket{f_i}=\ket{f_0}~\forall~i$. This means that all states
		$\ket{\psi}\in\mathcal{H}$ get transformed to factorized states under $\Lambda$. Hence, $\Lambda$ cannot be faithful.
		
		\item dim(span $\{\ket{f_i}\}_{i=1}^d) = D > 1$: Since $M_{d+1}$ has a rank of 1, there must exist an orthonormal basis $\{\ket{g_j}\}_{j=1}^D$ of span$\{\ket{f_i}\}_{i=1}^d$ with 
		$M_{d+1}\ket{g_1}\ne 0$ and $M_{d+1}\ket{g_j}= 0$ for $j=2,...,D$. We expand 
		\begin{align}
			\ket{f_i}=\sum_{j=1}^{D} f_{i,j} \ket{g_j}
		\end{align}
		and write
		\begin{align}
			0=&M_{d+1}\ket{g_j}=\sum_{i=1}^{d} \sum_{l=1}^{D} \alpha_i f^*_{i,l} \ket{e_i}\braket{g_l|g_j}\nonumber \\
			=&\sum_{i=1}^{d} \alpha_i f^*_{i,j} \ket{e_i}~\forall j=2,...,d.
		\end{align}
		This equation implies that $\{\ket{e_i}\}_{i=1}^d$ must be linearly dependent. The only case where this conclusion is not obvious is when $f_{1,j}=f_{2,j}=0$ for all $j=2, ..., d$. However, in this case we would have $\ket{f_1}=\ket{f_2}=\ket{g_1}$, and all superpositions in the subspace span$\{\ket{c_1},\ket{c_2}\}$ will get converted to product states by $\Lambda$,
		\begin{align}
			\Lambda (a\ket{c_1} + b\ket{c_2}) = (a\ket{e_1} + b\ket{e_2}) \otimes \ket{g_1},
		\end{align}
		which either violates the condition that $\Lambda$ is faithful or the condition that the free states are countable. Thus we conclude in all cases that $\{\ket{e_i}\}_{i=1}^d$ are linearly dependent.
	\end{itemize}
	Now choose a state $\ket{\psi}$ with classical rank $d$. Since $\{\ket{e_i}\}_{i=1}^d$ are linearly dependent, they span a subspace of dimension $D<d$.
	The Schmidt rank of $\Lambda\ket{\psi}$ can therefore only be as high as $D$, and thus strictly less than the 
	classical rank of $|\psi\rangle$. 
	Thus, $\Lambda$ cannot be faithful. 
	
	The second part of the theorem is proven in \cite{vogel2014unified}. Here we provide an alternative proof in line with our theory. 	
	Therefore we use that a beam splitter $B$ converts the set of free states $\{\ket{\alpha_i}\}_{i=1}^d$ into separable states,
	\begin{align}
		B \ket{\alpha_i} \otimes \ket{0} =\ket{\frac{\alpha_i}{\sqrt{2}}} \otimes \ket{\frac{\alpha_i}{\sqrt{2}}}.
	\end{align}
	A state $\ket{\psi}=\sum_{i=1}^{n} c_i \ket{\alpha_i}$ with classical rank $n\le d$ is thus transformed according to
	\begin{align}
		B \ket{\psi} \otimes \ket{0}= \sum_{i=1}^{n} c_i \ket{\frac{\alpha_i}{\sqrt{2}}} \otimes \ket{\frac{\alpha_i}{\sqrt{2}}}=:\ket{\phi}
	\end{align}
	and the Schmidt rank of $\ket{\phi}$ is at most $n$. The set of states $\{\ket{\frac{\alpha_i}{\sqrt{2}}}\}_{i=1}^d$ is linearly independent as well. Using the linear operator $V$ introduced in the main text with the property
	\begin{align}
		\ket{\frac{\alpha_i}{\sqrt{2}}}=V \ket{i}
	\end{align}
	and the fact that it is full rank, we can find a $p>0$ such that 
	\begin{align}
		p\left(V^{-1}\right)^\dagger V^{-1} \le \mathbbm{1}.
	\end{align} 
	Thus $\sqrt{p} V^{-1}$ can be seen as a Kraus operator describing a trace non-increasing quantum operation. Applying the local operation
	\begin{align}
		p V^{-1}\otimes V^{-1}  &\sum_{i=1}^{n} c_i \ket{\frac{\alpha_i}{\sqrt{2}}} \otimes \ket{\frac{\alpha_i}{\sqrt{2}}} \nonumber \\
		=& p\sum_{i=1}^{n} c_i \ket{i} \otimes \ket{i},
	\end{align}
	we transform $\ket{\phi}$ with probability $p^2>0$ into a state with Schmidt rank $n$. Since the probability to transform a state locally into a state with higher Schmidt rank is zero \cite{PhysRevLett.83.1046}, we showed that the Schmidt rank of $\ket{\phi}$ is $n$ and that the transformation is faithful. Note that the same arguments can be extended to any lossless non-trivial beam splitter and that there might exist other faithful conversions as well. 
\end{proof}
Note that we could still have non-faithful conversions. In addition, if the free states are uncountable, there exist cases in which faithful conversions are possible.
As a simple example,
if the free classical states are the (continuous) separable states and the non-classical
states are the entangled states, then obviously the Schmidt and classical ranks are
equivalent and we can meet the faithful conversion condition with
$\Lambda=\mathbbm{1}$. 

\addtocounter {theorem} {1} 
\begin{proposition} 
	$\mathcal{MFO}$ is strictly larger than $\mathcal{FO}$. This is also valid in the special case of coherence theory.
\end{proposition}
\begin{proof}
	In this proof we make use of the explicit representation of the superposition-free states introduced in the section above and the operation $\Phi$ defined in definition \ref{def:Phi}. 
	Lemma \ref{LemmaQubitMaxNonClCand} states that $\Phi$ maps superposition-free states to superposition-free states. Now it will be shown that $\Phi$ cannot be decomposed into superposition-free Kraus operators.
	Assume there would be such a decomposition. From theorem \ref{theoremKrausOperators}, we know how superposition-free Kraus operators are obtained from incoherent ones. In equations (\ref{qubitIncoherentKrausOperators}), the four different types of incoherent Kraus operators are given. With the help of equation (\ref{bloch}) allowing to construct the matrix $V$, we can obtain the following four types of superposition-free qubit Kraus operators
	\begin{alignat}{1} \label{classicalKrausExplicit}
		&K_1=C\begin{pmatrix}  
			A\alpha-a\beta & -a\alpha+A\beta \\
			a\alpha-B\beta& -B\alpha +a \beta
		\end{pmatrix}, \qquad  \nonumber \\
		&K_2=C\begin{pmatrix}  
			A\gamma-B\delta & a(-\gamma+\delta) \\
			-a(-\gamma+\delta)& -B\gamma+A\delta
		\end{pmatrix}, \nonumber \\
		&K_3=C\begin{pmatrix}  
			a\mu-B\nu & -B\mu +a\nu \\
			A\mu-a\nu& -a\mu +A\nu
		\end{pmatrix}, \qquad \nonumber \\
		&K_4=C\begin{pmatrix}  
			a(\epsilon-\xi) & A\xi-B\epsilon \\
			-B\xi+A\epsilon& -a(\epsilon-\xi) 
		\end{pmatrix} 
	\end{alignat}
	with 
	\begin{align} \label{A,B}
		&A=1+\sqrt{1-a^2}, \qquad B=1-\sqrt{1-a^2}, \nonumber \\
		& C=\frac{1}{2\sqrt{1-a^2}}.
	\end{align}
	From equation (\ref{bloch}), we find
	\begin{align}
		\ket{c_1}\bra{c_1}&=\frac{1}{2}\begin{pmatrix}  
			1+\sqrt{1-a^2} & a\\
			a & 1-\sqrt{1-a^2} \\
		\end{pmatrix}, \nonumber \\
		\ket{c_2}\bra{c_2}&=\frac{1}{2}\begin{pmatrix}  
			1-\sqrt{1-a^2} & a\\
			a & 1+\sqrt{1-a^2} \\
		\end{pmatrix}.
	\end{align}
	All superposition-free states are mapped to the same superposition-free state 
	\begin{align}
		\rho_c=&\frac{1}{2}\begin{pmatrix}  
			1+\frac{c_\theta}{2}(1-a) & a\\
			a & 1-\frac{c_\theta}{2}(1-a) \\
		\end{pmatrix} \nonumber \\
		=&p \ket{c_1}\bra{c_1}+q\ket{c_2}\bra{c_2}
	\end{align}
	
	with
	\begin{align}
		p=\frac{1}{2}+\frac{c_\theta}{4} \sqrt{\frac{1-a}{1+a}}, \nonumber \\
		q=\frac{1}{2}-\frac{c_\theta}{4} \sqrt{\frac{1-a}{1+a}}.
	\end{align}
	First only one Kraus operator of each type will be taken into consideration.
	Using 
	\begin{alignat}{2} \label{classicalOnClassical}
		&K_1\ket{c_1}=\alpha \ket{c_1}, \qquad &&
		K_1\ket{c_2}=\beta \ket{c_1}, \nonumber \\
		&K_2\ket{c_1}=\gamma\ket{c_1}, &&
		K_2\ket{c_2}=\delta \ket{c_2}, \nonumber \\
		&K_3\ket{c_1}=\mu \ket{c_2}, &&
		K_3\ket{c_2}=\nu \ket{c_2}, \nonumber \\
		&K_4\ket{c_1}=\epsilon \ket{c_2}, &&
		K_4\ket{c_2}=\xi \ket{c_1} 
	\end{alignat}
	and
	\begin{equation} 
		\sum_n K_n \ket{c_i}\bra{c_i} K_n^\dagger = \rho_c 
	\end{equation}
	leads to
	\begin{alignat}{2} \label{condPandQ}
		&|\alpha|^2+|\gamma|^2=p, \qquad \qquad&&	|\mu|^2+|\epsilon|^2=q,  \nonumber \\
		&|\beta|^2+|\xi|^2=p,  \qquad \qquad&&	|\nu|^2+|\delta|^2=q.
	\end{alignat}
	Using equation (\ref{phiOnOrthogonalStates}), we can obtain the additional constraints
	\begin{align}
		\frac{1}{2}\left(1+\frac{c_\theta}{2}\right)\overset{!}{=}&\bra{1}\left(\sum_n K_n \ket{1}\bra{1} K_n^\dagger\right)\ket{1} \nonumber \\
		=&\frac{1}{4(1-a^2)}\left(|A\alpha-a\beta|^2+|A\gamma-B\delta|^2\nonumber\right. \\
		&\left.+a^2|\epsilon-\xi|^2+|a\mu-B\nu|^2\right), \nonumber \\
		\frac{1}{2}\left(1-\frac{c_\theta}{2}\right)\overset{!}{=}&\bra{2}\left(\sum_n K_n \ket{1}\bra{1} K_n^\dagger\right)\ket{2} \nonumber \\
		=&\frac{1}{4(1-a^2)}\left(|a\alpha-B\beta|^2+|a(\gamma-\delta)|^2+\right.\nonumber\\
		&\left.|A\epsilon-B\xi|^2+|A\mu-a\nu|^2\right). 
	\end{align}
	With the help of equations (\ref{qubitKrausConditions}) and (\ref{condPandQ}) they can be simplified to
	\begin{align}
		2\left(1-\right.&\left. a^2\right)\left(1+\frac{c_\theta}{2}\right)\nonumber \\
		\overset{!}{=}&2+c_\theta(1-a)-2a^2B-2a\left(\alpha^*\beta +\alpha \beta^*\right)\sqrt{1-a^2}\nonumber\\
		&-\left(\gamma^*\delta+\gamma \delta^*+\xi^*\epsilon+\xi \epsilon^*\right)a^2\sqrt{1-a^2}, \nonumber \\
		2\left(1-\right.&\left.a^2\right)\left(1-\frac{c_\theta}{2}\right)\nonumber \\
		\overset{!}{=}&2+c_\theta(1-a)-2a^2A+2a\left(\alpha^*\beta +\alpha \beta^*\right)\sqrt{1-a^2}\nonumber\\
		&+\left(\gamma^*\delta+\gamma \delta^*+\xi^*\epsilon+\xi \epsilon^*\right)a^2\sqrt{1-a^2}.
	\end{align}
	Finally, adding these two equations leads to 
	\begin{equation}
		c_\theta(1-a)=0  \qquad \forall \theta
	\end{equation}
	which is for fixed $0\le a <1$ and $\theta \ne \frac{\pi}{2}$ a contradiction to the assumption. 
	Now consider the case that different superposition-free Kraus operators of the same type are used. Thus we use the Kraus operators introduced in equation (\ref{multipleKrausOfSameType}). By linearity, for example $\alpha^*\beta$ will just be replaced by $\sum_{i}\alpha_i^*\beta_i$ in the equations above. In the step where the summation is done, the sums of coefficients will cancel out and the same contradiction is obtained.
	By definition, all superposition-free quantum operations map superposition-free states to superposition-free states. Thus we have proven  that the set of quantum operations mapping superposition-free states to superposition-free states is greater than the set of superposition-free quantum operations. This is even true in the case of qubits. In the limit of $a=0$, coherence theory is recovered and thus the same result holds there. 
\end{proof}

\begin{theorem}
	A Kraus operator $K_n$ is superposition-free if and only if it is of the form
	\begin{align} 
		K_n&=\sum_{k} c_{k,n}\ket{c_{f_n(k)}}\bra{c_k^\perp},
	\end{align}
	where the $c_{k,n}\in\mathbb{C}$ and the $f_n(k)$ are index functions.
\end{theorem}

\begin{proof}
	Since both $\{\ket{c_i}\}_{i=1}^{d}$ and $\{\ket{c_i^\perp}\}_{i=1}^{d}$ form a basis, every Kraus operator $L$ can be expanded as 
	\begin{equation}
		L=\sum_{ij}L_{ij} \ket{c_i}\bra{c_j^\perp}.
	\end{equation}
	Applying such a general Kraus operator to an arbitrary superposition-free state $\ket{c_k}$ results in
	\begin{equation}
		L\ket{c_k}=\sum_{i} L_{ik}\ket{c_i}.
	\end{equation}
	If $L$ is superposition-free, $L_{ik}$ can be non-zero for at most one $i$ by definition. Thus $L$, in order to be free, has to be of the form
	\begin{equation} 
		L=\sum_{k} c_{k}\ket{c_{f(k)}}\bra{c_k^\perp},
	\end{equation} 
	where $\{c_{k}\}$ are complex valued coefficients and $f(k)$ is an index function. If we have a Kraus operator of this form and apply it to a superposition-free state $\rho=\sum_{i} \rho_{i} \ket{c_i}\bra{c_i}$, we find
	\begin{align}
		L\rho L^\dagger&=\sum_{kli} c_k \ket{c_{f(k)}}\bra{c_k^\perp} \rho_{i} \ket{c_i}\bra{c_i} c_l^* \ket{c_l^\perp}\bra{c_{f(l)}} \nonumber \\
		&=\sum_{i} |c_i|^2 \rho_{i} \ket{c_{f(i)}}   \bra{c_{f(i)}}
	\end{align}
	which is again superposition-free. 
\end{proof}

\begin{proposition}
	If both $\sigma_B$ and all $K_n$ are free,
	the quantum operation $\Phi(\rho_A)=\tr_B\sum_n K_n \rho_A \otimes \sigma_B K_n^\dagger$ is free.
\end{proposition} 
\begin{proof}
	In order to prove the proposition, let us consider
	\begin{align}
		\tr_BL\rho_A\otimes \sigma_BL^\dagger,
	\end{align}
	where both $L$ and $\sigma_B$ are superposition-free. Thus 
	\begin{align}
		L=&\sum_{i,j} c_{i,j} \ket{c_{g(i,j)}\ c_{h(i,j)}}\bra{c_i^\perp\ c_j^\perp}, \nonumber \\
		\sigma_B=&\sum_m \sigma_m \ket{c_m}\bra{c_m}_B\
	\end{align}
	according to theorem \ref{theoremKrausOperators}. 
	Let $\{\ket{x}_B\}$ be an orthonormal basis of system $B$.
	Since both the pure superposition-free and the reciprocal states form a basis, we can expand 
	\begin{align}
		\rho_A=&\sum_{s,t} \rho_{s,t} \ket{c_s}_A\bra{c_t}, \nonumber \\
		\ket{x}_B=&\sum_l d_{x,l}\ket{c_l^\perp}_B.
	\end{align}
	Thus we find
	\begin{widetext}
		\begin{align}
			\tr_BL\rho_A\otimes \sigma_BL^\dagger=&\sum_{i,j,k} c_{i,j} c_{k,j}^* \rho_{i,k}\sigma_j \ket{c_{g(i,j)}}_A \bra{\ c_{g(k,j)}}\sum_x \braket{x|c_{h(i,j)}}_B\braket{\ c_{h(k,j)}|x} \nonumber \\
			=&\sum_{i,j,k,x,l,m} c_{i,j} c_{k,j}^*\rho_{i,k}\sigma_j \ket{c_{g(i,j)}}_A \bra{\ c_{g(k,j)}}\delta_{l,h(i,j)} \delta_{m,h(k,j)} d_{x,l}^* d_{x,m}.
		\end{align}
	\end{widetext}	
	For $n=(j,x)$,  we introduce superposition-free Kraus operators 
	\begin{align}
		F_n=F_{(j,x)}=\sum_i k_{i,n} \ket{c_{f_n(i)}}\bra{c_i^\perp}
	\end{align}
	with
	\begin{alignat}{2}
		k_{i,n}=&k_{i,(j,x)}&&=c_{i,j} \sum_l\delta_{l,h(i,j)} d_{x,l}^*, \nonumber \\
		f_n(i)=&f_{(j,x)}(i)&&=g(i,j).
	\end{alignat}
	These Kraus operators have the property
	\begin{align}
		\sum_{j,x}F_{j,x} \rho_A F_{j,x}^\dagger=\tr_BL\rho_A\otimes \sigma_BL^\dagger
	\end{align}
	which finishes the proof by linearity. 
\end{proof}

\begin{theorem} 
	Assume we have an (incomplete) set of Kraus operators $\{K_m\}$ such that $\sum_m K_m^\dagger K_m \le \mathbbm{1}$. Then there always exist superposition-free Kraus operators $\{F_n\}$ with $\sum_m K_m^\dagger K_m +\sum_n F_n^\dagger F_n=\mathbbm{1}$.
\end{theorem}
\begin{proof}
	Let the assumptions hold. Then there exists an orthonormal basis $\{\ket{n}\}_{n=1}^d$ with
	\begin{equation}
		\sum_m K_m^\dagger K_m=\sum_{n=1}^{d} (1-p_n) \ket{n} \bra{n}\ :0\le p_n \le 1.
	\end{equation}
	Remember the reciprocal vectors ${\ket{c_i^\perp}}$ which form a basis as well. So we can expand $\ket{n}$ in this basis
	\begin{equation}
		\ket{n}=\sum_{j} d_{j,n}\ket{c_j^\perp}
	\end{equation}
	and write what is missing for our operation to be trace preserving as
	\begin{align}
		\sigma_r&=\mathbbm{1}-\sum_m K_m^\dagger K_m=\sum_{n=1}^{d} p_n \ket{n} \bra{n}\nonumber \\
		&=\sum_{k,l=1}^{d} \left(\sum_{n=1}^{d} p_n d_{k,n} d_{l,n}^*\right) \ket{c_k^\perp} \bra{c_l^\perp}.
	\end{align}
	Now define additional superposition-free Kraus operators $ \{F_n\}_{n=1}^d$ by
	\begin{equation}
		F_n=\sum_{k=1}^{d}\sqrt{p_n} d_{k,n}^*\ket{c_1}\bra{c_k^\perp}
	\end{equation}
	with $p_n, d_{k,n}$ from above. Since
	\begin{align}
		\sum_{n=1}^{d}F_n^\dagger F_n & = \sum_{k,l,n=1}^{d}\sqrt{p_n}d_{k,n} \sqrt{p_n}d_{l,n}^* \ket{c_k^\perp}	\braket{c_1|c_1}\bra{c_l^\perp} \nonumber \\
		&=\sum_{k,l=1}^{d} \left(\sum_{n=1}^{d} p_n d_{k,n} d_{l,n}^*\right) \ket{c_k^\perp} \bra{c_l^\perp}\nonumber \\
		&=\sigma_r,
	\end{align}
	we have $\sum_m K_m^\dagger K_m +\sum_n F_n^\dagger F_n=\mathbbm{1}$. 
\end{proof}

\addtocounter {theorem} {1} 
\begin{proposition}
	The following functions are superposition measures as defined in Definition \ref{C}.
	\begin{enumerate}[label=\textbf{\arabic*.},ref=\ref{prop:Measures}.\arabic*,leftmargin=0cm,itemindent=.5cm,labelwidth=\itemindent,labelsep=0cm,align=left]
		\litem{The relative entropy of superposition} \label{propMeasure:Ent}
		\begin{align}
			M_{\mathrm{rel.ent}}(\rho)&=\min_{\sigma \in \mathcal{F}} S(\rho||\sigma),
		\end{align} \\
		where $S(\rho||\sigma)=\mathrm{tr}\left[\rho \log \rho\right]-\mathrm{tr}\left[\rho \log \sigma\right]$ denotes the quantum relative entropy. See \cite{baumgratz2014quantifying} for the case of coherence theory.
		\litem{The $l_1$-measure of superposition} \label{propMeasure:l1}
		\begin{equation}
			M_{l_1} (\rho)=\sum_{i\ne j} |\rho_{ij}| 
		\end{equation} 
		for $\rho=\sum_{ij} \rho_{ij} \ket{c_i} \bra{c_j}$. See again \cite{baumgratz2014quantifying} for the case of coherence theory. 
		\litem{The rank-measure of superposition} \label{propMeasure:Ran}
		\begin{align}
			&M_\text{rank}(\ket{\psi})=\log(r_S(\ket{\psi})), \nonumber \\
			&M_\text{rank}(\rho)=\min_{\rho=\sum_i\lambda_i\ \ket{\psi_i}\bra{\psi_i}}\sum_i \lambda_i M_\text{rank}(\ket{\psi_i}).
		\end{align} 
		\litem{The robustness of superposition} \label{propMeasure:Rob}
		\begin{align}
			M_R(\rho)&=\min_{\tau \text{  density matrix}} \left\lbrace s\ge0 : \frac{\rho+s\tau}{1+s}\in\mathcal{F} \right\rbrace.
		\end{align}
		This quantity has an operational interpretation in the limit of coherence theory: the robustness
		of coherence quantifies the advantage enabled by a quantum state in a phase discrimination task \cite{napoli2016robustness}.
	\end{enumerate}
\end{proposition}

\noindent\textbf{Proof of Proposition \ref{propMeasure:Ent}:}
The quantum relative entropy has some useful properties derived for example in \cite{wehrl1978general}. For quantum states $\rho$ and $\sigma$,
\begin{equation}
	S(\rho||\sigma) \ge 0,
\end{equation} 
where equality holds if and only if $\rho=\sigma$. Thus property $(S1)$ is proven. In addition, the relative entropy is jointly convex. Therefore, as stated in the main text, $M_{\mathrm{rel.ent}}(\rho)$ 
satisfies $(S3)$. Property $(S2b)$ can be proved as in the Supplemental Material of \cite{baumgratz2014quantifying} for coherence theory. In \cite{vedral1998Entanglement}, it has been shown that the relative entropy satisfies 
certain conditions (called (F1)-(F5) there). Thus we can apply their theorem 2 telling us
\begin{equation}
	\sum_{n} S(L_n \rho L_n^\dagger||L_n \delta L_n^\dagger)\le S(\rho||\delta)
\end{equation}
for any CPTP map with Kraus operator decomposition $\{L_n\}$. With their condition (F4) stating 
\begin{align}
	\sum_n \text{tr}\left[L_n \rho L_n^\dagger\right]& S\left(\frac{L_n\rho L_n^\dagger}{\text{tr}\left[L_n \rho L_n^\dagger\right]}\right|\left|\frac{L_n\delta L_n^\dagger}{\text{tr}\left[L_n \delta L_n^\dagger\right]}\right) \nonumber \\
	& \le \sum_{n} S\left(L_n \rho L_n^\dagger\right|\left|L_n \delta L_n^\dagger\right)
\end{align} 
we find 
\begin{align}
	\sum_n \text{tr}\left[L_n \rho L_n^\dagger\right] &S\left(\frac{L_n\rho L_n^\dagger}{\text{tr}\left[L_n \rho L_n^\dagger\right]}\right|\left|\frac{L_n\delta L_n^\dagger}{\text{tr}\left[L_n \delta L_n^\dagger\right]}\right) \nonumber \\
	&\le S(\rho||\delta)
\end{align} 
again for any CPTP map with Kraus operator decomposition $\{L_n\}$. Now assume $p_n$ and $K_n$ as in $(S2b)$.
For a superposition-free state $\delta^\star$ minimizing the relative entropy with respect to $\rho$ we then find
\begin{align}
	M_{\mathrm{rel.ent}}(\rho)&=S(\rho||\delta^\star) \nonumber \\
	&\ge \sum_n p_n S(\rho_n||K_n\delta^\star K_n^\dagger/\text{tr}\left[K_n \delta^\star K_n^\dagger\right]) \nonumber \\
	&\ge \sum_n p_n \min_{\sigma_n \in \mathcal{F}}S\left(\rho_n||\sigma_n\right) \nonumber \\
	&= \sum_n p_n M_{\mathrm{rel.ent}}(\rho_n),
\end{align}
where we have used that $K_n \delta^\star K_n^\dagger \in \mathcal{F}$.\hfill$\blacksquare$\newline

\noindent\textbf{Proof of Proposition \ref{propMeasure:l1}:} 	
Obviously $M_{l_1}$ maps all quantum states to the positive real numbers and $(S1)$ is fulfilled. 
To prove that $M_{l_1}$ satisfies property $(S3)$ is straight forward.
With the notation
\begin{equation}
	\rho_n =\sum_{kl} \rho_{n_{kl}} \ket{c_k}\bra{c_l}
\end{equation}
we have
\begin{align}
	M_{l_1}\left(\sum_n p_n \rho_n \right)&=\sum_{i \ne j} \left| \left(\sum_n p_n \rho_n \right)_{ij}\right| \nonumber \\
	&=\sum_{i \ne j} \left| \left( \sum_{kl}\left(\sum_n p_n \rho_{n_{kl}}\right) \ket{c_k} \bra{c_l} \right)_{ij}\right| \nonumber \\
	&=\sum_{i \ne j} \left| \sum_n p_n \rho_{n_{ij}} \right| \nonumber \\
	&\le\sum_{i \ne j} \sum_n \left| p_n  \rho_{n_{ij}}\right| \nonumber \\
	&=\sum_n p_n \sum_{i \ne j} \left|   \rho_{n_{ij}}\right| \nonumber \\
	&=\sum_n p_n M_{l_1}(\rho_n).
\end{align}			
The proof of property $(S2b)$ is a bit more involved and inspired by the proof in the Supplemental Material of \cite{baumgratz2014quantifying} for the case of coherence theory. We write again  
\begin{equation}
	\rho_n =\sum_{kl} \rho_{n_{kl}} \ket{c_k}\bra{c_l}
\end{equation}
and 
\begin{equation}
	K_n=\sum_{kl}K_{n_{kl}} \ket{c_k} \bra{c_l}
\end{equation}
alike.
With this notation at hand we start with
\begin{align}\label{startEq}
	\sum_{n} p_n M_{l_1}(\rho_n) &= \sum_{n} p_n \sum_{i \ne j}|\rho_{n_{ij}}|=\sum_{n} \sum_{i \ne j}\left|(K_n\rho K_n^\dagger)_{ij}\right|.
\end{align}
\begin{widetext}
	Next step is to write down the summands $\left|(K_n\rho K_n^\dagger)_{ij}\right|$ explicitly,
	\begin{align} \label{summand}
		\left|(K_n\rho K_n^\dagger)_{ij}\right|&=\left|\left(\sum_{klstxy} K_{n_{kl}} \ket{c_k} \bra{c_l} \rho_{st} \ket{c_s} \bra{c_t} K^*_{n_{xy}}\ket{c_y}\bra{c_x}\right)_{ij}\right| \nonumber \\
		&=\left|\sum_{lsty} K_{n_{il}} K^*_{n_{jy}} \rho_{st} \braket{c_l| c_s} \braket{c_t|c_y}\right| \nonumber \\
		&=\left|\sum_{lsy} K_{n_{il}} K^*_{n_{jy}} \rho_{ss} \braket{c_l| c_s} \braket{c_s|c_y} +\sum_{s\ne t} \sum_{ly} K_{n_{il}} K^*_{n_{jy}} \rho_{st} \braket{c_l| c_s} \braket{c_t|c_y} \right|. 
	\end{align}
	With the general representation of superposition-free states
	\begin{equation}
		\rho_{\text{cl}}=\sum_{i} \rho_{ii} \ket{c_i} \bra{c_i}
	\end{equation}
	and the fact that superposition-free Kraus operators map free states to free states, we find 
	\begin{align} \label{equalSummand}
		\sum_{lsy} K_{n_{il}} K^*_{n_{jy}} \rho_{ss} \braket{c_l| c_s} \braket{c_s|c_y}= \left(K_n \rho_{\text{cl}} K_n^\dagger\right)_{ij}=\delta_{ij}\left(K_n \rho_{\text{cl}} K_n^\dagger\right)_{ii}.
	\end{align}
	Now we plug equations (\ref{summand}) and (\ref{equalSummand}) subsequently back into equation (\ref{startEq}),
	\begin{align}\label{vorCauchy}
		\sum_{n} p_n M_{l_1}(\rho_n) &\underset{(\text{\ref{startEq}})}{=} \sum_{n} \sum_{i \ne j}\left|(K_n\rho K_n^\dagger)_{ij}\right| \nonumber \\
		&\underset{(\text{\ref{summand})}}{=}\sum_{n} \sum_{i \ne j}\left|\sum_{lsy} K_{n_{il}} K^*_{n_{jy}} \rho_{ss} \braket{c_l| c_s} \braket{c_s|c_y} +\sum_{s\ne t} \sum_{ly} K_{n_{il}} K^*_{n_{jy}} \rho_{st} \braket{c_l| c_s} \braket{c_t|c_y} \right|\nonumber\\
		&\underset{(\text{\ref{equalSummand}})}{=}\sum_{n} \sum_{i \ne j}\left|\delta_{ij}\left(K_n \rho_{\text{cl}} K_n^\dagger\right)_{ii} +\sum_{s\ne t} \sum_{ly} K_{n_{il}} K^*_{n_{jy}} \rho_{st} \braket{c_l| c_s} \braket{c_t|c_y} \right|\nonumber\\
		&=\sum_{n} \sum_{i\ne j}\left|\sum_{s\ne t} \rho_{st} \sum_{ly} K_{n_{il}} K^*_{n_{jy}} \braket{c_l|c_s}\braket{c_t|c_y}\right| \nonumber \\
		&\le \sum_{s\ne t} |\rho_{st}| \sum_{n} \sum_{i\ne j} \left| \sum_{ly} K_{n_{il}} K^*_{n_{jy}} \braket{c_l|c_s} \braket{c_t|c_y}\right|.
	\end{align}
	Regarding the last part of this expression, we find
	\begin{align}\label{Cauchy}
		\sum_{n} \sum_{i\ne j} &\left| \sum_{ly} K_{n_{il}} K^*_{n_{jy}} \braket{c_l|c_s} \braket{c_t|c_y}\right| \nonumber \\
		&=\sum_{n} \sum_{i\ne j} \left| \left(\sum_{l} K_{n_{il}} \braket{c_l|c_s} \right) \left(\sum_y K^*_{n_{jy}} \braket{c_t|c_y} \right)\right| \nonumber \\
		&\le \sum_n \sum_i \left|\sum_{l} K_{n_{il}} \braket{c_l|c_s}\right| \sum_j \left|\sum_y K^*_{n_{jy}} \braket{c_t|c_y} \right| \nonumber \\
		&= \sum_n \sum_i \left|\sum_{l} K_{n_{il}} \braket{c_l|c_s}\right| \sum_j \left|\sum_y K_{n_{jy}} \braket{c_y|c_t} \right| \nonumber \\
		&\le \sqrt{\left(\sum_n \left( \sum_i \left|\sum_{l} K_{n_{il}} \braket{c_l|c_s}\right| \right)^2 \right) \left( \sum_n \left( \sum_j \left|\sum_{y} K_{n_{jy}} \braket{c_y|c_t}\right| \right)^2\right)} ,
	\end{align}
	where the Cauchy-Schwarz inequality has been used in the last line.
	Again we will simplify a part of this expression,
	\begin{align} \label{unterWurzel}
		\sum_n \left( \sum_i \left|\sum_{l} K_{n_{il}} \braket{c_l|c_s}\right| \right)^2 
		&=\sum_n \sum_{ij} \left|\sum_{l} K_{n_{il}} \braket{c_l|c_s}\right| \left|\sum_{k} K_{n_{jk}} \braket{c_k|c_s}\right| \nonumber \\
		&=\sum_n \sum_{ij} \left|\sum_{l} K_{n_{il}} \braket{c_l|c_s}\right| \left|\sum_{k} K^*_{n_{jk}} \braket{c_s|c_k}\right| \nonumber \\
		&=\sum_n \sum_{ij} \left|\sum_{lk} K_{n_{il}} K^*_{n_{jk}} \braket{c_l|c_s} \braket{c_s|c_k}\right|. 
	\end{align}
	Using 
	\begin{align}
		K_n\ket{c_s}\bra{c_s}K_n^\dagger=\sum_{iljk} K_{n_{il}} K^*_{n_{jk}} \ket{c_i}\braket{c_l|c_s} \braket{c_s|c_k} \bra{c_j},
	\end{align}
	we can write
	\begin{align} \label{Kroneckers}
		\left|\sum_{lk} K_{n_{il}} K^*_{n_{jk}} \braket{c_l|c_s} \braket{c_s|c_k}\right| &= \left|\left(K_n\ket{c_s}\bra{c_s}K_n^\dagger\right)_{ij}\right|\nonumber \\ 
		&= \left|\left(\sum_{kl} c_{kn}\ket{c_{f_n(k)}}\braket{c_k^\perp|c_s}\braket{c_s|c_l^\perp}\bra{c_{f_n(l)}}c_{ln}^* \right)_{ij}\right| \nonumber \\
		&= \left|\left(|c_{sn}|^2 \ket{c_{f_n(s)}}\bra{c_{f_n(s)}} \right)_{ij}\right| \nonumber \\
		&= \delta_{i,j} \delta_{i,f_n(s)} |c_{sn}|^2,
	\end{align}
	where the explicit representation of the Kraus operators from equation (\ref{classicalKrausOperator}) has been used.
	Writing
	\begin{equation}
		\tilde{p}_n=\mathrm{tr}(K_n \ket{c_s}\bra{c_s}K_n^\dagger)=|c_{sn}|^2 ,
	\end{equation}
	we can interpret $\tilde{p}_n$ as the probability of outcome $n$ when applying the operation to the state $\ket{c_s}\bra{c_s}$. Thus we have 
	\begin{equation} \label{sumOne}
		1=\sum_n \tilde{p}_n= \sum_n |c_{sn}|^2.
	\end{equation}
	Putting the pieces together leads to 
	\begin{align} \label{teileCauchy}
		\sum_n \left( \sum_i \left|\sum_{l} K_{n_{il}} \braket{c_l|c_s}\right| \right)^2&\underset{(\text{\ref{unterWurzel}})}{=}\sum_n \sum_{ij} \left|\sum_{lk} K_{n_{il}} K^*_{n_{jk}} \braket{c_l|c_s} \braket{c_s|c_k}\right| \nonumber \\ 
		&\underset{(\text{\ref{Kroneckers}})}{=}\sum_{nij} \delta_{i,j}\delta_{i,f_n(s)}|c_{sn}|^2\nonumber\\
		&=\sum_{n} |c_{sn}|^2
		\underset{(\text{\ref{sumOne})}}{=}1.
	\end{align}
	Finally we can finish the proof through
	\begin{align}
		\sum_{n} p_n M_{l_1}( \rho_n) &\underset{(\text{\ref{vorCauchy}})}{\le}\sum_{s\ne t} |\rho_{st}| \sum_{n} \sum_{i\ne j} \left| \sum_{ly} K_{n_{il}} K^*_{n_{jy}} \braket{c_l|c_s} \braket{c_t|c_y}\right| \nonumber \\
		&\underset{(\text{\ref{Cauchy}})}{\le} \sum_{s\ne t} |\rho_{st}| \sqrt{\left(\sum_n \left( \sum_i \left|\sum_{l} K_{n_{il}} \braket{c_l|c_s}\right| \right)^2 \right) \left( \sum_n \left( \sum_j \left|\sum_{y} K_{n_{jy}} \braket{c_y|c_t}\right| \right)^2\right)} \nonumber \\
		&\underset{(\text{\ref{teileCauchy}})}{=}
		\sum_{s\ne t} |\rho_{st}|=M_{l_1}(\rho).
	\end{align}
	\hfill$\blacksquare$\newline
\end{widetext}

\noindent\textbf{Proof of Proposition \ref{propMeasure:Ran}:}	
It can be seen directly that $(S1)$ is fulfilled. 
To show $(S2b)$, let $p_n$, $\rho_n$ and $K_n$ be as in the definition. Assume without loss of generality $p_n\ne 0\ \forall n$. Define for every state $\rho$ a decomposition into pure states by $\rho=\sum_i\tilde{\lambda}_i \ket{\tilde{\psi}_i}\bra{\tilde{\psi}_i}$ such that 
\begin{equation}
	M_\text{rank}(\rho)=\sum_i \tilde{\lambda}_i\ M_\text{rank}(\ket{\tilde{\psi}_i}).
\end{equation}
Further define
\begin{equation}
	\ket{\tilde{\psi}_i^{(n)}}=\frac{K_n \ket{\tilde{\psi}_i}}{\sqrt{p_n}}.
\end{equation}
Now notice that 
\begin{align}
	\rho_n&=\frac{K_n \rho K_n^\dagger}{p_n}=\sum_i \frac{\tilde{\lambda}_i}{p_n}K_n\ket{\tilde{\psi}_i}\bra{\tilde{\psi}_i}K_n^\dagger\nonumber \\
	&=\sum_i \tilde{\lambda}_i \ket{\tilde{\psi}_i^{(n)}}\bra{\tilde{\psi}_i^{(n)}}.
\end{align}
Remember that the superposition rank can never increase under the action of a superposition-free Kraus operator. Since $log(r_S)$ decreases if $r_S$ decreases, we can finish the proof of $(S2b)$ by
\begin{align}
	M_\text{rank}(\rho)&=\sum_i \tilde{\lambda}_i\ M_\text{rank}(\ket{\tilde{\psi}_i}) \nonumber \\
	&=\sum_n p_n \sum_i \tilde{\lambda}_i\ M_\text{rank}(\ket{\tilde{\psi}_i})\nonumber \\
	& \ge \sum_n p_n \sum_i \tilde{\lambda}_i\ M_\text{rank}(\ket{\tilde{\psi}_i^{(n)}}) \nonumber \\
	&  \ge \sum_n p_n \min \sum_i \lambda_i^{(n)}\ M_\text{rank}(\ket{\psi_i^{(n)}}) \nonumber \\
	&=\sum_n p_n\ M_\text{rank}(\rho_n),
\end{align}
where the minimization in the second last line runs over all decompositions $\rho_n=\sum_i\lambda_i^{(n)}\ket{\psi_i^{(n)}}\bra{\psi_i^{(n)}}$.
In order to show $(S3)$, follow the same spirit. Again, define  $ \sigma_n=\sum_i\tilde{\lambda}_i^{(n)} \ket{\tilde{\psi}_i^{(n)}}\bra{\tilde{\psi}_i^{(n)}}$ such that 
\begin{equation}
	M_\text{rank}(\sigma_n)=\sum_i \tilde{\lambda}_i^{(n)}\ M_\text{rank}(\ket{\tilde{\psi}_i^{(n)}})
\end{equation}
and note that
\begin{equation}
	\sum_n p_n \sigma_n =\sum_{n,i}p_n \tilde{\lambda}_i^{(n)} \ket{\tilde{\psi}_i^{(n)}}\bra{\tilde{\psi}_i^{(n)}}.
\end{equation}
Hence
\begin{align}
	M_\text{rank}\left(\sum_n p_n \sigma_n\right)&=\min_{\sum_n p_n \sigma_n=\lambda_i \ket{\psi_i}\bra{\psi_i}}\sum_i\lambda_i\ M_\text{rank}( \ket{\psi_i})\nonumber \\
	&\le \sum_{n,i}p_n \tilde{\lambda}_i^{(n)} M_\text{rank}(\ket{\tilde{\psi}_i^{(n)}}) \nonumber \\
	&=\sum_n p_n\ M_\text{rank} (\sigma_n).
\end{align}\hfill$\blacksquare$\newline

\noindent\textbf{Proof of Proposition \ref{propMeasure:Rob}:}	
Property $(S1)$ is given by definition. To prove $(S2b)$ decompose
\begin{equation}
	\rho=(1+\tilde{s}) \tilde{\delta}- \tilde{s}\tilde{\tau}
\end{equation}
such that $\tilde{s}=M_R(\rho), \tilde{\delta} \in \mathcal{F}$ and $\tilde{\tau}$ a density matrix.
Hence
\begin{align}
	K_n\rho K_n^\dagger&=(1+\tilde{s})K_n \tilde{\delta}K_n^\dagger- \tilde{s}K_n\tilde{\tau}K_n^\dagger, \nonumber\\
	p_n&=\mathrm{tr}\left(K_n\rho K_n^\dagger\right)\nonumber \\
	&=(1+\tilde{s})\ \mathrm{tr}\left(K_n \tilde{\delta}K_n^\dagger\right)- \tilde{s}\ \mathrm{tr}\left(K_n\tilde{\tau}K_n^\dagger\right).
\end{align}
Assume without loss of generality $p_n \ne 0,\mathrm{tr}\left(K_n \tilde{\delta}K_n^\dagger\right)\ne 0$ and $ \mathrm{tr}\left(K_n\tilde{\tau}K_n^\dagger\right)\ne 0$.
Then
\begin{align}
	\rho_n=&\frac{K_n\rho K_n^\dagger}{p_n} \nonumber \\
	=&\frac{1+\tilde{s}}{p_n}\ \mathrm{tr}(K_n\tilde{\delta}K_n^\dagger) \frac{K_n\tilde{\delta}K_n^\dagger}{\mathrm{tr}(K_n\tilde{\delta}K_n^\dagger)}\nonumber \\
	&-\frac{\tilde{s}}{p_n}\ \mathrm{tr}(K_n\tilde{\tau}K_n^\dagger) \frac{K_n\tilde{\tau}K_n^\dagger}{\mathrm{tr}(K_n\tilde{\tau}K_n^\dagger)} \nonumber \\
	=&(1+s_n)\tilde{\delta}^{(n)}-s_n \tilde{\tau}^{(n)}
\end{align}
for $s_n=\frac{\tilde{s} \ \mathrm{tr}(K_n\tilde{\tau}K_n^\dagger)}{p_n} \ge 0,\  \tilde{\delta}^{(n)}=\frac{K_n\tilde{\delta}K_n^\dagger}{\mathrm{tr}(K_n\tilde{\delta}K_n^\dagger)} \in \mathcal{F}$ and a density matrix $\tilde{\tau}^{(n)}= \frac{K_n\tilde{\tau}K_n^\dagger}{\mathrm{tr}(K_n\tilde{\tau}K_n^\dagger)}$. Thus $M_R(\rho_n)\le s_n$ and
\begin{align}
	\sum_n p_n &M_R(\rho_n)\le \sum_n p_n s_n=\sum_n \tilde{s}\ \mathrm{tr}(K_n\tilde{\tau}K_n^\dagger) \nonumber \\
	&=\tilde{s}\  \mathrm{tr}\left(\sum_n K_n\tilde{\tau}K_n^\dagger\right)=\tilde{s}=M_R(\rho).
\end{align}			
In order to prove $(S3)$, decompose again two density matrices $\rho_1, \rho_2$ into
\begin{equation}
	\rho_i=(1+\tilde{s}_i) \tilde{\delta}_i- \tilde{s}_i\tilde{\tau}_i
\end{equation}
such that $\tilde{s}_i=M_R(\rho_i), \tilde{\delta}_i \in \mathcal{F}$ and $\tilde{\tau}_i$ a density matrix. For $0\le p \le 1$ and without loss of generality $\tilde{s}_i \ne 0$, 
\begin{align}
	p \rho_1 &+(1-p) \rho_2\nonumber \\
	=&p\left[(1+\tilde{s}_1) \tilde{\delta}_1- \tilde{s}_1\tilde{\tau}_1\right] +(1-p)\left[(1+\tilde{s}_2) \tilde{\delta}_2- \tilde{s}_2\tilde{\tau}_2\right] \nonumber \\
	=&\left[1+p \tilde{s}_1+(1-p)\tilde{s}_2\right] \frac{p(1+\tilde{s}_1)\tilde{\delta}_1+(1-p)(1+\tilde{s}_2)\tilde{\delta}_2}{1+p \tilde{s}_1+(1-p)\tilde{s}_2} \nonumber \\
	&-\left[p\tilde{s}_1+(1-p)\tilde{s}_2\right] \frac{p\tilde{s}_1\tilde{\tau}_1+(1-p)\tilde{s}_2\tilde{\tau}_2}{p\tilde{s}_1+(1-p)\tilde{s}_2} \nonumber \\
	=&(1+s) \delta-s\tau
\end{align}
for $s=p\tilde{s}_1+(1-p)\tilde{s}_2 \ge 0,\ \delta=\frac{p(1+\tilde{s}_1)\tilde{\delta}_1+(1-p)(1+\tilde{s}_2)\tilde{\delta}_2}{1+p \tilde{s}_1+(1-p)\tilde{s}_2} \in \mathcal{F}$ and a density matrix 
$\tau=\frac{p\tilde{s}_1\tilde{\tau}_1+(1-p)\tilde{s}_2\tilde{\tau}_2}{p\tilde{s}_1+(1-p)\tilde{s}_2}$.
Hence 
\begin{align}
	M_R&(p\rho_1+(1-p)\rho_2)\nonumber \\
	=&\min \left\lbrace t\ge0 : p\rho_1+(1-p)\rho_2=(1+t)\delta-t\tau:\right.\nonumber \\
	&\left.\delta\in \mathcal{F}, \tau \text{  density matrix} \right\rbrace \nonumber \\
	\le& s \nonumber \\
	=&p\tilde{s}_1+(1-p)\tilde{s}_2\nonumber \\ 
	=&pM_R(\rho_1)+(1-p)M_R(\rho_2)
\end{align}
which proves convexity.\hfill$\blacksquare$\newline

\begin{proposition}
	For qubit systems with $\braket{c_1|c_2}\ne 0$, there exists a single state with maximal superposition.
	For higher dimensions, there exists no state with maximal superposition in general.
\end{proposition}

\begin{proof}
	To prove this proposition, we will use again the representation introduced in section \ref{Free operations on qubits} and the four different types of superposition-free Kraus operators given in equation (\ref{classicalKrausExplicit}). To show that $\ket{m_2}$, the candidate from lemma \ref{LemmaQubitMaxNonClCand}, has indeed maximal superposition, we will explicitly construct a superposition-free operation that generates a target state 
	\begin{equation}
		\ket{\psi_t}=\begin{pmatrix}
			c_{\theta/2} \\
			e^{i \phi}s_{\theta/2}
		\end{pmatrix}
	\end{equation} from $\ket{m_2}$ with certainty. In order to achieve this, we choose  one superposition-free Kraus operator of each type with
	\begin{align}
		\delta&=\frac{1}{2(1+a)}\left[(B+a)c_{\theta/2}-(a+A)e^{i\phi}s_{\theta/2}\right], \nonumber \\
		\gamma&=\frac{1}{2(1+a)} \left[(A+a)c_{\theta/2}-(a+B)e^{i\phi}s_{\theta/2}\right], \nonumber \\
		\epsilon&=-\frac{1}{2(1+a)}\left[(B+a)c_{\theta/2}-(a+A)e^{i\phi}s_{\theta/2}\right], \nonumber \\
		\xi&=-\frac{1}{2(1+a)}\left[(A+a)c_{\theta/2}-(a+B)e^{i\phi}s_{\theta/2}\right], \nonumber \\
		\alpha&=\beta=\mu=\nu=\sqrt{\frac{a}{2(1+a)}(1+c_\phi s_\theta)},
	\end{align}
	where $A$ and $B$ are defined as in equation (\ref{A,B}).
	With the help of equation (\ref{qubitKrausConditions}), it is easy to check that they form a trace preserving operation, i.e. $\sum_{n=1}^{4}K_n^\dagger K_n=\mathbbm{1}$.
	Remember that in the representation chosen, $\ket{m_2}$ is given by	
	\begin{equation}
		\ket{m_2}=\begin{pmatrix}
			1 \\
			-1
		\end{pmatrix}.
	\end{equation}
	Thus one can see directly that the four Kraus operators have the property 
	\begin{align}
		K_2 \ket{m_2}&=K_4\ket{m_2} =\frac{1}{\sqrt{2}}\ket{\psi_t}, \nonumber \\
		K_1 \ket{m_2}&=K_3\ket{m_2}=0.
	\end{align} 
	With the help of theorem \ref{Theorem:CompletionOfMaps}, the explicit construction of $K_1$ and $K_3$ would not have been necessary \footnote{For $a=0$, we have $K_1=K_3=0$ and we need only two superposition-free Kraus operators (see also the Supplemental Material of \cite{baumgratz2014quantifying}). For $a\ne0$, in general we need four superposition-free Kraus operators to make the entire operation trace preserving. However, the actual transformation is still done by $K_2$ and $K_4$. We will encounter a similar case in the proof of theorem \ref{theorem:UnitaryReali}}. 
	The generation of mixed target states $\rho_M$ is again possible due to linearity. Since $\rho_M$ can be decomposed into pure states through $\rho_M=\sum_{i} p_i \ket{\phi_i}\bra{\phi_i}$, we can just apply the operation $\Phi_M=\sum_{i} p_i \Phi_i$ to $\ket{m_2}$ where $\Phi_i$ generates $\ket{\phi_i}$ from $\ket{m_2}$. As stated in lemma \ref{LemmaQubitMaxNonClCand}, $\ket{m_2}$ is the only qubit state with maximal superposition for $\braket{c_1|c_2}\ne 0$.
	
	To prove the second part of the proposition, we show with the help of the $l_1$-measure of superposition and some tools from optimization theory (described for example in \cite{boyd2004convex}) that for a specific superposition-free basis in $d$=3, there exists no state with maximal superposition.
	
	The main idea is that a state with maximal superposition  has to maximize the $l_1$-measure of superposition since a superposition measure cannot increase under superposition-free operations. 
	First step is to define the superposition-free states in a fixed dimension $d$. Then all states maximizing $M_{l_1}$ have to be determined. Once these candidate states are found, one can use the optimization problem from the main text to calculate the maximal transformation probability to a given target state. If the solution is smaller than one, the considered candidate state cannot have maximal superposition. If all candidate states are ruled out, one concludes that there is no state with maximal superposition (for this set of superposition-free states). 
	For specific choices of superposition-free states, this can be done analytically using the concept of duality. In the case of $d=3$, we choose the pure superposition-free states to be represented by
	\begin{align}
		\ket{c_1}=\frac{1}{\sqrt{2}}\begin{pmatrix}
			0 \\
			1\\
			1\\
		\end{pmatrix}, \ 
		\ket{c_2}=\frac{1}{\sqrt{2}}\begin{pmatrix}
			1 \\
			0\\
			1\\
		\end{pmatrix}, \ 
		\ket{c_3}=\frac{1}{\sqrt{2}}\begin{pmatrix}
			1 \\
			1\\
			0\\
		\end{pmatrix}. 
	\end{align}
	Then the (not normalized) reciprocal vectors introduced in the main text are given by
	\begin{align}
		&\ket{c_1^\perp}=\frac{1}{\sqrt{2}}\begin{pmatrix}
			-1 \\
			1\\
			1\\
		\end{pmatrix}, \ 
		\ket{c_2^\perp}=\frac{1}{\sqrt{2}}\begin{pmatrix}
			1 \\
			-1\\
			1\\
		\end{pmatrix}, \nonumber \\
		&\ket{c_3^\perp}=\frac{1}{\sqrt{2}}\begin{pmatrix}
			1 \\
			1\\
			-1\\
		\end{pmatrix}.
	\end{align}
	It is easy to check that
	\begin{align}
		\braket{c_i|c_j}=\frac{1}{2}\left( 1+\delta_{i,j}\right).
	\end{align}
	Because of this symmetry, every candidate state has to be of the form
	\begin{align}
		\ket{m_d}=\mathcal{N}\sum_i e^{i \phi_i}\ket{c_i},
	\end{align}
	where $\mathcal{N}$ is positive and $\phi_3=0$ by choice. The $l_1$ measure of superposition is then given by
	\begin{align}
		M_{l_1}\left(\ket{m_d}\right)=  3(3-1)\mathcal{N}^2=6\mathcal{N}^2.
	\end{align} 
	Thus to find proper candidate states maximizing $M_{l_1}$, we have to solve the following optimization problem
	\begin{alignat}{2}\label{opti1}
		&\text{maximize} \qquad&&M_{l_1}\left(\ket{m_d}\right)=6 \mathcal{N}^2 \nonumber \\
		&\text{subject to} &&1=\braket{m_d|m_d} \nonumber \\
		& &&=\mathcal{N}^2\left[ 3+s_{\phi_1} s_{\phi_2} +c_{\phi_1} c_{\phi_2}+c_{\phi_1}+c_{\phi_2}   \right] .
	\end{alignat}
	This is equivalent to minimizing $s_{\phi_1} s_{\phi_2} +c_{\phi_1} c_{\phi_2}+c_{\phi_1}+c_{\phi_2}$. We choose $\phi_{1,2} \in [0,2 \pi)$ and do the change of variables
	\begin{align}
		\alpha=\frac{\phi_1-\phi_2}{2} \in (-\pi,\pi], \qquad \beta=\frac{\phi_1+\phi_2}{2}\in [0,2\pi).
	\end{align}
	This transforms the initial optimization problem (\ref{opti1}) into the problem
	\begin{alignat}{2}
		&\text{minimize} \qquad&& c_{2\alpha}+2c_\alpha c_\beta \nonumber \\
		&\text{subject to} &&\alpha \in (-\pi,\pi] \qquad\nonumber \\
		& && \beta\in [0,2\pi),
	\end{alignat}
	which can be solved considering three separate cases. 
	
	\textit{First case ($c_\alpha>0$):}
	In this case, in order to minimize $c_{2\alpha}+2c_\alpha c_\beta$, we need $c_\beta=-1 \Leftrightarrow \beta=\pi$. The problem reduces to 
	\begin{alignat}{2}
		&\text{minimize} \qquad&& c_{2\alpha}-2 c_\alpha \nonumber \\
		&\text{subject to} &&\alpha \in \left(-\frac{\pi}{2},\frac{\pi}{2}\right).
	\end{alignat}
	The solution is $-\frac{3}{2}$ for $\alpha=\pm \frac{\pi}{3}$ which leads to 
	\begin{alignat}{3}
		&\mathcal{N}^{(1)}=\sqrt{\frac{2}{3}}, \qquad&& \phi_1^{(1)}=\frac{4\pi}{3},\qquad && \phi_2^{(1)}=\frac{2\pi}{3}, \nonumber \\
		&\mathcal{N}^{(2)}=\sqrt{\frac{2}{3}}, \qquad&& \phi_1^{(2)}=\frac{2\pi}{3},\qquad && \phi_2^{(2)}=\frac{4\pi}{3}. 
	\end{alignat}
	
	\textit{Second case ($c_\alpha<0$):}
	In order to minimize $c_{2\alpha}+2c_\alpha c_\beta$, we now need $c_\beta=1 \Leftrightarrow \beta=0$ and the remaining problem is
	\begin{alignat}{2}
		&\text{minimize} \qquad&& c_{2\alpha}+2 c_\alpha \nonumber \\
		&\text{subject to} &&\alpha \in \left(-\pi,-\frac{\pi}{2}\right)\cup \left(-\frac{\pi}{2},\pi\right).
	\end{alignat}
	Again, the solution is $-\frac{3}{2}$, this time for $\alpha=\pm \frac{2\pi}{3}$ which leads to 
	\begin{alignat}{3}
		&\mathcal{N}^{(3)}=\sqrt{\frac{2}{3}}, \qquad&& \phi_1^{(3)}=\frac{2\pi}{3},\qquad && \phi_2^{(3)}=-\frac{2\pi}{3}, \nonumber \\
		&\mathcal{N}^{(4)}=\sqrt{\frac{2}{3}}, \qquad&& \phi_1^{(4)}=-\frac{2\pi}{3},\qquad && \phi_2^{(4)}=\frac{2\pi}{3}. 
	\end{alignat}
	
	\textit{Third case ($c_\alpha=0 \Leftrightarrow \alpha=\pm\frac{\pi}{2}$):}
	This case is trivial since then $c_{2\alpha}+2c_\alpha c_\beta$ is  equal to -1 for all $\beta$ and no more global minima are found.
	
	Thus we remain with four candidate states which maximize $M_{l_1}$. They are given by
	\begin{align}
		&\ket{m_d^{(1)}}=\sqrt{\frac{2}{3}}\left[e^{i\frac{4\pi}{3}}\ket{c_1}+e^{i\frac{2\pi}{3}}\ket{c_2}+\ket{c_3}\right], \nonumber \\
		&\ket{m_d^{(2)}}=\sqrt{\frac{2}{3}}\left[e^{i\frac{2\pi}{3}}\ket{c_1}+e^{i\frac{4\pi}{3}}\ket{c_2}+\ket{c_3}\right], \nonumber \\
		&\ket{m_d^{(3)}}=\sqrt{\frac{2}{3}}\left[e^{i\frac{2\pi}{3}}\ket{c_1}+e^{-i\frac{2\pi}{3}}\ket{c_2}+\ket{c_3}\right], \nonumber \\
		&\ket{m_d^{(4)}}=\sqrt{\frac{2}{3}}\left[e^{-i\frac{2\pi}{3}}\ket{c_1}+e^{i\frac{2\pi}{3}}\ket{c_2}+\ket{c_3}\right] .
	\end{align}
	Next step is to show that none of these states can be transformed to every other state with certainty using superposition-free Kraus operators. Therefore we will focus on the transformation to target states of the form $\ket{\psi}=\sum_{k=1}^{3} \psi_k \ket{c_k}$ with $\psi_k\ne 0\ \forall k$ since then we can use the semidefinite program introduced in the main text:
	\begin{alignat}{2}
		&\text{maximize} \qquad&& \sum_n p_n \nonumber \\
		&\text{subject to} &&\sum_n p_n F_n^\dagger F_n \le \mathbbm{1} \nonumber \\
		& &&p_n\ge0 \qquad \forall n.
	\end{alignat}
	Note that $\{F_n\}$ depends on the choice of the target states and on the candidate state we want to rule out.
	This problem is equivalent to
	\begin{alignat}{2}
		&\text{minimize} \qquad&& -\sum_n p_n \nonumber \\
		&\text{subject to} &&\sum_n p_n F_n^\dagger F_n - \mathbbm{1} \le 0  \nonumber \\
		& &&p_n\ge0 \qquad \forall n.
	\end{alignat}
	If we interpret the condition $p_n\ge0 \ \forall n$ as a restriction of the domain of the objective function, the Lagrangian is given by
	\begin{align}
		L(p_n,\Lambda)&=-\sum_n p_n +\mathrm{tr}\left(\Lambda\left(\sum_n p_n F_n^\dagger F_n-\mathbbm{1}\right)\right) \nonumber \\
		&=\mathrm{tr}\left(-\frac{1}{d}\mathbbm{1}\sum_n p_n + \Lambda\left(\sum_n p_n F_n^\dagger F_n-\mathbbm{1}\right)\right)
	\end{align}
	and the dual function by
	\begin{align}
		g(\Lambda)&=\inf_{p_n \ge 0} L(p_n,\Lambda) \nonumber \\
		&=-\mathrm{tr}\left(\Lambda\right) +\inf_{p_n \ge 0} \sum_n p_n \mathrm{tr}\left(-\frac{1}{d}\mathbbm{1} + \Lambda F_n^\dagger F_n\right) \nonumber \\
		&=\begin{cases} -\infty &\text{if }  \exists n: \mathrm{tr}\left(-\frac{1}{d} \mathbbm{1}+\Lambda F_n^\dagger F_n \right) <0 \\ -\mathrm{tr}(\Lambda) &\text{if }  \mathrm{tr}\left(-\frac{1}{d} \mathbbm{1}+\Lambda F_n^\dagger F_n \right) \ge 0\ \forall n \end{cases}.
	\end{align}
	Hence the dual problem is 
	\begin{alignat}{2} \label{dualProblem}
		&\text{maximize} \qquad&& -\mathrm{tr}\left(\Lambda\right) \nonumber \\
		&\text{subject to} &&\mathrm{tr}\left(\Lambda F_n^\dagger F_n\right)\ge 1 \qquad \forall n \nonumber \\
		& &&\Lambda\ge0 .
	\end{alignat}
	By duality, we have
	\begin{equation}
		- \mathrm{tr}\Lambda\le -\sum_n p_n
	\end{equation}
	or 
	\begin{equation}
		\sum_n p_n \le 	\mathrm{tr}\Lambda
	\end{equation}
	for all feasible $\Lambda$. Thus if one can find a feasible $\Lambda$ with tr$(\Lambda)<1$, it is shown that there exists no deterministic free transformation between the candidate and the target state.
	For $\ket{\psi}=\sum_i\psi_i\ket{c_i}$ we have $\psi_i=\braket{c_i^\perp|\psi}$.
	Now we choose as a target state 
	\begin{equation}
		\ket{\psi}=\begin{pmatrix}
			1 \\ 0 \\ 0
		\end{pmatrix}=\frac{1}{\sqrt{2}}\bigg(-\ket{c_1}+\ket{c_2}+\ket{c_3}\bigg).
	\end{equation}	
	A direct calculation (using for example Mathematica) shows that in this case, we have tr$(F_n^\dagger F_n)=\frac{51}{16}$ for all $n$ and for all candidate states. If we choose 
	\begin{equation}
		\Lambda=\frac{t}{3}\ \mathbbm{1}_3,
	\end{equation}
	we have tr$\Lambda=t$ and $\Lambda\ge 0$ iff $t\ge 0$. Thus
	\begin{align}
		\mathrm{tr}\left(\Lambda F_n^\dagger F_n\right)=\frac{t}{3}\mathrm{tr}\left(F_n^\dagger F_n\right)=t\frac{17}{16} \qquad \forall n, \ \forall \ket{m_d^{(i)}}.
	\end{align}
	Now we can choose $t=\frac{16}{17}$ for $\Lambda$ to be feasible and bound the maximal conversion probability from above by $p_{\text{max}}\le \mathrm{tr}\Lambda=t=\frac{16}{17}<1$.
\end{proof}

\begin{theorem}
	Any unitary operation $U$ on a qubit can be implemented by means of $\mathcal{FO}$ and the consumption of an additional qubit state with maximal superposition $\ket{m_2}$ provided both qubits posses the same superposition-free basis. This means that for every $U$ there exists a fixed $\Psi\in \mathcal{FO}$ independent of $\rho_s$ acting on two qubits such that 
	\begin{equation}
		\Psi\left(\rho_s\otimes\ket{m_2}\bra{m_2}\right)=\left(U\rho_s U^\dagger\right)\otimes \rho_h,
	\end{equation}
	where $\rho_h$ is a superposition-free qubit state.
\end{theorem}
\begin{proof}
	In this proof, we will again make use of the computational basis introduced in section \ref{Free operations on qubits}. A matrix representation of a unitary $U$ with respect to this basis can be represented as
	\begin{equation}\label{unitary}
		U=\begin{pmatrix}  
			u_{00} & u_{01}\\
			u_{10} & u_{11} \\
		\end{pmatrix}=e^{i \phi_g}
		\begin{pmatrix}  
			e^{i \phi_1} c_\theta &  e^{i \phi_2} s_\theta\\
			- e^{-i \phi_2} s_\theta & e^{-i \phi_1} c_\theta \\
		\end{pmatrix},
	\end{equation}
	where $\phi_g$ is a (physically) unimportant global phase. Therefore we will drop it from now on. 
	Consider the two Kraus operators 
	\begin{align}
		F_0=&c_{00} \ket{c_1 c_1}\bra{c_1^\perp c_1^\perp}+c_{10} \ket{c_2 c_1}\bra{c_1^\perp c_2^\perp}\nonumber\\&+c_{01} \ket{c_1 c_1}\bra{c_2^\perp c_1^\perp}+c_{11} \ket{c_2 c_1}\bra{c_2^\perp c_2^\perp}, \nonumber\\
		F_1=&d_{00} \ket{c_1 c_2}\bra{c_1^\perp c_2^\perp}+d_{10} \ket{c_2 c_2}\bra{c_1^\perp c_1^\perp}\nonumber\\&+d_{01} \ket{c_1 c_2}\bra{c_2^\perp c_2^\perp}+d_{11} \ket{c_2 c_2}\bra{c_2^\perp c_1^\perp}.
	\end{align}
	We fix the coefficients by
	\begingroup
	\allowdisplaybreaks
	\begin{align}
		c_{00}&=\frac{A u_{00}+a(u_{01}-u_{10})-B u_{11}}{2 \sqrt{1+a}}, \nonumber \\
		c_{01}&=\frac{A u_{01}+a(u_{00}-u_{11})-B u_{10}}{2 \sqrt{1+a}}, \nonumber \\
		c_{10}&=\frac{B u_{01}+a(u_{00}-u_{11})-A u_{10}}{2 \sqrt{1+a}}, \nonumber \\
		c_{11}&=\frac{B u_{00}+a(u_{01}-u_{10})-A u_{11}}{2 \sqrt{1+a}}, \nonumber \\
		d_{00}&=\frac{B u_{11}+a(u_{10}-u_{01})-A u_{00}}{2 \sqrt{1+a}}, \nonumber \\
		d_{01}&=\frac{B u_{10}+a(u_{11}-u_{00})-A u_{01}}{2 \sqrt{1+a}}, \nonumber \\
		d_{10}&=\frac{A u_{10}+a(u_{11}-u_{00})-B u_{01}}{2 \sqrt{1+a}}, \nonumber \\
		d_{11}&=\frac{A u_{11}+a(u_{10}-u_{01})-B u_{00}}{2 \sqrt{1+a}}.
	\end{align}
	\endgroup
	Thus the coefficients depend on the unitary transformation $U$ and on the overlap $a$ of the pure superposition-free states (also through $A$ and $B$ which are defined as in equation (\ref{A,B})). The two Kraus operators $F_0$ and $F_1$ are superposition-free because they satisfy theorem \ref{theoremKrausOperators}. In addition, we have for every qubit state $\ket{s}$
	\begin{align}
		F_0 \ket{s} \otimes \ket{m_2}&= \frac{1}{\sqrt{2}}(U\ket{s})\otimes \ket{c_1}, \nonumber \\
		F_1 \ket{s} \otimes \ket{m_2}&= \frac{1}{\sqrt{2}}(U\ket{s})\otimes \ket{c_2}.
	\end{align}
	Making use of the explicit representation of $U$, the eigenvalues of $F_0^\dagger F_0+F_1^\dagger F_1$ can be calculated to be $(1,1,1,(\frac{1-a}{1+a})^2)$. With the help of theorem \ref{Theorem:CompletionOfMaps}, we  know that there exist additional Kraus operators $\{L_i\}$ such that $F_0^\dagger F_0+F_1^\dagger F_1+\sum_i L_i^\dagger L_i = \mathbbm{1}$ and $L_i\ket{s} \otimes \ket{m_2}=0$. 
	By linearity, this finishes the proof. 
\end{proof}

\section{On superposition transformations}\label{appendix:NonClassTrafoQubit} In this section, we have a short look at superposition-free transformations of qubit states. Therefore we use again the representation introduced in section \ref{Free operations on qubits}, the semidefinite program from the main text and its dual problem (\ref{dualProblem}).
For qubits, there are only two different useful Kraus operators contributing to the transformation except in the case where we transform to the two pure superposition-free states. 
Now we define the Bloch representation of the initial state $\ket{\psi}$ and the target state $\ket{\phi}$ by
\begin{align}
	\ket{\psi}&=c_{w/2} \ket{1}+s_{w/2}\ e^{i y} \ket{2}, \nonumber \\
	\ket{\phi}&=c_{x/2} \ket{1}+s_{x/2}\ e^{i z} \ket{2}. 
\end{align}
Further we consider
\begin{equation}
	\Lambda=\frac{t}{2}\ \mathbbm{1}_2
\end{equation}
in the dual problem.
Thus tr$\Lambda=t$ and $\Lambda\ge 0$ iff $t\ge 0$.
For $a=1/2,\ w=\pi/2,\ y=0$, one finds  
\begin{align}
	\mathrm{tr}\left(\Lambda F_1^\dagger F_1\right)=\mathrm{tr}\left(\Lambda F_2^\dagger F_2\right)=t\left[3-2c_z s_x\right] \ge t.
\end{align}
Remember $x \in \left[0,\pi\right]$ and $z\in \left[0,2\pi\right)$ due to the definition of the Bloch representation. Thus we only have equality for $x=\pi/2$ and $z=0$ which is equivalent to $\ket{\phi}=\ket{\psi}$. In case the above expression is strictly larger than $t$, we can always choose $t<1$ such that $\mathrm{tr}\left(\Lambda F_1^\dagger F_1\right)=1$. Then $\Lambda$ is feasible and the optimal probability of successfully transforming the initial state to the target state is smaller than one.
For transformation to the pure superposition-free states, we have to consider additional Kraus operators.
Thus finally we can conclude that there are only the three trivial pure states to which $\ket{\psi}=\frac{1}{\sqrt{2}}\left(\ket{1}+ \ket{2}\right)=\frac{1}{\sqrt{3}}\left(\ket{c_1}+ \ket{c_2}\right)$ can be transformed by $\mathcal{FO}$ with certainty: to itself and to the pure superposition-free states. 

This is surprising because as one can see easily with the help of the $l_1$-measure of superposition, the initial state under consideration contains a large amount of superposition and there are other states with less superposition to which this state cannot be transformed by $\mathcal{FO}$ with certainty nevertheless.
In contrast, for $a=0$, it can be shown with the help of \cite{du2015conditions} that a pure state can be transformed deterministically to all other pure states that are closer or equally close to the $z$-axis of the Bloch sphere (and have thus less superposition according to the $l_1$-measure of superposition).
A possible explanation to this difference could be that by breaking the symmetry on the Bloch sphere, one loses an entire class of superposition-free operations since rotations around the $z$-axis are no longer possible. Using duality, the maximal probability of conversion between qubit states was investigated numerically. The results are shown in figure \ref{sfig:QubitTrafo}.

\begin{figure*}
	\includegraphics[width=0.8\linewidth]{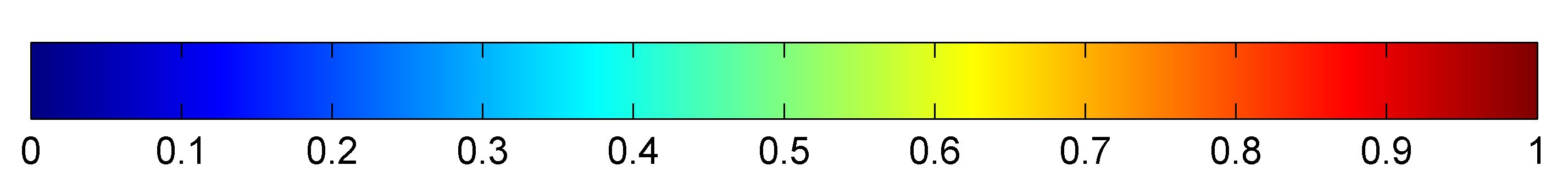}
	\subfloat[$a=1/2,\ \theta=\pi/4 ,\ \phi=\pi.$\label{sfig:a}]{%
		\includegraphics[trim = 60mm 40mm 60mm 35mm, clip, width=0.35\linewidth]{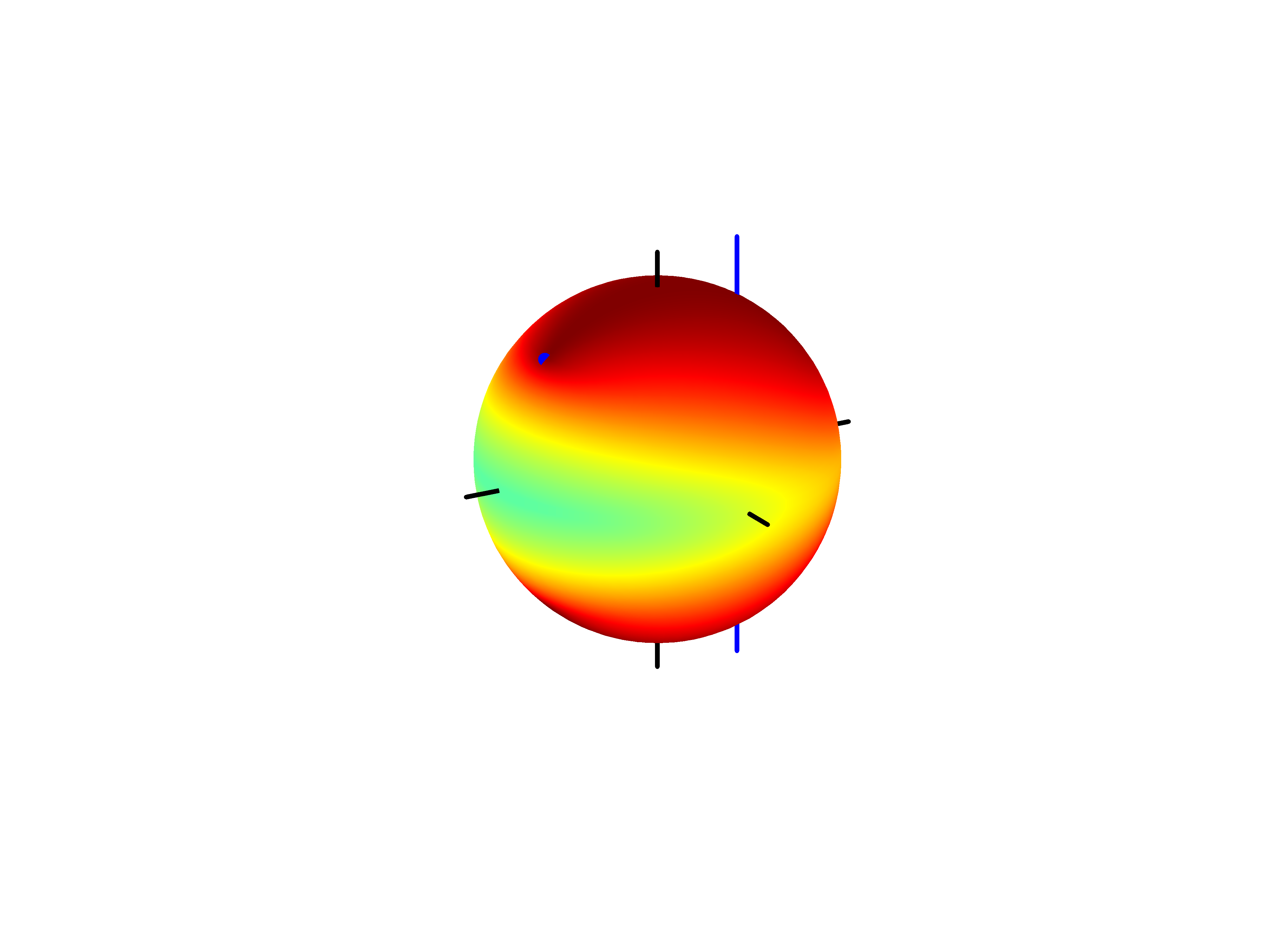}%
		\hfill
		\includegraphics[trim = 60mm 40mm 60mm 35mm, clip,width=0.35\linewidth]{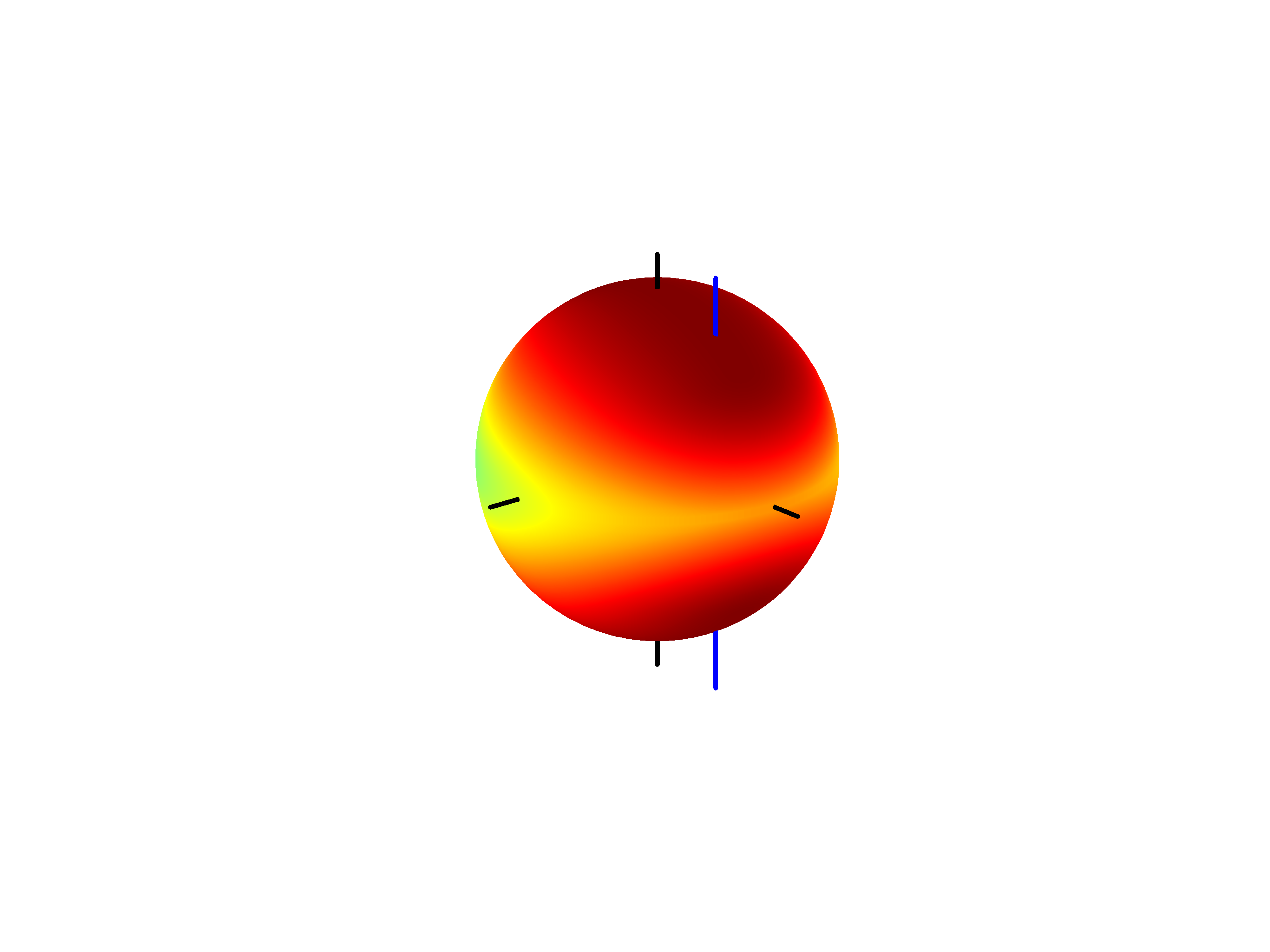}%
	}\vspace{0.5cm}%
	
	\subfloat[$a=1/2,\ \theta=\pi/4 ,\ \phi=5\pi/4. $\label{sfig:b}]{%
		\includegraphics[trim = 60mm 40mm 60mm 35mm, clip, width=0.35\linewidth]{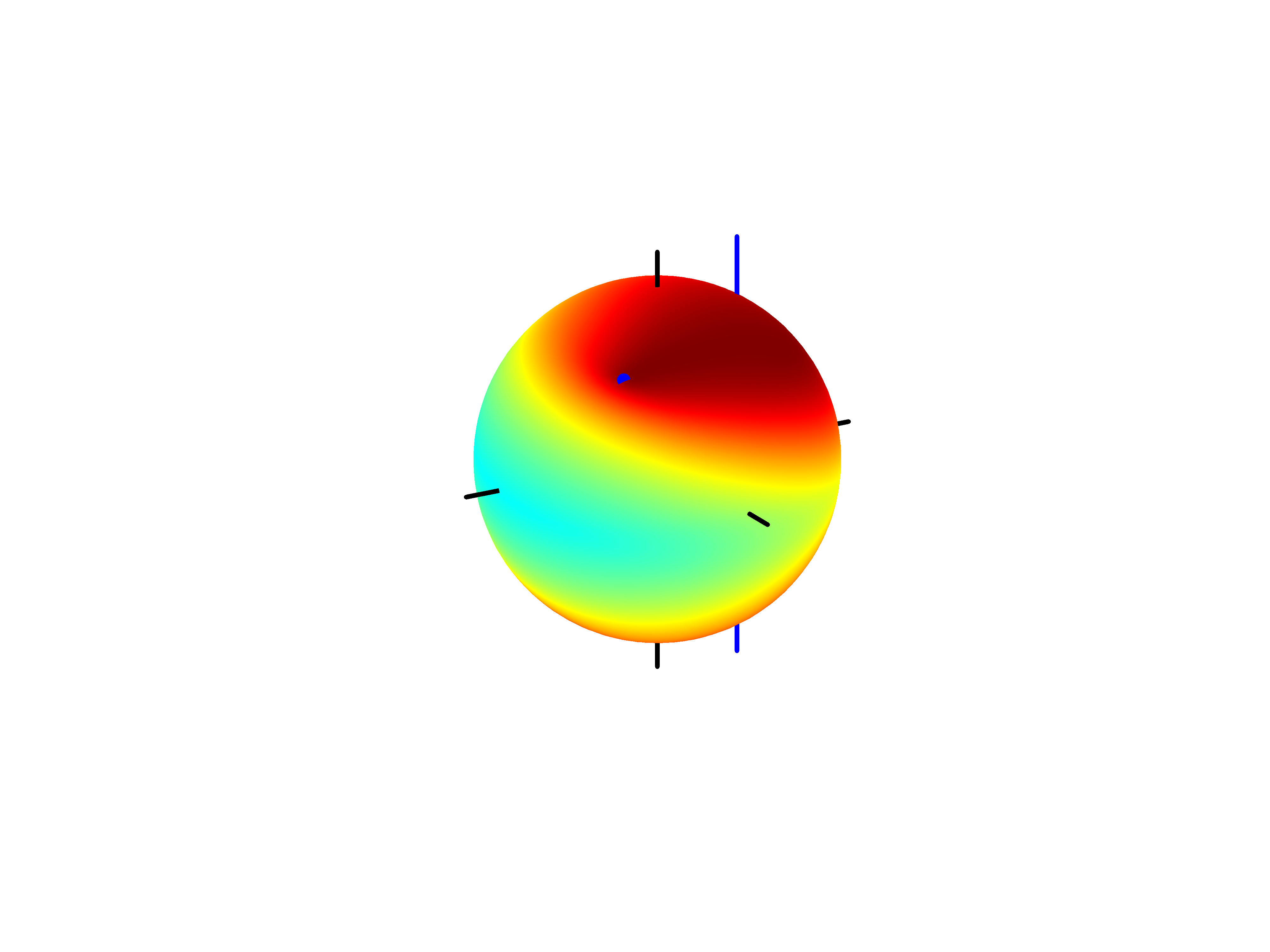}%
		\hfill
		\includegraphics[trim = 60mm 40mm 60mm 35mm, clip,width=0.35\linewidth]{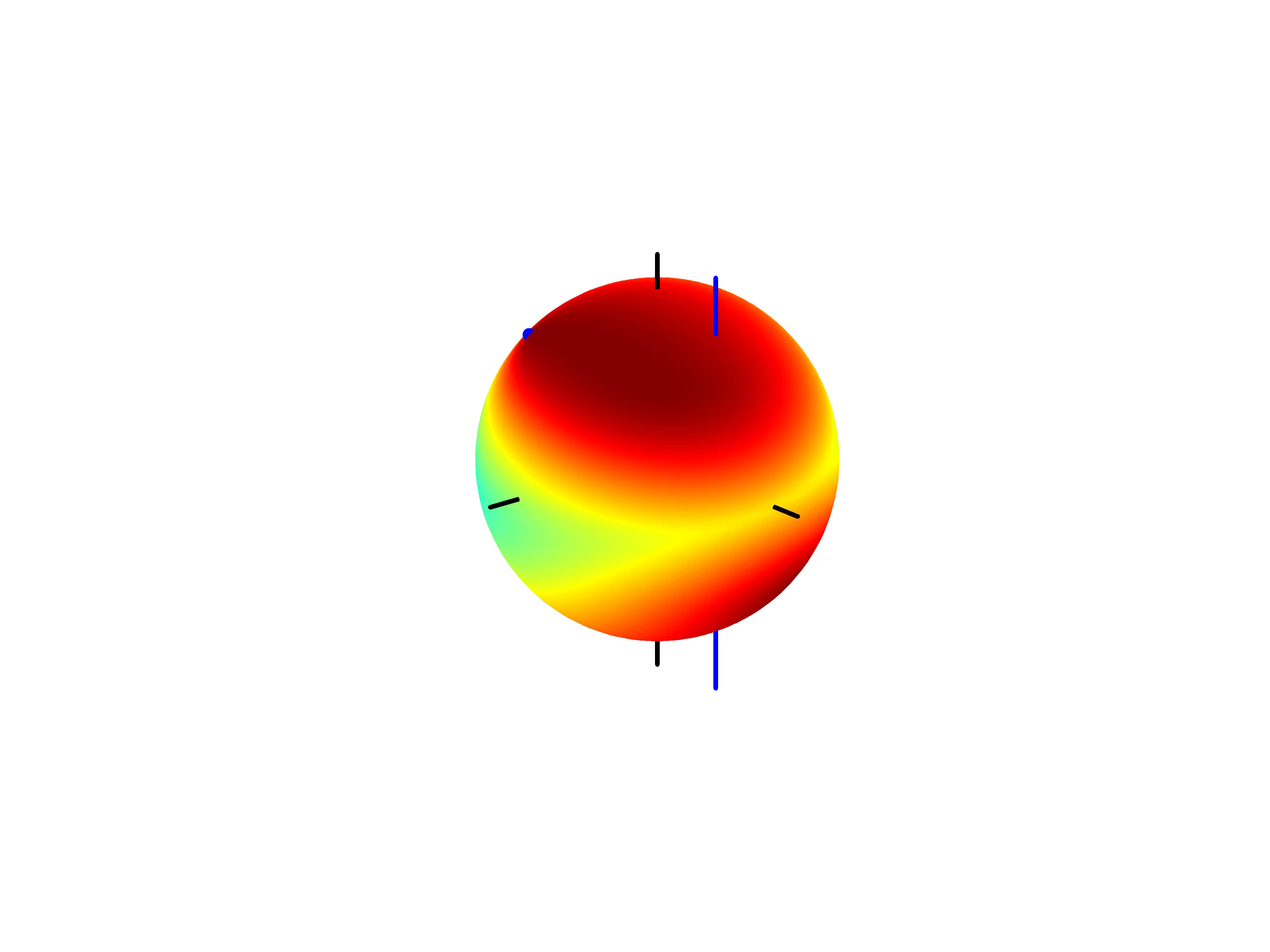}%
	}\vspace{0.5cm}%
	
	\subfloat[$a=1/2,\ \theta=\pi/2 ,\ \phi=0.$\label{sfig:c}]{%
		\includegraphics[trim = 60mm 40mm 60mm 35mm, clip, width=0.35\linewidth]{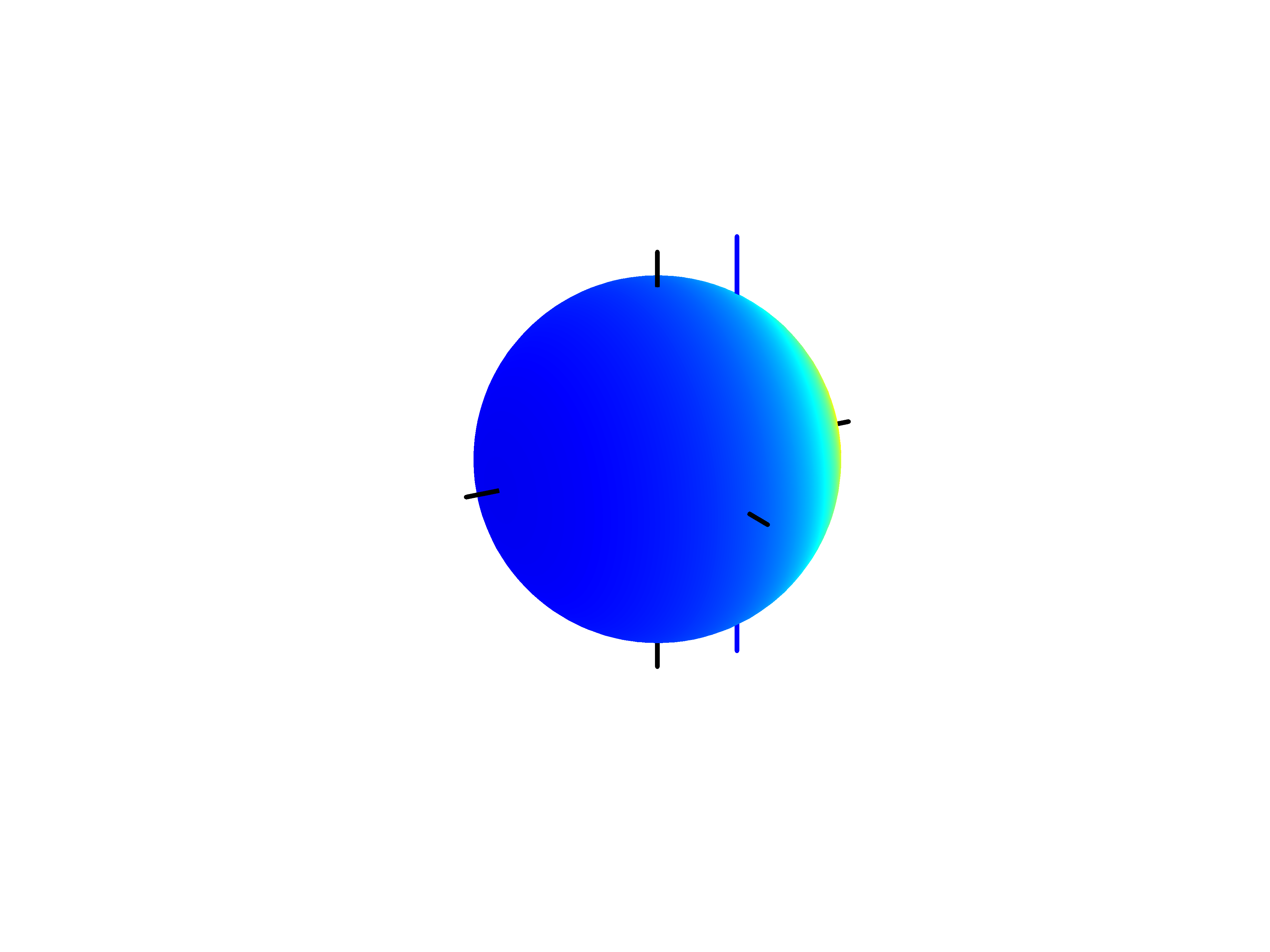}%
		\hfill
		\includegraphics[trim = 60mm 40mm 60mm 35mm, clip,width=0.35\linewidth]{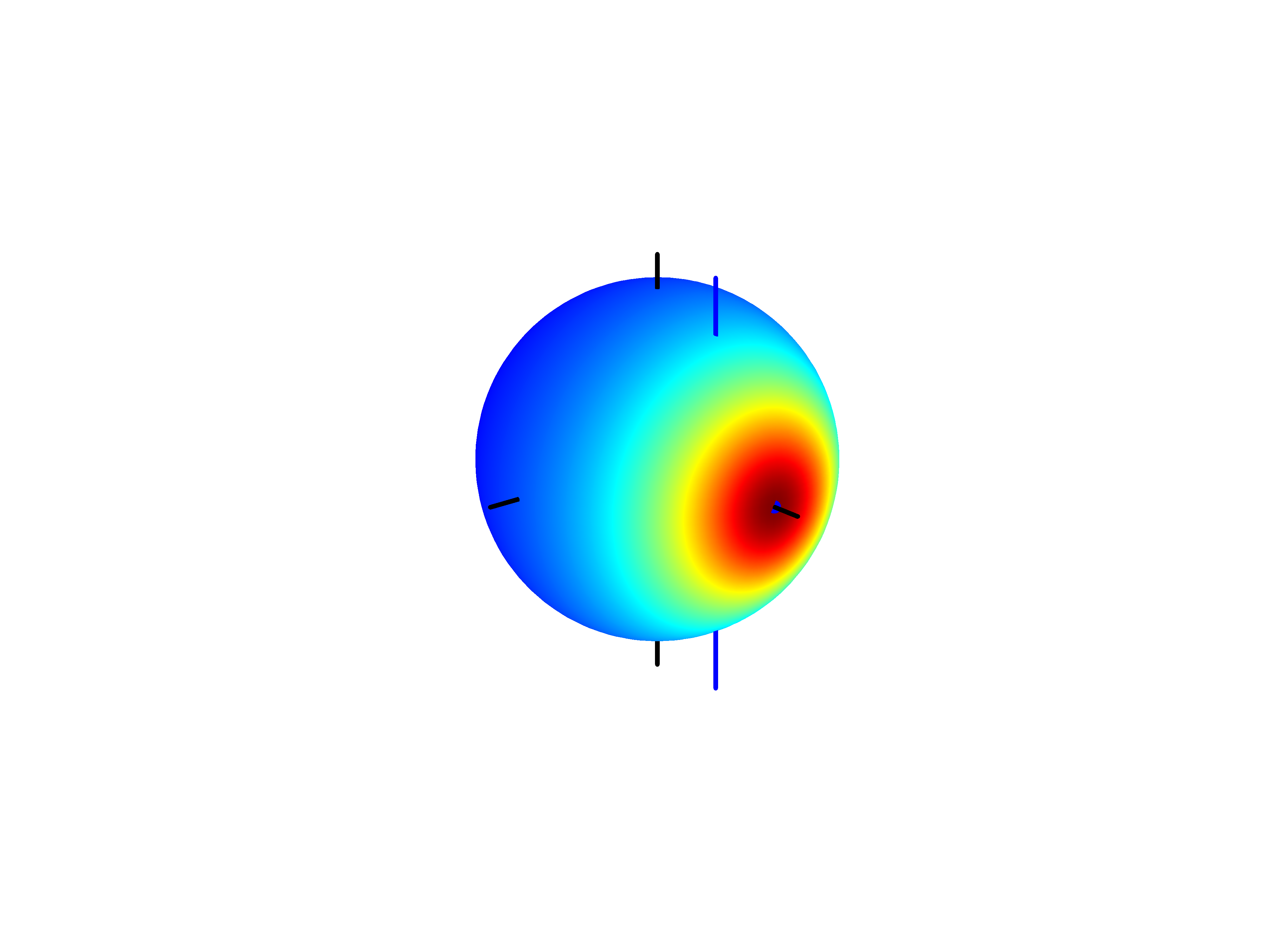}%
	}%

	\caption{Maximal probability to transform different initial qubit states into target states using superposition-free operations. The superposition-free states are given by the blue line. Below each Bloch sphere, the overlap $a=\braket{c_1|c_2}$ of the pure free states is given. The Bloch angles $\theta$ and $\phi$ describe the initial state in spherical coordinates visualized by a blue dot. The colours on the Bloch sphere show the maximal probability to transform the initial state to this state by means of superposition-free operations.}
	\label{sfig:QubitTrafo}
\end{figure*}

\endgroup
\end{appendices}


\begin{thebibliography}{58}%
	\makeatletter
	\providecommand \@ifxundefined [1]{%
		\@ifx{#1\undefined}
	}%
	\providecommand \@ifnum [1]{%
		\ifnum #1\expandafter \@firstoftwo
		\else \expandafter \@secondoftwo
		\fi
	}%
	\providecommand \@ifx [1]{%
		\ifx #1\expandafter \@firstoftwo
		\else \expandafter \@secondoftwo
		\fi
	}%
	\providecommand \natexlab [1]{#1}%
	\providecommand \enquote  [1]{``#1''}%
	\providecommand \bibnamefont  [1]{#1}%
	\providecommand \bibfnamefont [1]{#1}%
	\providecommand \citenamefont [1]{#1}%
	\providecommand \href@noop [0]{\@secondoftwo}%
	\providecommand \href [0]{\begingroup \@sanitize@url \@href}%
	\providecommand \@href[1]{\@@startlink{#1}\@@href}%
	\providecommand \@@href[1]{\endgroup#1\@@endlink}%
	\providecommand \@sanitize@url [0]{\catcode `\\12\catcode `\$12\catcode
		`\&12\catcode `\#12\catcode `\^12\catcode `\_12\catcode `\%12\relax}%
	\providecommand \@@startlink[1]{}%
	\providecommand \@@endlink[0]{}%
	\providecommand \url  [0]{\begingroup\@sanitize@url \@url }%
	\providecommand \@url [1]{\endgroup\@href {#1}{\urlprefix }}%
	\providecommand \urlprefix  [0]{URL }%
	\providecommand \Eprint [0]{\href }%
	\providecommand \doibase [0]{http://dx.doi.org/}%
	\providecommand \selectlanguage [0]{\@gobble}%
	\providecommand \bibinfo  [0]{\@secondoftwo}%
	\providecommand \bibfield  [0]{\@secondoftwo}%
	\providecommand \translation [1]{[#1]}%
	\providecommand \BibitemOpen [0]{}%
	\providecommand \bibitemStop [0]{}%
	\providecommand \bibitemNoStop [0]{.\EOS\space}%
	\providecommand \EOS [0]{\spacefactor3000\relax}%
	\providecommand \BibitemShut  [1]{\csname bibitem#1\endcsname}%
	\let\auto@bib@innerbib\@empty
	\bibitem [{\citenamefont {Ekert}(1991)}]{ekert1991quantum}%
	\BibitemOpen
	\bibfield  {author} {\bibinfo {author} {\bibfnamefont {A.~K.}\ \bibnamefont
			{Ekert}},\ }\href {\doibase 10.1103/PhysRevLett.67.661} {\bibfield  {journal}
		{\bibinfo  {journal} {Phys. Rev. Lett.}\ }\textbf {\bibinfo {volume} {67}},\
		\bibinfo {pages} {661} (\bibinfo {year} {1991})}\BibitemShut {NoStop}%
	\bibitem [{\citenamefont {Bennett}\ and\ \citenamefont
		{Wiesner}(1992)}]{bennett1992communication}%
	\BibitemOpen
	\bibfield  {author} {\bibinfo {author} {\bibfnamefont {C.~H.}\ \bibnamefont
			{Bennett}}\ and\ \bibinfo {author} {\bibfnamefont {S.~J.}\ \bibnamefont
			{Wiesner}},\ }\href {\doibase 10.1103/PhysRevLett.69.2881} {\bibfield
		{journal} {\bibinfo  {journal} {Phys. Rev. Lett.}\ }\textbf {\bibinfo
			{volume} {69}},\ \bibinfo {pages} {2881} (\bibinfo {year}
		{1992})}\BibitemShut {NoStop}%
	\bibitem [{\citenamefont {Bennett}\ \emph {et~al.}(1993)\citenamefont
		{Bennett}, \citenamefont {Brassard}, \citenamefont {Cr\'epeau}, \citenamefont
		{Jozsa}, \citenamefont {Peres},\ and\ \citenamefont
		{Wootters}}]{bennett1993teleporting}%
	\BibitemOpen
	\bibfield  {author} {\bibinfo {author} {\bibfnamefont {C.~H.}\ \bibnamefont
			{Bennett}}, \bibinfo {author} {\bibfnamefont {G.}~\bibnamefont {Brassard}},
		\bibinfo {author} {\bibfnamefont {C.}~\bibnamefont {Cr\'epeau}}, \bibinfo
		{author} {\bibfnamefont {R.}~\bibnamefont {Jozsa}}, \bibinfo {author}
		{\bibfnamefont {A.}~\bibnamefont {Peres}}, \ and\ \bibinfo {author}
		{\bibfnamefont {W.~K.}\ \bibnamefont {Wootters}},\ }\href {\doibase
		10.1103/PhysRevLett.70.1895} {\bibfield  {journal} {\bibinfo  {journal}
			{Phys. Rev. Lett.}\ }\textbf {\bibinfo {volume} {70}},\ \bibinfo {pages}
		{1895} (\bibinfo {year} {1993})}\BibitemShut {NoStop}%
	\bibitem [{\citenamefont {Plenio}\ and\ \citenamefont
		{Virmani}(2007)}]{plenio2007introduction}%
	\BibitemOpen
	\bibfield  {author} {\bibinfo {author} {\bibfnamefont {M.~B.}\ \bibnamefont
			{Plenio}}\ and\ \bibinfo {author} {\bibfnamefont {S.}~\bibnamefont
			{Virmani}},\ }\href {http://www.rintonpress.com/xqic7/qic-7-12/001-051.pdf}
	{\bibfield  {journal} {\bibinfo  {journal} {Quant. Inf. Comp.}\ }\textbf
		{\bibinfo {volume} {7}},\ \bibinfo {pages} {1} (\bibinfo {year}
		{2007})}\BibitemShut {NoStop}%
	\bibitem [{\citenamefont {Gour}\ and\ \citenamefont
		{Spekkens}(2008)}]{gour2008resource}%
	\BibitemOpen
	\bibfield  {author} {\bibinfo {author} {\bibfnamefont {G.}~\bibnamefont
			{Gour}}\ and\ \bibinfo {author} {\bibfnamefont {R.~W.}\ \bibnamefont
			{Spekkens}},\ }\href {http://stacks.iop.org/1367-2630/10/i=3/a=033023}
	{\bibfield  {journal} {\bibinfo  {journal} {New J. Phys.}\ }\textbf {\bibinfo
			{volume} {10}},\ \bibinfo {pages} {033023} (\bibinfo {year}
		{2008})}\BibitemShut {NoStop}%
	\bibitem [{\citenamefont {Brand{\~a}o}\ and\ \citenamefont
		{Plenio}(2008)}]{brandao2008entanglement}%
	\BibitemOpen
	\bibfield  {author} {\bibinfo {author} {\bibfnamefont {F.~G. S.~L.}\
			\bibnamefont {Brand{\~a}o}}\ and\ \bibinfo {author} {\bibfnamefont {M.~B.}\
			\bibnamefont {Plenio}},\ }\href {\doibase 10.1038/nphys1100} {\bibfield
		{journal} {\bibinfo  {journal} {Nat. Phys.}\ }\textbf {\bibinfo {volume}
			{4}},\ \bibinfo {pages} {873} (\bibinfo {year} {2008})}\BibitemShut {NoStop}%
	\bibitem [{\citenamefont {Horodecki}\ and\ \citenamefont
		{Oppenheim}(2013)}]{horodecki2013quantumness}%
	\BibitemOpen
	\bibfield  {author} {\bibinfo {author} {\bibfnamefont {M.}~\bibnamefont
			{Horodecki}}\ and\ \bibinfo {author} {\bibfnamefont {J.}~\bibnamefont
			{Oppenheim}},\ }\href {\doibase 10.1142/S0217979213450197} {\bibfield
		{journal} {\bibinfo  {journal} {Int. J. Mod. Phys. B}\ }\textbf {\bibinfo
			{volume} {27}},\ \bibinfo {pages} {1345019} (\bibinfo {year}
		{2013})}\BibitemShut {NoStop}%
	\bibitem [{\citenamefont {Matera}\ \emph {et~al.}(2016)\citenamefont {Matera},
		\citenamefont {Egloff}, \citenamefont {Killoran},\ and\ \citenamefont
		{Plenio}}]{MEKP.16}%
	\BibitemOpen
	\bibfield  {author} {\bibinfo {author} {\bibfnamefont {J.~M.}\ \bibnamefont
			{Matera}}, \bibinfo {author} {\bibfnamefont {D.}~\bibnamefont {Egloff}},
		\bibinfo {author} {\bibfnamefont {N.}~\bibnamefont {Killoran}}, \ and\
		\bibinfo {author} {\bibfnamefont {M.~B.}\ \bibnamefont {Plenio}},\ }\href
	{http://stacks.iop.org/2058-9565/1/i=1/a=01LT01} {\bibfield  {journal}
		{\bibinfo  {journal} {Quantum Sci. Technol.}\ }\textbf {\bibinfo {volume}
			{1}},\ \bibinfo {pages} {01LT01} (\bibinfo {year} {2016})}\BibitemShut
	{NoStop}%
	\bibitem [{\citenamefont {Hillery}(2016{\natexlab{a}})}]{hillery2016coherence}%
	\BibitemOpen
	\bibfield  {author} {\bibinfo {author} {\bibfnamefont {M.}~\bibnamefont
			{Hillery}},\ }\href {\doibase 10.1103/PhysRevA.93.012111} {\bibfield
		{journal} {\bibinfo  {journal} {Phys. Rev. A}\ }\textbf {\bibinfo {volume}
			{93}},\ \bibinfo {pages} {012111} (\bibinfo {year}
		{2016}{\natexlab{a}})}\BibitemShut {NoStop}%
	\bibitem [{\citenamefont {Adesso}\ \emph {et~al.}(2016)\citenamefont {Adesso},
		\citenamefont {Bromley},\ and\ \citenamefont
		{Cianciaruso}}]{adesso2016discord}%
	\BibitemOpen
	\bibfield  {author} {\bibinfo {author} {\bibfnamefont {G.}~\bibnamefont
			{Adesso}}, \bibinfo {author} {\bibfnamefont {T.~R.}\ \bibnamefont {Bromley}},
		\ and\ \bibinfo {author} {\bibfnamefont {M.}~\bibnamefont {Cianciaruso}},\
	}\href {http://stacks.iop.org/1751-8121/49/i=47/a=473001} {\bibfield
	{journal} {\bibinfo  {journal} {J. Phys. A}\ }\textbf {\bibinfo {volume}
		{49}},\ \bibinfo {pages} {473001} (\bibinfo {year} {2016})}\BibitemShut
{NoStop}%
\bibitem [{\citenamefont {Streltsov}\ \emph
	{et~al.}(2016{\natexlab{a}})\citenamefont {Streltsov}, \citenamefont
	{Adesso},\ and\ \citenamefont {Plenio}}]{streltsov2016quantum}%
\BibitemOpen
\bibfield  {author} {\bibinfo {author} {\bibfnamefont {A.}~\bibnamefont
		{Streltsov}}, \bibinfo {author} {\bibfnamefont {G.}~\bibnamefont {Adesso}}, \
	and\ \bibinfo {author} {\bibfnamefont {M.~B.}\ \bibnamefont {Plenio}},\
}\href {http://arxiv.org/abs/1609.02439} {\bibfield  {journal} {\bibinfo
	{journal} {arXiv:1609.02439}\ } (\bibinfo {year}
{2016}{\natexlab{a}})}\BibitemShut {NoStop}%
\bibitem [{\citenamefont {Killoran}\ \emph {et~al.}(2016)\citenamefont
	{Killoran}, \citenamefont {Steinhoff},\ and\ \citenamefont
	{Plenio}}]{killoran2016converting}%
\BibitemOpen
\bibfield  {author} {\bibinfo {author} {\bibfnamefont {N.}~\bibnamefont
		{Killoran}}, \bibinfo {author} {\bibfnamefont {F.~E.~S.}\ \bibnamefont
		{Steinhoff}}, \ and\ \bibinfo {author} {\bibfnamefont {M.~B.}\ \bibnamefont
		{Plenio}},\ }\href {\doibase 10.1103/PhysRevLett.116.080402} {\bibfield
	{journal} {\bibinfo  {journal} {Phys. Rev. Lett.}\ }\textbf {\bibinfo
		{volume} {116}},\ \bibinfo {pages} {080402} (\bibinfo {year}
	{2016})}\BibitemShut {NoStop}%
\bibitem [{\citenamefont {Regula}\ \emph {et~al.}(2017)\citenamefont {Regula},
	\citenamefont {Piani}, \citenamefont {Cianciaruso}, \citenamefont {Bromley},
	\citenamefont {Streltsov},\ and\ \citenamefont
	{Adesso}}]{regula2017converting}%
\BibitemOpen
\bibfield  {author} {\bibinfo {author} {\bibfnamefont {B.}~\bibnamefont
		{Regula}}, \bibinfo {author} {\bibfnamefont {M.}~\bibnamefont {Piani}},
	\bibinfo {author} {\bibfnamefont {M.}~\bibnamefont {Cianciaruso}}, \bibinfo
	{author} {\bibfnamefont {T.~R.}\ \bibnamefont {Bromley}}, \bibinfo {author}
	{\bibfnamefont {A.}~\bibnamefont {Streltsov}}, \ and\ \bibinfo {author}
	{\bibfnamefont {G.}~\bibnamefont {Adesso}},\ }\href
{https://arxiv.org/abs/1704.04153} {\bibfield  {journal} {\bibinfo  {journal}
		{arXiv:1704.04153}\ } (\bibinfo {year} {2017})}\BibitemShut {NoStop}%
\bibitem [{\citenamefont {Vogel}\ and\ \citenamefont
	{Sperling}(2014)}]{vogel2014unified}%
\BibitemOpen
\bibfield  {author} {\bibinfo {author} {\bibfnamefont {W.}~\bibnamefont
		{Vogel}}\ and\ \bibinfo {author} {\bibfnamefont {J.}~\bibnamefont
		{Sperling}},\ }\href {\doibase 10.1103/PhysRevA.89.052302} {\bibfield
	{journal} {\bibinfo  {journal} {Phys. Rev. A}\ }\textbf {\bibinfo {volume}
		{89}},\ \bibinfo {pages} {052302} (\bibinfo {year} {2014})}\BibitemShut
{NoStop}%
\bibitem [{\citenamefont {Sperling}\ and\ \citenamefont
	{Vogel}(2015)}]{sperling2015convex}%
\BibitemOpen
\bibfield  {author} {\bibinfo {author} {\bibfnamefont {J.}~\bibnamefont
		{Sperling}}\ and\ \bibinfo {author} {\bibfnamefont {W.}~\bibnamefont
		{Vogel}},\ }\href {http://stacks.iop.org/1402-4896/90/i=7/a=074024}
{\bibfield  {journal} {\bibinfo  {journal} {Phys. Scr.}\ }\textbf {\bibinfo
		{volume} {90}},\ \bibinfo {pages} {074024} (\bibinfo {year}
	{2015})}\BibitemShut {NoStop}%
\bibitem [{\citenamefont {Baumgratz}\ \emph {et~al.}(2014)\citenamefont
	{Baumgratz}, \citenamefont {Cramer},\ and\ \citenamefont
	{Plenio}}]{baumgratz2014quantifying}%
\BibitemOpen
\bibfield  {author} {\bibinfo {author} {\bibfnamefont {T.}~\bibnamefont
		{Baumgratz}}, \bibinfo {author} {\bibfnamefont {M.}~\bibnamefont {Cramer}}, \
	and\ \bibinfo {author} {\bibfnamefont {M.~B.}\ \bibnamefont {Plenio}},\
}\href {\doibase 10.1103/PhysRevLett.113.140401} {\bibfield  {journal}
{\bibinfo  {journal} {Phys. Rev. Lett.}\ }\textbf {\bibinfo {volume} {113}},\
\bibinfo {pages} {140401} (\bibinfo {year} {2014})}\BibitemShut {NoStop}%
\bibitem [{\citenamefont {Aberg}(2006)}]{aberg2006quantifying}%
\BibitemOpen
\bibfield  {author} {\bibinfo {author} {\bibfnamefont {J.}~\bibnamefont
		{Aberg}},\ }\href {http://arxiv.org/abs/quant-ph/0612146} {\bibfield
	{journal} {\bibinfo  {journal} {arXiv:quant-ph/0612146}\ } (\bibinfo {year}
	{2006})}\BibitemShut {NoStop}%
\bibitem [{\citenamefont {Hillery}(2016{\natexlab{b}})}]{PhysRevA.93.012111}%
\BibitemOpen
\bibfield  {author} {\bibinfo {author} {\bibfnamefont {M.}~\bibnamefont
		{Hillery}},\ }\href {\doibase 10.1103/PhysRevA.93.012111} {\bibfield
	{journal} {\bibinfo  {journal} {Phys. Rev. A}\ }\textbf {\bibinfo {volume}
		{93}},\ \bibinfo {pages} {012111} (\bibinfo {year}
	{2016}{\natexlab{b}})}\BibitemShut {NoStop}%
\bibitem [{\citenamefont {Ma}\ \emph {et~al.}(2016)\citenamefont {Ma},
	\citenamefont {Yadin}, \citenamefont {Girolami}, \citenamefont {Vedral},\
	and\ \citenamefont {Gu}}]{PhysRevLett.116.160407}%
\BibitemOpen
\bibfield  {author} {\bibinfo {author} {\bibfnamefont {J.}~\bibnamefont
		{Ma}}, \bibinfo {author} {\bibfnamefont {B.}~\bibnamefont {Yadin}}, \bibinfo
	{author} {\bibfnamefont {D.}~\bibnamefont {Girolami}}, \bibinfo {author}
	{\bibfnamefont {V.}~\bibnamefont {Vedral}}, \ and\ \bibinfo {author}
	{\bibfnamefont {M.}~\bibnamefont {Gu}},\ }\href {\doibase
	10.1103/PhysRevLett.116.160407} {\bibfield  {journal} {\bibinfo  {journal}
		{Phys. Rev. Lett.}\ }\textbf {\bibinfo {volume} {116}},\ \bibinfo {pages}
	{160407} (\bibinfo {year} {2016})}\BibitemShut {NoStop}%
\bibitem [{\citenamefont {Napoli}\ \emph {et~al.}(2016)\citenamefont {Napoli},
	\citenamefont {Bromley}, \citenamefont {Cianciaruso}, \citenamefont {Piani},
	\citenamefont {Johnston},\ and\ \citenamefont
	{Adesso}}]{napoli2016robustness}%
\BibitemOpen
\bibfield  {author} {\bibinfo {author} {\bibfnamefont {C.}~\bibnamefont
		{Napoli}}, \bibinfo {author} {\bibfnamefont {T.~R.}\ \bibnamefont {Bromley}},
	\bibinfo {author} {\bibfnamefont {M.}~\bibnamefont {Cianciaruso}}, \bibinfo
	{author} {\bibfnamefont {M.}~\bibnamefont {Piani}}, \bibinfo {author}
	{\bibfnamefont {N.}~\bibnamefont {Johnston}}, \ and\ \bibinfo {author}
	{\bibfnamefont {G.}~\bibnamefont {Adesso}},\ }\href {\doibase
	10.1103/PhysRevLett.116.150502} {\bibfield  {journal} {\bibinfo  {journal}
		{Phys. Rev. Lett.}\ }\textbf {\bibinfo {volume} {116}},\ \bibinfo {pages}
	{150502} (\bibinfo {year} {2016})}\BibitemShut {NoStop}%
\bibitem [{\citenamefont {Misra}\ \emph {et~al.}(2016)\citenamefont {Misra},
	\citenamefont {Singh}, \citenamefont {Bhattacharya},\ and\ \citenamefont
	{Pati}}]{PhysRevA.93.052335}%
\BibitemOpen
\bibfield  {author} {\bibinfo {author} {\bibfnamefont {A.}~\bibnamefont
		{Misra}}, \bibinfo {author} {\bibfnamefont {U.}~\bibnamefont {Singh}},
	\bibinfo {author} {\bibfnamefont {S.}~\bibnamefont {Bhattacharya}}, \ and\
	\bibinfo {author} {\bibfnamefont {A.~K.}\ \bibnamefont {Pati}},\ }\href
{\doibase 10.1103/PhysRevA.93.052335} {\bibfield  {journal} {\bibinfo
		{journal} {Phys. Rev. A}\ }\textbf {\bibinfo {volume} {93}},\ \bibinfo
	{pages} {052335} (\bibinfo {year} {2016})}\BibitemShut {NoStop}%
\bibitem [{\citenamefont {Asb\'oth}\ \emph {et~al.}(2005)\citenamefont
	{Asb\'oth}, \citenamefont {Calsamiglia},\ and\ \citenamefont
	{Ritsch}}]{PhysRevLett.94.173602}%
\BibitemOpen
\bibfield  {author} {\bibinfo {author} {\bibfnamefont {J.~K.}\ \bibnamefont
		{Asb\'oth}}, \bibinfo {author} {\bibfnamefont {J.}~\bibnamefont
		{Calsamiglia}}, \ and\ \bibinfo {author} {\bibfnamefont {H.}~\bibnamefont
		{Ritsch}},\ }\href {\doibase 10.1103/PhysRevLett.94.173602} {\bibfield
	{journal} {\bibinfo  {journal} {Phys. Rev. Lett.}\ }\textbf {\bibinfo
		{volume} {94}},\ \bibinfo {pages} {173602} (\bibinfo {year}
	{2005})}\BibitemShut {NoStop}%
\bibitem [{\citenamefont {Tan}\ \emph {et~al.}(2017)\citenamefont {Tan},
	\citenamefont {Volkoff}, \citenamefont {Kwon},\ and\ \citenamefont
	{Jeong}}]{tan2017quantifying}%
\BibitemOpen
\bibfield  {author} {\bibinfo {author} {\bibfnamefont {K.~C.}\ \bibnamefont
		{Tan}}, \bibinfo {author} {\bibfnamefont {T.}~\bibnamefont {Volkoff}},
	\bibinfo {author} {\bibfnamefont {H.}~\bibnamefont {Kwon}}, \ and\ \bibinfo
	{author} {\bibfnamefont {H.}~\bibnamefont {Jeong}},\ }\href
{http://arxiv.org/abs/1703.01067} {\bibfield  {journal} {\bibinfo  {journal}
		{arXiv:1703.01067}\ } (\bibinfo {year} {2017})}\BibitemShut {NoStop}%
\bibitem [{\citenamefont {Chefles}(1998)}]{chefles1998unambiguous}%
\BibitemOpen
\bibfield  {author} {\bibinfo {author} {\bibfnamefont {A.}~\bibnamefont
		{Chefles}},\ }\href {\doibase 10.1016/S0375-9601(98)00064-4} {\bibfield
	{journal} {\bibinfo  {journal} {Phys. Lett. A}\ }\textbf {\bibinfo {volume}
		{239}},\ \bibinfo {pages} {339 } (\bibinfo {year} {1998})}\BibitemShut
{NoStop}%
\bibitem [{\citenamefont {Chefles}(2000)}]{chefles2000quantum}%
\BibitemOpen
\bibfield  {author} {\bibinfo {author} {\bibfnamefont {A.}~\bibnamefont
		{Chefles}},\ }\href {\doibase 10.1080/00107510010002599} {\bibfield
	{journal} {\bibinfo  {journal} {Contemp. Phys.}\ }\textbf {\bibinfo {volume}
		{41}},\ \bibinfo {pages} {401} (\bibinfo {year} {2000})}\BibitemShut
{NoStop}%
\bibitem [{\citenamefont {Chitambar}\ and\ \citenamefont
	{Gour}(2016)}]{chitambar2016critical}%
\BibitemOpen
\bibfield  {author} {\bibinfo {author} {\bibfnamefont {E.}~\bibnamefont
		{Chitambar}}\ and\ \bibinfo {author} {\bibfnamefont {G.}~\bibnamefont
		{Gour}},\ }\href {\doibase 10.1103/PhysRevLett.117.030401} {\bibfield
	{journal} {\bibinfo  {journal} {Phys. Rev. Lett.}\ }\textbf {\bibinfo
		{volume} {117}},\ \bibinfo {pages} {030401} (\bibinfo {year}
	{2016})}\BibitemShut {NoStop}%
\bibitem [{\citenamefont {Marvian}\ and\ \citenamefont
	{Spekkens}(2016)}]{marvian2016quantify}%
\BibitemOpen
\bibfield  {author} {\bibinfo {author} {\bibfnamefont {I.}~\bibnamefont
		{Marvian}}\ and\ \bibinfo {author} {\bibfnamefont {R.~W.}\ \bibnamefont
		{Spekkens}},\ }\href {\doibase 10.1103/PhysRevA.94.052324} {\bibfield
	{journal} {\bibinfo  {journal} {Phys. Rev. A}\ }\textbf {\bibinfo {volume}
		{94}},\ \bibinfo {pages} {052324} (\bibinfo {year} {2016})}\BibitemShut
{NoStop}%
\bibitem [{\citenamefont {Winter}\ and\ \citenamefont
	{Yang}(2016)}]{winter2016operational}%
\BibitemOpen
\bibfield  {author} {\bibinfo {author} {\bibfnamefont {A.}~\bibnamefont
		{Winter}}\ and\ \bibinfo {author} {\bibfnamefont {D.}~\bibnamefont {Yang}},\
}\href {\doibase 10.1103/PhysRevLett.116.120404} {\bibfield  {journal}
{\bibinfo  {journal} {Phys. Rev. Lett.}\ }\textbf {\bibinfo {volume} {116}},\
\bibinfo {pages} {120404} (\bibinfo {year} {2016})}\BibitemShut {NoStop}%
\bibitem [{\citenamefont {Yao}\ \emph {et~al.}(2015)\citenamefont {Yao},
	\citenamefont {Xiao}, \citenamefont {Ge},\ and\ \citenamefont
	{Sun}}]{yao2015quantum}%
\BibitemOpen
\bibfield  {author} {\bibinfo {author} {\bibfnamefont {Y.}~\bibnamefont
		{Yao}}, \bibinfo {author} {\bibfnamefont {X.}~\bibnamefont {Xiao}}, \bibinfo
	{author} {\bibfnamefont {L.}~\bibnamefont {Ge}}, \ and\ \bibinfo {author}
	{\bibfnamefont {C.~P.}\ \bibnamefont {Sun}},\ }\href {\doibase
	10.1103/PhysRevA.92.022112} {\bibfield  {journal} {\bibinfo  {journal} {Phys.
			Rev. A}\ }\textbf {\bibinfo {volume} {92}},\ \bibinfo {pages} {022112}
	(\bibinfo {year} {2015})}\BibitemShut {NoStop}%
\bibitem [{Note1()}]{Note1}%
\BibitemOpen
\bibinfo {note} {This can happen in the case of entanglement theory. Let
	$\protect \{\mathinner {|{\psi _i}\delimiter "526930B }=\mathinner {|{\phi
			_i}\delimiter "526930B }\otimes \mathinner {|{\xi _i}\delimiter "526930B
	}\protect \}_i$ be an unextendable separable product basis \cite
	{bennett1999unextendible} of a bipartite system and define a trace-decreasing
	separable operation $\Lambda [\rho ]=\DOTSB \sum@ \slimits@ _i \mathinner
	{|{0}\delimiter "526930B }\mathinner {\delimiter "426830A {\phi _i}|}\otimes
	\mathinner {|{0}\delimiter "526930B }\mathinner {\delimiter "426830A {\xi
			_i}|} \rho \mathinner {|{\phi _i}\delimiter "526930B }\mathinner {\delimiter
		"426830A {0}|}\otimes \mathinner {|{\xi _i}\delimiter "526930B }\mathinner
	{\delimiter "426830A {0}|}$ where $\mathinner {|{0}\delimiter "526930B }$ is
	an arbitrary reference state. This operation cannot be completed by separable
	Kraus operators by construction.}\BibitemShut {Stop}%
\bibitem [{\citenamefont {Vedral}\ \emph {et~al.}(1997)\citenamefont {Vedral},
	\citenamefont {Plenio}, \citenamefont {Rippin},\ and\ \citenamefont
	{Knight}}]{QuantifyingEntanglement}%
\BibitemOpen
\bibfield  {author} {\bibinfo {author} {\bibfnamefont {V.}~\bibnamefont
		{Vedral}}, \bibinfo {author} {\bibfnamefont {M.~B.}\ \bibnamefont {Plenio}},
	\bibinfo {author} {\bibfnamefont {M.~A.}\ \bibnamefont {Rippin}}, \ and\
	\bibinfo {author} {\bibfnamefont {P.~L.}\ \bibnamefont {Knight}},\ }\href
{\doibase 10.1103/PhysRevLett.78.2275} {\bibfield  {journal} {\bibinfo
		{journal} {Phys. Rev. Lett.}\ }\textbf {\bibinfo {volume} {78}},\ \bibinfo
	{pages} {2275} (\bibinfo {year} {1997})}\BibitemShut {NoStop}%
\bibitem [{\citenamefont {Wehrl}(1978)}]{wehrl1978general}%
\BibitemOpen
\bibfield  {author} {\bibinfo {author} {\bibfnamefont {A.}~\bibnamefont
		{Wehrl}},\ }\href {\doibase 10.1103/RevModPhys.50.221} {\bibfield  {journal}
	{\bibinfo  {journal} {Rev. Mod. Phys.}\ }\textbf {\bibinfo {volume} {50}},\
	\bibinfo {pages} {221} (\bibinfo {year} {1978})}\BibitemShut {NoStop}%
\bibitem [{\citenamefont {Boyd}\ and\ \citenamefont
	{Vandenberghe}(2004)}]{boyd2004convex}%
\BibitemOpen
\bibfield  {author} {\bibinfo {author} {\bibfnamefont {S.}~\bibnamefont
		{Boyd}}\ and\ \bibinfo {author} {\bibfnamefont {L.}~\bibnamefont
		{Vandenberghe}},\ }\href@noop {} {\emph {\bibinfo {title} {Convex
			optimization}}}\ (\bibinfo  {publisher} {Cambridge university press},\
\bibinfo {year} {2004})\BibitemShut {NoStop}%
\bibitem [{\citenamefont {Jarre}\ and\ \citenamefont
	{Stoer}(2013)}]{jarre2013optimierung}%
\BibitemOpen
\bibfield  {author} {\bibinfo {author} {\bibfnamefont {F.}~\bibnamefont
		{Jarre}}\ and\ \bibinfo {author} {\bibfnamefont {J.}~\bibnamefont {Stoer}},\
}\href@noop {} {\emph {\bibinfo {title} {Optimierung}}}\ (\bibinfo
{publisher} {Springer-Verlag},\ \bibinfo {year} {2013})\BibitemShut {NoStop}%
\bibitem [{Note2()}]{Note2}%
\BibitemOpen
\bibinfo {note} {We can express the unitary operation as a matrix with
	respect to the orthonormal basis obtained when applying the Gram-Schmidt
	process on the pure superposition-free states. As shown in \cite
	{reck1994experimental}, we can then decompose the unitary into unitaries
	$U_2$ acting on two-dimensional subspaces spanned by two pure free states.
	With the help of a qubit state with maximal superposition (with respect to
	the two free states spanning the two-dimensional subspace under
	consideration) every $U_2$ can be implemented.}\BibitemShut {Stop}%
\bibitem [{\citenamefont {Jonathan}\ and\ \citenamefont
	{Plenio}(1999)}]{jonathan1999entanglement}%
\BibitemOpen
\bibfield  {author} {\bibinfo {author} {\bibfnamefont {D.}~\bibnamefont
		{Jonathan}}\ and\ \bibinfo {author} {\bibfnamefont {M.~B.}\ \bibnamefont
		{Plenio}},\ }\href {\doibase 10.1103/PhysRevLett.83.3566} {\bibfield
	{journal} {\bibinfo  {journal} {Phys. Rev. Lett.}\ }\textbf {\bibinfo
		{volume} {83}},\ \bibinfo {pages} {3566} (\bibinfo {year}
	{1999})}\BibitemShut {NoStop}%
\bibitem [{\citenamefont {Du}\ \emph {et~al.}(2015)\citenamefont {Du},
	\citenamefont {Bai},\ and\ \citenamefont {Guo}}]{du2015conditions}%
\BibitemOpen
\bibfield  {author} {\bibinfo {author} {\bibfnamefont {S.}~\bibnamefont
		{Du}}, \bibinfo {author} {\bibfnamefont {Z.}~\bibnamefont {Bai}}, \ and\
	\bibinfo {author} {\bibfnamefont {Y.}~\bibnamefont {Guo}},\ }\href {\doibase
	10.1103/PhysRevA.91.052120} {\bibfield  {journal} {\bibinfo  {journal} {Phys.
			Rev. A}\ }\textbf {\bibinfo {volume} {91}},\ \bibinfo {pages} {052120}
	(\bibinfo {year} {2015})}\BibitemShut {NoStop}%
\bibitem [{\citenamefont {\AA{}berg}(2014)}]{aaberg2014catalytic}%
\BibitemOpen
\bibfield  {author} {\bibinfo {author} {\bibfnamefont {J.}~\bibnamefont
		{\AA{}berg}},\ }\href {\doibase 10.1103/PhysRevLett.113.150402} {\bibfield
	{journal} {\bibinfo  {journal} {Phys. Rev. Lett.}\ }\textbf {\bibinfo
		{volume} {113}},\ \bibinfo {pages} {150402} (\bibinfo {year}
	{2014})}\BibitemShut {NoStop}%
\bibitem [{\citenamefont {Duarte}\ \emph {et~al.}(2016)\citenamefont {Duarte},
	\citenamefont {Drumond},\ and\ \citenamefont {Cunha}}]{duarte2016self}%
\BibitemOpen
\bibfield  {author} {\bibinfo {author} {\bibfnamefont {C.}~\bibnamefont
		{Duarte}}, \bibinfo {author} {\bibfnamefont {R.~C.}\ \bibnamefont {Drumond}},
	\ and\ \bibinfo {author} {\bibfnamefont {M.~T.}\ \bibnamefont {Cunha}},\
}\href {http://stacks.iop.org/1751-8121/49/i=14/a=145303} {\bibfield
{journal} {\bibinfo  {journal} {J. Phys. A}\ }\textbf {\bibinfo {volume}
	{49}},\ \bibinfo {pages} {145303} (\bibinfo {year} {2016})}\BibitemShut
{NoStop}%
\bibitem [{\citenamefont {Bu}\ \emph {et~al.}(2016)\citenamefont {Bu},
	\citenamefont {Singh},\ and\ \citenamefont {Wu}}]{bu2016catalytic}%
\BibitemOpen
\bibfield  {author} {\bibinfo {author} {\bibfnamefont {K.}~\bibnamefont
		{Bu}}, \bibinfo {author} {\bibfnamefont {U.}~\bibnamefont {Singh}}, \ and\
	\bibinfo {author} {\bibfnamefont {J.}~\bibnamefont {Wu}},\ }\href {\doibase
	10.1103/PhysRevA.93.042326} {\bibfield  {journal} {\bibinfo  {journal} {Phys.
			Rev. A}\ }\textbf {\bibinfo {volume} {93}},\ \bibinfo {pages} {042326}
	(\bibinfo {year} {2016})}\BibitemShut {NoStop}%
\bibitem [{\citenamefont {Renes}(2016)}]{renes2015relative}%
\BibitemOpen
\bibfield  {author} {\bibinfo {author} {\bibfnamefont {J.~M.}\ \bibnamefont
		{Renes}},\ }\href {\doibase 10.1063/1.4972295} {\bibfield  {journal}
	{\bibinfo  {journal} {J. Math. Phys.}\ }\textbf {\bibinfo {volume} {57}},\
	\bibinfo {pages} {122202} (\bibinfo {year} {2016})}\BibitemShut {NoStop}%
\bibitem [{\citenamefont {Chiribella}\ and\ \citenamefont
	{Yang}(2017)}]{chiribella2015optimal}%
\BibitemOpen
\bibfield  {author} {\bibinfo {author} {\bibfnamefont {G.}~\bibnamefont
		{Chiribella}}\ and\ \bibinfo {author} {\bibfnamefont {Y.}~\bibnamefont
		{Yang}},\ }\href {\doibase 10.1103/PhysRevA.96.022327} {\bibfield  {journal}
	{\bibinfo  {journal} {Phys. Rev. A}\ }\textbf {\bibinfo {volume} {96}},\
	\bibinfo {pages} {022327} (\bibinfo {year} {2017})}\BibitemShut {NoStop}%
\bibitem [{\citenamefont {Chitambar}\ and\ \citenamefont
	{Hsieh}(2016)}]{chitambar2015relating}%
\BibitemOpen
\bibfield  {author} {\bibinfo {author} {\bibfnamefont {E.}~\bibnamefont
		{Chitambar}}\ and\ \bibinfo {author} {\bibfnamefont {M.-H.}\ \bibnamefont
		{Hsieh}},\ }\href {\doibase 10.1103/PhysRevLett.117.020402} {\bibfield
	{journal} {\bibinfo  {journal} {Phys. Rev. Lett.}\ }\textbf {\bibinfo
		{volume} {117}},\ \bibinfo {pages} {020402} (\bibinfo {year}
	{2016})}\BibitemShut {NoStop}%
\bibitem [{\citenamefont {Chitambar}\ \emph {et~al.}(2016)\citenamefont
	{Chitambar}, \citenamefont {Streltsov}, \citenamefont {Rana}, \citenamefont
	{Bera}, \citenamefont {Adesso},\ and\ \citenamefont
	{Lewenstein}}]{chitambar2016assisted}%
\BibitemOpen
\bibfield  {author} {\bibinfo {author} {\bibfnamefont {E.}~\bibnamefont
		{Chitambar}}, \bibinfo {author} {\bibfnamefont {A.}~\bibnamefont
		{Streltsov}}, \bibinfo {author} {\bibfnamefont {S.}~\bibnamefont {Rana}},
	\bibinfo {author} {\bibfnamefont {M.~N.}\ \bibnamefont {Bera}}, \bibinfo
	{author} {\bibfnamefont {G.}~\bibnamefont {Adesso}}, \ and\ \bibinfo {author}
	{\bibfnamefont {M.}~\bibnamefont {Lewenstein}},\ }\href {\doibase
	10.1103/PhysRevLett.116.070402} {\bibfield  {journal} {\bibinfo  {journal}
		{Phys. Rev. Lett.}\ }\textbf {\bibinfo {volume} {116}},\ \bibinfo {pages}
	{070402} (\bibinfo {year} {2016})}\BibitemShut {NoStop}%
\bibitem [{\citenamefont {Streltsov}\ \emph {et~al.}(2017)\citenamefont
	{Streltsov}, \citenamefont {Rana}, \citenamefont {Bera},\ and\ \citenamefont
	{Lewenstein}}]{streltsov2015hierarchies}%
\BibitemOpen
\bibfield  {author} {\bibinfo {author} {\bibfnamefont {A.}~\bibnamefont
		{Streltsov}}, \bibinfo {author} {\bibfnamefont {S.}~\bibnamefont {Rana}},
	\bibinfo {author} {\bibfnamefont {M.~N.}\ \bibnamefont {Bera}}, \ and\
	\bibinfo {author} {\bibfnamefont {M.}~\bibnamefont {Lewenstein}},\ }\href
{\doibase 10.1103/PhysRevX.7.011024} {\bibfield  {journal} {\bibinfo
		{journal} {Phys. Rev. X}\ }\textbf {\bibinfo {volume} {7}},\ \bibinfo {pages}
	{011024} (\bibinfo {year} {2017})}\BibitemShut {NoStop}%
\bibitem [{\citenamefont {Streltsov}\ \emph
	{et~al.}(2016{\natexlab{b}})\citenamefont {Streltsov}, \citenamefont
	{Chitambar}, \citenamefont {Rana}, \citenamefont {Bera}, \citenamefont
	{Winter},\ and\ \citenamefont {Lewenstein}}]{streltsov2016entanglement}%
\BibitemOpen
\bibfield  {author} {\bibinfo {author} {\bibfnamefont {A.}~\bibnamefont
		{Streltsov}}, \bibinfo {author} {\bibfnamefont {E.}~\bibnamefont
		{Chitambar}}, \bibinfo {author} {\bibfnamefont {S.}~\bibnamefont {Rana}},
	\bibinfo {author} {\bibfnamefont {M.~N.}\ \bibnamefont {Bera}}, \bibinfo
	{author} {\bibfnamefont {A.}~\bibnamefont {Winter}}, \ and\ \bibinfo {author}
	{\bibfnamefont {M.}~\bibnamefont {Lewenstein}},\ }\href {\doibase
	10.1103/PhysRevLett.116.240405} {\bibfield  {journal} {\bibinfo  {journal}
		{Phys. Rev. Lett.}\ }\textbf {\bibinfo {volume} {116}},\ \bibinfo {pages}
	{240405} (\bibinfo {year} {2016}{\natexlab{b}})}\BibitemShut {NoStop}%
\bibitem [{\citenamefont {Veitch}\ \emph {et~al.}(2014)\citenamefont {Veitch},
	\citenamefont {Mousavian}, \citenamefont {Gottesman},\ and\ \citenamefont
	{Emerson}}]{1367-2630-16-1-013009}%
\BibitemOpen
\bibfield  {author} {\bibinfo {author} {\bibfnamefont {V.}~\bibnamefont
		{Veitch}}, \bibinfo {author} {\bibfnamefont {S.~A.~H.}\ \bibnamefont
		{Mousavian}}, \bibinfo {author} {\bibfnamefont {D.}~\bibnamefont
		{Gottesman}}, \ and\ \bibinfo {author} {\bibfnamefont {J.}~\bibnamefont
		{Emerson}},\ }\href {http://stacks.iop.org/1367-2630/16/i=1/a=013009}
{\bibfield  {journal} {\bibinfo  {journal} {New J. Phys.}\ }\textbf {\bibinfo
		{volume} {16}},\ \bibinfo {pages} {013009} (\bibinfo {year}
	{2014})}\BibitemShut {NoStop}%
\bibitem [{\citenamefont {Howard}\ and\ \citenamefont
	{Campbell}(2017)}]{PhysRevLett.118.090501}%
\BibitemOpen
\bibfield  {author} {\bibinfo {author} {\bibfnamefont {M.}~\bibnamefont
		{Howard}}\ and\ \bibinfo {author} {\bibfnamefont {E.}~\bibnamefont
		{Campbell}},\ }\href {\doibase 10.1103/PhysRevLett.118.090501} {\bibfield
	{journal} {\bibinfo  {journal} {Phys. Rev. Lett.}\ }\textbf {\bibinfo
		{volume} {118}},\ \bibinfo {pages} {090501} (\bibinfo {year}
	{2017})}\BibitemShut {NoStop}%
\bibitem [{\citenamefont {Bennett}\ \emph {et~al.}(1999)\citenamefont
	{Bennett}, \citenamefont {DiVincenzo}, \citenamefont {Mor}, \citenamefont
	{Shor}, \citenamefont {Smolin},\ and\ \citenamefont
	{Terhal}}]{bennett1999unextendible}%
\BibitemOpen
\bibfield  {author} {\bibinfo {author} {\bibfnamefont {C.~H.}\ \bibnamefont
		{Bennett}}, \bibinfo {author} {\bibfnamefont {D.~P.}\ \bibnamefont
		{DiVincenzo}}, \bibinfo {author} {\bibfnamefont {T.}~\bibnamefont {Mor}},
	\bibinfo {author} {\bibfnamefont {P.~W.}\ \bibnamefont {Shor}}, \bibinfo
	{author} {\bibfnamefont {J.~A.}\ \bibnamefont {Smolin}}, \ and\ \bibinfo
	{author} {\bibfnamefont {B.~M.}\ \bibnamefont {Terhal}},\ }\href {\doibase
	10.1103/PhysRevLett.82.5385} {\bibfield  {journal} {\bibinfo  {journal}
		{Phys. Rev. Lett.}\ }\textbf {\bibinfo {volume} {82}},\ \bibinfo {pages}
	{5385} (\bibinfo {year} {1999})}\BibitemShut {NoStop}%
\bibitem [{\citenamefont {Reck}\ \emph {et~al.}(1994)\citenamefont {Reck},
	\citenamefont {Zeilinger}, \citenamefont {Bernstein},\ and\ \citenamefont
	{Bertani}}]{reck1994experimental}%
\BibitemOpen
\bibfield  {author} {\bibinfo {author} {\bibfnamefont {M.}~\bibnamefont
		{Reck}}, \bibinfo {author} {\bibfnamefont {A.}~\bibnamefont {Zeilinger}},
	\bibinfo {author} {\bibfnamefont {H.~J.}\ \bibnamefont {Bernstein}}, \ and\
	\bibinfo {author} {\bibfnamefont {P.}~\bibnamefont {Bertani}},\ }\href
{\doibase 10.1103/PhysRevLett.73.58} {\bibfield  {journal} {\bibinfo
		{journal} {Phys. Rev. Lett.}\ }\textbf {\bibinfo {volume} {73}},\ \bibinfo
	{pages} {58} (\bibinfo {year} {1994})}\BibitemShut {NoStop}%
\bibitem [{\citenamefont {Horodecki}\ \emph {et~al.}(2009)\citenamefont
	{Horodecki}, \citenamefont {Horodecki}, \citenamefont {Horodecki},\ and\
	\citenamefont {Horodecki}}]{horodecki2009quantum}%
\BibitemOpen
\bibfield  {author} {\bibinfo {author} {\bibfnamefont {R.}~\bibnamefont
		{Horodecki}}, \bibinfo {author} {\bibfnamefont {P.}~\bibnamefont
		{Horodecki}}, \bibinfo {author} {\bibfnamefont {M.}~\bibnamefont
		{Horodecki}}, \ and\ \bibinfo {author} {\bibfnamefont {K.}~\bibnamefont
		{Horodecki}},\ }\href {\doibase 10.1103/RevModPhys.81.865} {\bibfield
	{journal} {\bibinfo  {journal} {Rev. Mod. Phys.}\ }\textbf {\bibinfo {volume}
		{81}},\ \bibinfo {pages} {865} (\bibinfo {year} {2009})}\BibitemShut
{NoStop}%
\bibitem [{\citenamefont {Nielsen}\ and\ \citenamefont
	{Chuang}(2010)}]{nielsen2010quantum}%
\BibitemOpen
\bibfield  {author} {\bibinfo {author} {\bibfnamefont {M.~A.}\ \bibnamefont
		{Nielsen}}\ and\ \bibinfo {author} {\bibfnamefont {I.~L.}\ \bibnamefont
		{Chuang}},\ }\href@noop {} {\emph {\bibinfo {title} {Quantum computation and
			quantum information}}}\ (\bibinfo  {publisher} {Cambridge university press},\
\bibinfo {year} {2010})\BibitemShut {NoStop}%
\bibitem [{\citenamefont {Pasieka}\ \emph {et~al.}(2009)\citenamefont
	{Pasieka}, \citenamefont {Kribs}, \citenamefont {Laflamme},\ and\
	\citenamefont {Pereira}}]{pasieka2009geometric}%
\BibitemOpen
\bibfield  {author} {\bibinfo {author} {\bibfnamefont {A.}~\bibnamefont
		{Pasieka}}, \bibinfo {author} {\bibfnamefont {D.~W.}\ \bibnamefont {Kribs}},
	\bibinfo {author} {\bibfnamefont {R.}~\bibnamefont {Laflamme}}, \ and\
	\bibinfo {author} {\bibfnamefont {R.}~\bibnamefont {Pereira}},\ }\href
{\doibase 10.1007/s10440-008-9423-z} {\bibfield  {journal} {\bibinfo
		{journal} {Acta Appl. Math.}\ }\textbf {\bibinfo {volume} {108}},\ \bibinfo
	{pages} {697} (\bibinfo {year} {2009})}\BibitemShut {NoStop}%
\bibitem [{\citenamefont {Choi}(1975)}]{choi1975completely}%
\BibitemOpen
\bibfield  {author} {\bibinfo {author} {\bibfnamefont {M.-D.}\ \bibnamefont
		{Choi}},\ }\href {\doibase 10.1016/0024-3795(75)90075-0} {\bibfield
	{journal} {\bibinfo  {journal} {Linear Algebra Appl.}\ }\textbf {\bibinfo
		{volume} {10}},\ \bibinfo {pages} {285 } (\bibinfo {year}
	{1975})}\BibitemShut {NoStop}%
\bibitem [{\citenamefont {Jiang}\ \emph {et~al.}(2013)\citenamefont {Jiang},
	\citenamefont {Luo},\ and\ \citenamefont {Fu}}]{jiang2013channel}%
\BibitemOpen
\bibfield  {author} {\bibinfo {author} {\bibfnamefont {M.}~\bibnamefont
		{Jiang}}, \bibinfo {author} {\bibfnamefont {S.}~\bibnamefont {Luo}}, \ and\
	\bibinfo {author} {\bibfnamefont {S.}~\bibnamefont {Fu}},\ }\href {\doibase
	10.1103/PhysRevA.87.022310} {\bibfield  {journal} {\bibinfo  {journal} {Phys.
			Rev. A}\ }\textbf {\bibinfo {volume} {87}},\ \bibinfo {pages} {022310}
	(\bibinfo {year} {2013})}\BibitemShut {NoStop}%
\bibitem [{\citenamefont {Vidal}(1999)}]{PhysRevLett.83.1046}%
\BibitemOpen
\bibfield  {author} {\bibinfo {author} {\bibfnamefont {G.}~\bibnamefont
		{Vidal}},\ }\href {\doibase 10.1103/PhysRevLett.83.1046} {\bibfield
	{journal} {\bibinfo  {journal} {Phys. Rev. Lett.}\ }\textbf {\bibinfo
		{volume} {83}},\ \bibinfo {pages} {1046} (\bibinfo {year}
	{1999})}\BibitemShut {NoStop}%
\bibitem [{\citenamefont {Vedral}\ and\ \citenamefont
	{Plenio}(1998)}]{vedral1998Entanglement}%
\BibitemOpen
\bibfield  {author} {\bibinfo {author} {\bibfnamefont {V.}~\bibnamefont
		{Vedral}}\ and\ \bibinfo {author} {\bibfnamefont {M.~B.}\ \bibnamefont
		{Plenio}},\ }\href {\doibase 10.1103/PhysRevA.57.1619} {\bibfield  {journal}
	{\bibinfo  {journal} {Phys. Rev. A}\ }\textbf {\bibinfo {volume} {57}},\
	\bibinfo {pages} {1619} (\bibinfo {year} {1998})}\BibitemShut {NoStop}%
\bibitem [{Note3()}]{Note3}%
\BibitemOpen
\bibinfo {note} {For $a=0$, we have $K_1=K_3=0$ and we need only two
	superposition-free Kraus operators (see also the Supplemental Material of
	\cite {baumgratz2014quantifying}). For $a\not =0$, in general we need four
	superposition-free Kraus operators to make the entire operation trace
	preserving. However, the actual transformation is still done by $K_2$ and
	$K_4$. We will encounter a similar case in the proof of theorem \ref
	{theorem:UnitaryReali}}\BibitemShut {NoStop}%
\end{thebibliography}
\end{document}